\documentclass[fleqn,usenatbib,useAMS]{mnras}

\usepackage[T1]{fontenc}

\DeclareRobustCommand{\VAN}[3]{#2}
\let\VANthebibliography\thebibliography
\def\thebibliography{\DeclareRobustCommand{\VAN}[3]{##3}\VANthebibliography}

\usepackage{afterpage}
\usepackage{amsmath}
\usepackage[english]{babel}
\usepackage{booktabs}
\usepackage{dcolumn}
\usepackage{enumerate}
\usepackage{epsfig}
\usepackage{graphicx}
\usepackage{float}
\usepackage{flushend}
\usepackage{indentfirst}
\usepackage[utf8]{inputenc}
\usepackage{longtable} 
\usepackage{lscape}
\usepackage{mathrsfs}
\usepackage{morefloats}
\usepackage{multirow}
\usepackage{natbib}
\usepackage{pdflscape}
\usepackage{rotating,caption}   
\usepackage{supertabular}
\usepackage{tabularx}
\usepackage[absolute]{textpos}
\usepackage{threeparttablex}
\usepackage{url}
\usepackage{xcolor}

\newcommand{\zem}{\ensuremath{z_\mathrm{em}}}

\newcommand{\msun}{\ifmmode \rm{M}_{\odot} \else M$_{\odot}$ \fi}
\newcommand{\zsun}{\ifmmode \rm{Z}_{\odot} \else Z$_{\odot}$ \fi}
\newcommand{\lsun}{\ifmmode \rm{L}_{\odot} \else L$_{\odot}$ \fi}
\newcommand{\rsun}{\ifmmode \rm{R}_{\odot} \else R$_{\odot}$ \fi}
\newcommand{\Hn}{\ifmmode \rm{H} \else H \fi}
\newcommand{\Hm}{\ifmmode \rm{H}^- \else H$^{-}$ \fi}
\newcommand{\Hp}{\ifmmode \rm{H}^+ \else H$^{+}$ \fi}
\newcommand{\Ht}{\ifmmode \rm{H}_2 \else H$_2$ \fi}
\newcommand{\Htp}{\ifmmode \rm{H}_2^+ \else H$_2^{+}$ \fi}
\newcommand{\Hte}{\ifmmode \rm{H}_2^* \else H$_2^{*}$ \fi}
\newcommand{\el}{\ifmmode \rm{e}^- \else e$^{-}$ \fi}
\newcommand{\kms}{\ifmmode \rm{km s}^{-1} \else km s$^{-1}$\fi}
\newcommand{\ergs}{\ifmmode \rm{erg s}^{-1} \else erg s$^{-1}$\fi}
\newcommand{\Lya}{Ly\ensuremath{\alpha}}

\newcommand{\HI}{\mbox{H\,{\sc i}}}

\newcommand{\OI}{\mbox{O\,{\sc i}}}

\newcommand{\OVI}{\mbox{O\,{\sc vi}}}

\newcommand{\CIV}{\mbox{C\,{\sc iv}}}

\newcommand{\CII}{\mbox{C\,{\sc ii}}}

\newcommand{\SiII}{\mbox{Si\,{\sc ii}}}

\newcommand{\SiIV}{\mbox{Si\,{\sc iv}}}

\newcommand{\NV}{\mbox{N\,{\sc v}}}

\newcommand{\RNum}[1]{\uppercase\expandafter{\romannumeral #1\relax}}

\newcommand{\ks}{Kolmogorov-Smirnov}
\newcommand{\ad}{Anderson-Darling}

\title[Reporting a Deficit...]{   
\vskip-0.5cm 
{Reporting a Deficit of Intrinsic {\NV} Absorbers in Core-dominated, Radio-loud Quasars}}

\author[C. Culliton et al.]
{
\parbox{\textwidth}{\vskip-0.5cm  
Chris Culliton$^{1}$,
Amber Roberts$^{1}$,
Bryan DeMarcy$^{1}$,
Sowgat Muzahid$^{1,2,3}$
Rajib Ganguly$^{4}$\thanks{E-mail: ganguly@umich.edu}, 
Jane Charlton$^{1}$, 
Michael Eracleous$^{1,5}$,
Toru Misawa$^{6}$
} 
\vspace*{4pt}\\   
$^{1}$Dept. of Astronomy \& Astrophysics, Penn State University, 525 Davey Lab, 251 Pollock Road, University Park, PA 16802, USA\\
$^{2}$Leiden Observatory, Leiden University, PO Box 9513, NL-2300 RA Leiden, The Netherlands\\
$^{3}$Inter-University Centre for Astronomy and Astrophysics (IUCAA), Post Bag 4, Ganeshkhind, Pune 411 007, India\\
$^{4}$Department of Applied and Engineering Physics, College of Innovation \& Technology, University of Michigan-Flint, \\
~~529 Murchie Science Building, 303 E. Kearsley Street, Flint, MI 48502-1950, USA\\
$^{5}$Institute for Gravitation and the Cosmos, Penn State University, University Park, PA 16802, USA\\
$^{6}$School of Gen. Education, Shinshu University, 3-1-1 Asahi, Matsumoto, Nagano, 390-8621, Japan
}

\date{Accepted XXX. Received YYY; in original form ZZZ}

\pubyear{\the\year{}}

\begin{document}
\label{firstpage}
\pagerange{\pageref{firstpage}--\pageref{lastpage}}
\maketitle

\begin{abstract}
\noindent 
We searched the Hubble Space Telescope Cosmic Origins Spectrograph archive for ultraviolet spectra of 428 AGN to identify intrinsic {\NV} absorption systems. We filtered out Type 2 AGN, blazars, and spectra that do not cover at least part of the velocity window from 5000\,\kms\ blueward to 5000\,\kms redward (hereafter, the ``associated'' region) of the {\NV} emission line. This yielded 175 Type 1 quasars, 34 radio-loud, 133 radio-quiet, and eight unconstrained. Our survey uncovered 77 associated {\NV} systems in the spectra of 48 of these low-redshift quasars. We consider the incidence of intrinsic absorbers as a function of quasar properties $\left(\textrm{optical, radio and X-ray}\right)$. We find a statistically significant dearth of intrinsic {\NV} systems in the spectra of the 34 radio-loud quasars (6\%), compared to 29\% of the 133 radio-quiet quasars containing at least one intrinsic system.  Assuming intrinsic systems are equally likely to occur in radio-loud and radio-quiet quasars and the orientations of the two subsamples are comparable, there is a $0.1\%$ probability of such a deficit occurring by chance in the radio-loud population.  We propose that this deficit of systems is caused by orientation effects.  FIRST radio images are available for 14 of the 33 radio-loud quasars. These show that only three of the 14 radio-loud quasars have lobe-dominated morphologies, whereas 11 of the 14 radio-loud quasars have compact radio morphologies, implying that these quasars are face on, and suggesting that clouds that produce {\NV} absorption are rarely found along the polar axis.
\end{abstract}



\vskip-1cm 
\section{Introduction}
\label{sec:intro}

Quasar outflows can take the form of bipolar jets and/or accretion disk winds, and have been identified as an essential component of galaxy evolution. The feedback from the winds can impart energy and momentum to the interstellar medium (ISM).  This depletes the host galaxy of gas and dust, which disrupts star formation within the galaxy \citep{2004ApJ...600..580G,2005Natur.433..604D,2005ApJ...620L..79S,2007ApJ...665.1038C}. The jets can also enrich the intergalactic medium (IGM) with heavy elements \citep{2004ApJ...608...62S}. To better understand the effects of the quasar outflows on galaxies, it is important to better understand the properties of the outflows.

Absorption produced by outflowing gas that originates from or is physically associated with the quasar is termed ``intrinsic absorption.'' These outflowing absorbers are often observed through the rest frame UV transitions. They can be used to probe many properties of the outflows, including the kinematics, ionization state, and chemical composition of the gas. There are multiple categories of intrinsic absorbers, based upon the width of the absorption trough \citep[see, for example,][]{2004ASPC..311..203H}: broad absorption lines (BALs; $\Delta v>2000$~km s$^{-1}$ by definition), mini-BALs (widths 500~km s$^{-1}<v<2000$~km s$^{-1}$), and narrow absorption lines (NALs; widths $<500$~km s$^{-1}$). Both BALs and mini-BALs are likely to arise from intrinsic structures as intervening structures intercepted by the line of sight to the quasar tend to have small velocity dispersions At the narrow extreme, intrinsic NALs are almost indistinguishable from intervening absorbers, but some can still be shown to be intrinsic in a variety of ways \citep[e.g.,][]{1997ASPC..128...13B}. 

The association between ``intrinsic absorbers'' and quasar outflows suggests that their incidence rate and absorbing properties (e.g., column density, kinematics, ionization) should be connected to the physical properties of the outflows. In turn, the properties of the outflows are likely linked to physical properties of the quasar, like black hole mass and accretion rate, and potentially other observed properties \citep[e.g.,][]{1995ApJ...451..498M,1995ApJ...454L.105M,1999MNRAS.310..476P,2000ApJ...543..686P}.

One historically-curious connection ties the radio-loudness of the quasars \citep{1994AJ....108.1163K,1989AJ.....98.1195K}  to the incidence and kinematics of intrinsic absorbers. For example, \citet{1999ApJ...513..576R} and \citet{2001ApJS..133...53R} found that narrow \CIV\ absorption lines were more common at large velocity offsets ($>5000$~km s$^{-1}$) in the spectra of medium- to high-redshift radio-quiet quasars than in radio-loud quasars.

Furthermore, there is a long history connecting the presence of absorption systems near the quasar redshift to quasar radio-loudness and other radio properties \citep{1979ApJ...234...33W, 1986ApJ...307..504F, 1987AJ.....94..278A, 1987ApJ...320L..75M, 1997ASPC..128...25A, 1997ASPC..128...48B, 1999ApJ...513..576R, 2000ApJ...538...72B, 2000ApJ...544..142G, 2001ApJS..135..227B, 2001ApJ...547..635R, 2001ApJS..133...53R, 2001ApJ...549..133G, 2002AJ....124.2575B, 2003ApJ...599..116V, 2006ApJ...641..210G, 2007ApJS..171....1M, 2008ApJS..177...39T, 2009ApJ...702..911M, 2010ApJS..189...83D, 2013MNRAS.435.1233G, 2013MNRAS.433.1778R, 2019MNRAS.488.5916S}. Hereafter, we call any NAL that appears at an apparent velocity offset less than 5000~km s$^{-1}$\ blueward of the emission redshift an associated absorber. These earlier studies focused on species ({\CIV}, {\OVI}) that could conceivably arise in the host galaxy or in the environment of the host galaxy. The current study focuses on {\NV} absorption, where there is a higher likelihood that the gas is directly and physically connected with the quasar central engine, and the vast majority of them can be demonstrated to be intrinsic \citep[e.g. ][]{2007ApJS..171....1M, cc15,2013MNRAS.435.1233G} rather than intervening like most {\CIV} or {\SiIV} NALs. Exceptions can occur when intervening {\NV} absorbers are produced due to the transverse proximity effect \citep{2007A&A...473..805W}. They can also arise from the circumgactic medium (CGM) of intervening galaxies with flickering AGN activities \citep{2017MNRAS.471.1026S}.  Having strong intervening {\NV} absorbers in low-z is very difficult, although not impossible \citep{2015ApJ...811..132M}. Other than these rare exceptions, almost all {\NV} NALs are in the associated region of the quasar spectra and some can be shown to be intrinsic because of demonstrated partial covering of the quasar continuum source and/or broad emission line region \citep{2003ApJ...598..922G, 2007ApJS..171....1M, 2008ApJ...672..102G, 2010ApJ...722..997W, 2016MNRAS.462.3285P, cc15}.  Their proximity to the quasar is also supported by their extreme supersolar metallicities needed to explain the relatively weak {\Lya} profiles (relative to metal line transitions) as demonstrated by \citet{2010ApJ...722..997W}. Although most {\NV} NALs are intrinsic, {\NV} absorption is not a requirement for intrinsic absorbers.  It should be noted that intrinsic {\NV} NALs may trace specific regions of the quasar accretion disk wind where the appropriate physical conditions exist \citep{cc15}.

With the above considerations in mind, we survey the quasars within the Hubble Space Telescope (HST) Cosmic Origins Spectrograph (COS) archive for {\NV} absorption systems.  Our goal is to create a sample of intrinsic absorption systems in order to probe the outflows of quasars by comparing the properties of the outflows, such as the outflow velocity and equivalent width, to the host quasar properties, such as the X-ray, optical, and radio luminosities.  Once these comparisons have been made, we can investigate how the absorption systems are affected by the host quasar, as well as determine if any property of the quasar affects the likelihood of finding an intrinsic system. This is a follow-up to previous studies using HST/STIS \citep{2013MNRAS.435.1233G} and HST/FOS \citep{2001ApJ...549..133G,1999ApJ...521..572C}.

In {\S}\ref{sec:Sample}, we define the sample.  In {\S}\ref{sec:Methodology}, we explain our method of surveying intrinsic {\NV} systems. In {\S}\ref{sec:Results}, we describe the survey results and correlations found, and compare them to previous samples. We consider potential geometric constraints on the distribution of {\NV}-bearing gas around the quasar central engine in {\S}\ref{sec:discussion}, and summarize the results in {\S}\ref{sec:Summary}.  In this paper, we assume a flat cosmology and use WMAP cosmological parameters of $H_0=71$ km s$^{-1}$ Mpc$^{-1}$, $\Omega _M=0.266$, $\Omega _k=0$, and $\Omega _{\Lambda }=0.734$. In the Appendices (available online) we present all the detailed information of the quasar observations and the quasar properties.

\section{Sample Properties}
\label{sec:Sample}

\begin{figure*}
\includegraphics[width=1.0\textwidth, trim = 100 30 130 80, clip=true]{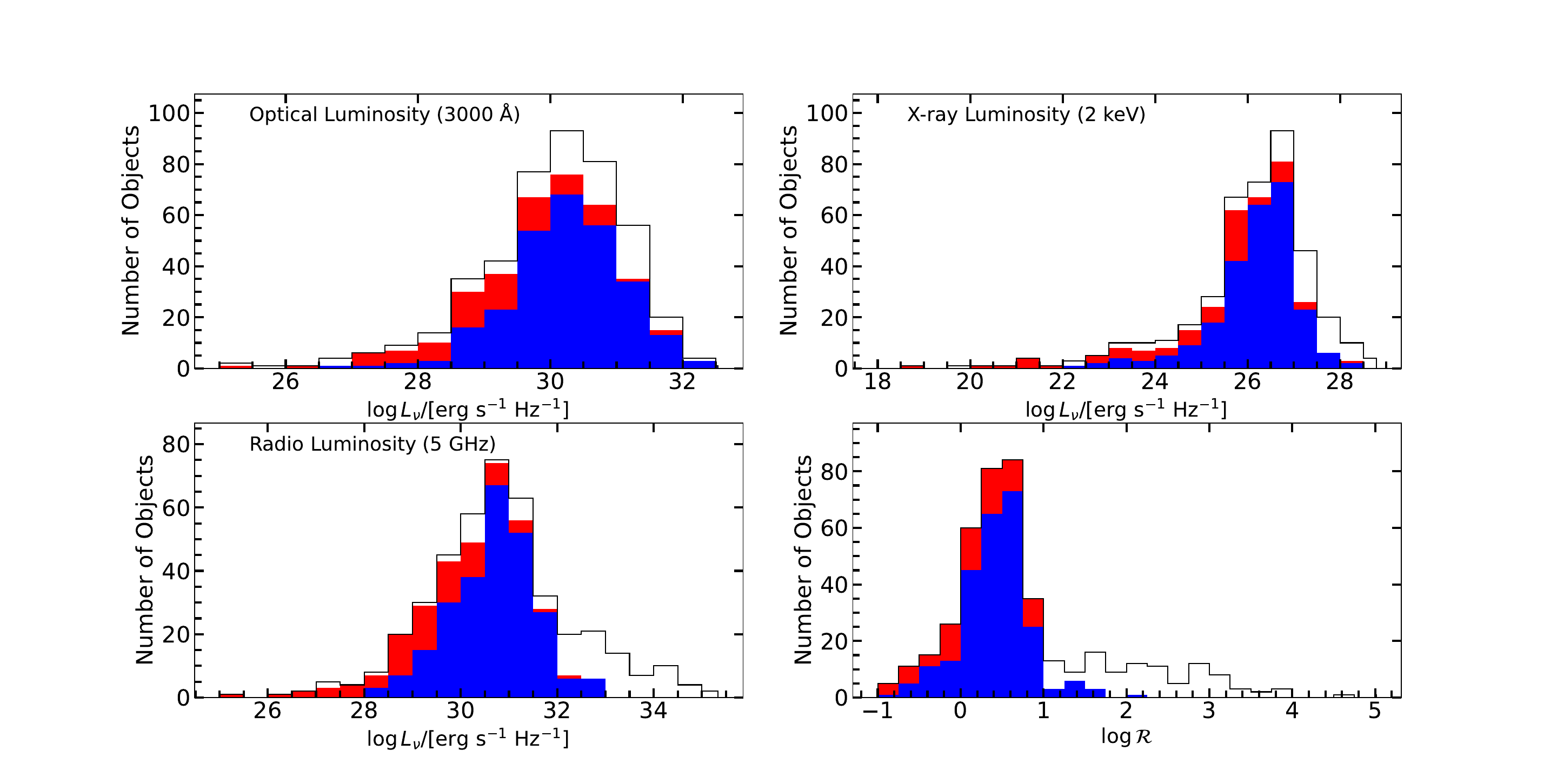} 
\caption{Distribution of optical ($3000$~$\mbox{\AA}$), radio (5~GHz), and X-ray (2~keV) luminosities, and the radio-loudness parameter for the 428 quasars.  The unshaded regions represent radio-loud quasars.  The red shaded regions represent radio-quiet quasars.  The blue shaded regions represent quasars that did not have a radio detection, and so only have an upper limit on $L_{\nu }$(5 GHz).  The three regions are not cumulative.  The combination of the three regions represents all quasars. \label{fig:QuasarProperties3}}
\end{figure*}

The Cosmic Origins Spectrograph (COS) was installed on the Hubble Space Telescope (HST) in May 2009.  We obtained all publicly available spectra of redshift $z<2$ quasars observed by COS between the time of installation and August 2015. The sample of 428 low-redshift quasars in which we searched for {\NV} absorbers had the following selection criteria: (1) archived HST/COS observations obtained with either the G130M and/or the G160M echelle gratings, (2) spectra with signal-to-noise (S/N) ratio per resolution element $\ge$3, and (3) observations released prior to August 2015. The properties of COS and its in-flight operations are characterized by \citet{2011Ap&SS.335..257O} and \citet{2012ApJ...744...60G}. We retrieved the flux-calibrated spectra from the HST/COS archive and reduced them using the STScI CALCOS v.2.12 pipeline. The individual G130M and G160M grating integrations were combined together in flux units weighted by their respective exposure time using the the software developed by the COS GTO team\footnote{http://casa.colorado.edu/$\sim$danforth/science/cos/costools.html}. A full description of this routine is given in \citet{2010ApJ...710..613D}.  The procedures followed for data reduction are described in greater detail in \citet{2011ApJ...730...15N}. 

The COS instrument, with its G130M and G160M gratings, provides medium resolution ($R=20,000$) spectra in a wavelength range of 1150$-$1800~$\mbox{\AA}$, allowing us to search for {\NV}~$\lambda \lambda $1239, 1243~$\mbox{\AA}$ doublets at $z<0.4$.  Due to limitations of HST/COS resolution (the LSF of the COS spectra is not Gaussian, components blend together, etc.), as well as the low signal-to-noise of some of the spectra, and contamination by blends affecting the {\NV} doublets, a partial coverage analysis was not feasible (however, see \citet{2012MNRAS.424L..59M} and \citet{2013MNRAS.431.2885M} for such analysis of mini-BALs detected in HST/COS spectra).  However, based on previous work \citep{2007ApJS..171....1M, 2013MNRAS.435.1233G, cc15} and the fact that the vast majority of {\NV} absorbers are in the associated region, it is safe to assume the intrinsic nature of the NALs in our sample.  

Observational details of the quasars can be found in Table~\ref{table:QuasarObservations}, which includes the original observer, the quasar emission redshift, the wavelengths covered, and the redshifts and velocities covered with respect to the {\NV} emission line. Some of the quasars have gaps in their wavelength coverage. Therefore, each line in the table represents a different wavelength range covered by COS for the given quasar, as well as the associated redshift and velocity range covered. The final columns in the table provide the offset velocity range for which the spectra cover the \NV\ $\lambda\lambda$1238.821, 1242.804 doublet. (Positive offset velocities indicate blueshifted absorption relative to the emission-line redshift.)

In order to determine if the host quasar properties affect either the incidence rate or properties of the intrinsic NALs, we compile monochromatic luminosities in the optical, radio, and soft X-ray bands. The monochromatic luminosities were derived from monochromatic fluxes in the optical, radio, and soft X-ray bands (see below) and the luminosity distance corresponding to the quasar redshift \citep[see, for example, Equation 24 of][]{1999astro.ph..5116H}: \((\nu L_\nu)_\mathrm{rest} = 4 \pi d_L^2 (\nu F_\nu)_\mathrm{obs}\). We make extensive use of Astroquery \citep{2019AJ....157...98G},  a Python package affiliated with The Astropy Project \citep{astropy:2013,astropy:2018,astropy:2022}.

Table~\ref{table:QuasarProperties3} lists the emission redshift (column 2), $z_{em}$, the fraction of the host galaxy contribution to the 3000\,\AA\ luminosity (column 3), the optical luminosity, $L_{\nu}$(3000\,\AA) (column 4), the rest frequency at which the radio luminosity was measured (column 5), the radio spectral index at that frequency (column 6), followed by the 5\,GHz luminosity (column 7), the radio-loudness (defined below, column 8), the soft X-ray spectral index (column 9), the soft X-ray luminosity at 2\,keV (column 10), and the references for the photometry used (column 11).

In the optical band, the monochromatic luminosities are computed typically using the $V$\ or Sloan-$g$\ magnitudes, as compiled by the NASA/IPAC Extragalactic Database (NED). From the optical photometry compiled by NED, we linearly interpolate in \((\log \nu, \log L_\nu )\) to get the rest-frame 3000\,\AA\ luminosity density when possible. Otherwise, we use the photometric point closest to that rest-frame wavelength and adopted a power law spectral index of $\alpha_\circ = -0.5$\ ($L_\nu \propto \nu^{\alpha_\circ}$), consistent with typical values for quasars \citep[$\alpha_\circ = -0.44 \pm 0.1$][]{2001AJ....122..549V}.

In principle, the optical light from low-redshift AGN can be contaminated by a contribution from the host galaxy. To address this possibility, we downloaded the SDSS spectra, when available, which cover most of the Balmer series. We then fit a linear combination of the DESI QSO and galaxy spectral principal components used by the Redrock\footnote{\url{https://github.com/desihub/redrock/}} spectral fitter \citep{2024AJ....168..124A}. While SDSS spectra are not available for all of our targets, the results for those that are available (see Table~\ref{table:QuasarProperties3}) confirm that the contribution to the \(3000\)\,\AA\ luminosity, blueward of the Balmer edge, is not significant.

For the radio band, we take the same approach as the optical band, linearly interpolating the radio luminosity density measurements below rest-frame 30\,GHz in \((\log \nu, \log L_\nu)\). The flux densities are compiled primarily from the \emph{Faint Images of the Sky at 20-cm} survey \citep[FIRST, ][ covering 73\%\ of sources]{1996AJ....112..407G}. When FIRST data are not available, we supplement the compilation with data from the NRAO VLA Sky Survey \citep[NVSS, ][ covering an additional 18\%\ of sources]{1998AJ....115.1693C} the Sydney University Molonglo Sky Survey \citep{1999AJ....117.1578B}, the Parkes catalog \citep{1990PKS...C......0W}, and the Parkes-MIT-NRAO survey \citep{1993AJ....105.1666G}. The frequency of the monochromatic flux is listed in the table (column 5). When juxtaposing fluxes are available, we also report the implied radio spectral index. When there is only one measurement available, we assume a radio spectral index of $\alpha _r=-0.7$\ if the flux measurement is at a frequency below rest-frame 5\,GHz or $\alpha _r=-0.3$\ if the flux measurement is at a frequency above rest-frame 5\,GHz.

Similar to the radio and optical bands, when flux density measurement juxtaposing rest-frame 2\,keV is available through NED, we linearly interpolate those measurements in \((\log \nu, \log L_\nu)\) and report the soft X-ray spectral index, $\alpha_{x}$\ in Table~\ref{table:QuasarProperties3}. When not available, we supplement this through queries of the Chandra XAssist Source List \citep{2003ASPC..295..465P}, the \emph{Chandra} Source Catalog \citep{2024ApJS..274...22E}, the \emph{XMM-Newton} Serendipitous Survey \citep{2020A&A...641A.136W}, the SRG/eROSITA All-Sky Survey \citep{2024A&A...682A..34M}, the NWAY AllWISE Counterparts and Gaia Matches to \emph{ROSAT}/2RXS X-Ray Sources \citep{2018MNRAS.473.4937S}, \emph{ROSAT} All-Sky Survey Bright Source Catalog \citep{1996IAUC.6420....2V} or the WGACAT \citep{1994IAUC.6100....1W}. We convert the reported integrated magnitudes to monochromatic magnitudes assuming a power-law spectrum with a photon index as adopted by the survey. The 2\,keV luminosity densities and the soft X-ray spectral indices are reported in Table~\ref{table:QuasarProperties3}.

To determine which quasars are radio-loud we use the radio-loudness parameter, $\mathcal R$, which is defined as $\mathcal R=f_{\nu }(5$ GHz$)/f_{\nu }(4400$ \AA $)$ \citep{1989AJ.....98.1195K,1994AJ....108.1163K}. The criterion for radio-loudness adopted by \citet{1989AJ.....98.1195K, 1994AJ....108.1163K} is $\mathcal R\geq 10$. However, as we were not always able to obtain the flux at 4400~$\mbox{\AA}$, we instead used a value for radio-loudness parameter of $R=f_{\nu }(5$ GHz$)/f_{\nu }(3000$ \AA $)$.  The criterion for radio-loudness was unchanged: $R \geq 10$. Table~\ref{table:QuasarProperties3} lists $R$\ for each quasar. The optical (3000~$\mbox{\AA}$), radio (5~GHz), and X-ray (2~keV) luminosities and the radio-loudness parameter distributions are presented as histograms in Fig.~\ref{fig:QuasarProperties3}.  The red shaded regions represent radio-quiet quasars, while the blue shaded regions represent quasars that did not have a radio detection, and so only have upper limits on $L_{\nu }$(5 GHz) and $\mathcal R$. Unshaded regions represent definitely radio-loud quasars.  Although most of the quasars without a radio detection can safely be called radio-quiet quasars, there are some quasars without a radio detection whose upper limit to their radio luminosity is such that the radio-loudness parameter upper limit is above ten, making it impossible to determine whether the quasar is a radio-loud or radio-quiet quasar.

\section{Method For Surveying Absorption Systems}
\label{sec:Methodology}

We normalized the quasar spectra using a spline fit on user-defined points, then searched the normalized COS spectra for {\NV} systems over the full wavelength coverage up to $10,000$~{\kms} redward of the peak of the {\NV} emission line.  Our criteria for identifying absorption line systems follows that of \citet{2013MNRAS.435.1233G}.
\begin{enumerate}[1.]
\item The stronger member of the {\NV} doublet (1238.821\,\AA) is detected at $\ge$5$\sigma$ confidence using the unresolved feature detection algorithm of \citet{1993ApJS...87...45S} and \citet{1999ApJS..120...51C}. (This method is optimal for detecting unresolved features, but may miss some broad, shallow features.)
\item The weaker member of the same {\NV} system (1242.804\,\AA) is detected at $\ge$3$\sigma$ confidence.
\item The doublets have very similar kinematic profiles, and a physically consistent ratio of equivalent widths, even when taking potential partial coverage into account.
\item The \HI\ \Lya\ transition, when covered, corroborates the existence of the system.
\end{enumerate}

Although not required, additional lines such as {\CIV}, {\SiIV}, {\OVI}, etc., are used to help further solidify the candidacy of a system when the wavelength coverage permits.  Such corroboration used in conjunction with criteria (3) and (4) can improve the likelihood of the positive identification even if it has kinematic profiles that greatly differ between the {\NV} 1238.821\,\AA\ and 1242.804\,\AA\ lines in the case of one or both of the profiles being blended. Such criteria were particularly useful in the case of line-locked systems.  Line-locked systems occur when the {\NV}~$\lambda$1238.821 transition of one system and the {\NV}~$\lambda$1242.804 transition of a second, slightly higher redshift, system coincide in observed wavelength.  Line-locked systems can exist in chains, with both transitions of at least one system overlapping with the opposite transition of two other systems.  Multiple line-locked systems were seen in our sample, including a chain of four systems in Mrk\,1298.  This is another reason criterion (4) is useful to corroborate an absorption system.

For measurements of the kinematic structure (i.e., absorber velocities), we take an approach of not asserting that the substructure is composed of symmetric (e.g., Gaussian, Voigt) optical depth profiles. Instead, we identify distinct profiles that are separated by regions of unabsorbed flux in the {\NV} \(\lambda1238.821\) line and compute equivalent widths and equivalent-width-weighted velocities for these absorption \emph{systems}, rather than components. Define the equivalent width in a single bin (denoted \(i\)) as \(w_i = (1 - I_i/C_i) \Delta \lambda_i\), where \(C_i\) is the unabsorbed flux and \(I_i\) is the transmitted flux through (or around in the event of partial covering) the absorber, and \(\Delta \lambda_i\) is the width of the bin. Then the total equivalent width and equivalent-width-weighted velocity is:
\[
W_\lambda = \sum_{i_\mathrm{lo}}^{i_\mathrm{hi}} w_i = \sum_{i_\mathrm{lo}}^{i_\mathrm{hi}} \left ( 1 - \frac{I_i}{C_i} \right ) \Delta \lambda_i
\]
\[
\frac{v}{c} = \frac{\bar{\lambda}}{(1 + \zem) \lambda_r} - 1 = \frac{\sum_{i_\mathrm{lo}}^{i_\mathrm{hi}} w_i \lambda_i}{(1 + \zem) \lambda_r W_\lambda} - 1
\]
where \(i_\mathrm{lo}\) and \(i_\mathrm{hi}\) indicate the lower and upper bounds of the integration. Note, as mentioned previously, with this definition of velocity, positive velocities indicate redshifts relative to the emission lines. Detailed notes regarding the results of this survey of systems can be found in Appendix\,\ref{sec:notes}.

\section{Survey Results}
\label{sec:Results}

As mentioned in {\S}\ref{sec:Methodology}, although we conducted a search throughout the entire available spectra to the blue of the quasar redshift and up to $10,000$~{\kms} to the red of the quasar redshift, we found 11 {\NV} systems outside of $\left|v_{offset}\right|\le 5000$~{\kms} relative to the quasar redshift. Of those systems that we found outside the associated region, 10 of the 11 systems were NALs. Most NALs (7 of 10) were high velocity clouds (HVCs) of the Milky Way. Three others were at redshifts consistent with known foreground galaxies, making them unlikely to be physically associated with the
\onecolumn
\begin{longtable}{ c c c c r @{$\pm$} r c}
\caption[Intrinsic System Properties]{Intrinsic System Properties\label{table:SystemProperties3}} \\[-7pt] \hline
&  &  & $v_{offset}^\mathrm{a}$ & \multicolumn{2}{c}{$EW_{N V}^\mathrm{b}$} & [$v_{-}$,$v_{+}$]\\
QSO                              & $z_{\mathrm{em}}$ & $z_{\mathrm{sys}}$ & (km s$^{-1}$)        & \multicolumn{2}{c}{(m\AA)} & (km s$^{-1}$) \\ 
\endfirsthead
\multicolumn{2}{c}{\tablename\ \thetable\ -- continued from previous page} \\
\hline
\endhead
\hline
\multicolumn{7}{r}{Continued on next page} \\
\endfoot
\multicolumn{7}{l}{[a] Negative velocity offsets indicate that the absorber is to the blue of the emission redshift.}\\
\multicolumn{7}{l}{[b] Equivalent width of the $\lambda 1238.821$\ transition.} 
\endlastfoot
\hline
QSO\,J$0012-1022$                & 0.228191 & 0.229368 &     287 &  404 &  86 & [   10,  465] \\
2MASS\,J$0036230.0+431640.2$     & 0.123870 & 0.121177 &    -718 & 2259 &  15 & [-1134, -196] \\
2MASX\,J$004818.99+394111.8$     & 0.134000 & 0.110314 &   -6262 &  117 &   5 & [-6490,-6030] \\
2MASX\,J$004818.99+394111.8$     & 0.134000 & 0.131504 &    -660 &   82 &   6 & [ -850, -530] \\
2MASX\,J$004818.99+394111.8$     & 0.134000 & 0.135213 &     321 &  234 &   7 & [   90,  500] \\
6dFGS\,gJ$015530.0-085704$       & 0.164427 & 0.163306 &    -289 &  609 &  13 & [ -500, -162] \\
6dFGS\,gJ$015530.0-085704$       & 0.164427 & 0.164155 &     -70 &  324 &  11 & [ -162,   45] \\
2MASS\,J$02121832-0737198$       & 0.173920 & 0.174077 &      40 &  457 &  27 & [ -286,  255] \\
Mrk\,1044                        & 0.016450 & 0.012617 &   -1131 &  240 &   5 & [-1200,-1043] \\
Mrk\,595                         & 0.026982 & 0.020359 &   -1933 &  257 &  14 & [-2100,-1809] \\
Mrk\,595                         & 0.026982 & 0.025396 &    -463 &  166 &  14 & [ -610, -280] \\
Mrk\,595                         & 0.026982 & 0.028718 &     507 &   34 &   9 & [  460,  540] \\
LB\,1727                         & 0.104000 & 0.103346 &    -178 &  554 &   2 & [ -300,    0] \\
TON\,0951                        & 0.064000 & 0.064525 &     148 &  277 &   9 & [   60,  250] \\
2XMM\,J$100420.0+051300$         & 0.160116 & 0.136362 &   -6138 & 3777 &  48 & [-8800,-3800] \\
2XMM\,J$100420.0+051300$         & 0.160116 & 0.160853 &     190 &  -58 &  14 & [  -50,  450] \\
TON\,0488                        & 0.256074 & 0.254796 &    -305 &  637 &  16 & [ -470, -180] \\
2MASS\,J$10155924-2748289$       & 0.242200 & 0.242840 &     154 &  427 &  40 & [   -7,  395] \\
Mrk\,1253                        & 0.049986 & 0.048988 &    -285 &   70 &  14 & [ -390, -210] \\
PG\,$1049-005$                   & 0.359900 & 0.341605 &   -4033 &  300 &  20 & [-4160,-3900] \\
{[}VV2006{]}\,J$110313.0+414154$ & 0.403095 & 0.391213 &   -2539 &  118 &  24 & [-2600,-2500] \\
{[}VV2006{]}\,J$110313.0+414154$ & 0.403095 & 0.392633 &   -2235 &  350 &  39 & [-2380,-2110] \\
{[}VV2006{]}\,J$110313.0+414154$ & 0.403095 & 0.399784 &    -708 &  361 &  38 & [ -839, -592] \\
NGC\,3516                        & 0.008836 & 0.007983 &    -253 & 1810 &  10 & [ -603,   30] \\
2MASSi\,J$1118302+402553$        & 0.154338 & 0.154631 &      76 &  216 &   7 & [  -80,  200] \\
ESO\,$265-$G$023$                & 0.056576 & 0.055369 &    -343 &  128 &  13 & [ -500, -200] \\
Mrk\,1298                        & 0.059801 & 0.045770 &   -3969 & 1763 &  32 & [-4735,-3075] \\
Mrk\,1298                        & 0.059801 & 0.052845 &   -1968 & 2683 &  35 & [-2450,-1455] \\
Mrk\,1298                        & 0.059801 & 0.059501 &     -85 & 1583 &  32 & [ -513,  370] \\
6dF\,J$1157588-002221$           & 0.259812 & 0.256537 &    -779 &  415 &  24 & [ -878, -677] \\
NGC\,4051                        & 0.002336 & 0.001080 &    -376 & 2125 &  11 & [ -890,   30] \\
ESO\,$267$                       & 0.014880 & 0.013860 &    -301 & 1293 &  32 & [ -550, -100] \\
2MASS\,J$12072101+2624292$       & 0.322492 & 0.321199 &    -293 &  710 &  43 & [ -537, -131] \\
2MASS\,J$12305003+0115226$       & 0.117000 & 0.099994 &   -4564 &  335 &   3 & [-4750,-4400] \\
2MASS\,J$12305003+0115226$       & 0.117000 & 0.105785 &   -3010 & 1466 &   6 & [-3450,-2600] \\
2MASS\,J$12305003+0115226$       & 0.117000 & 0.109118 &   -2115 & 1485 &   4 & [-2500,-1850] \\
2MASS\,J$12305003+0115226$       & 0.117000 & 0.117261 &      70 &  437 &   3 & [ -100,  300] \\
{[}HB89{]}\,$1339+053$           & 0.265710 & 0.265332 &     -90 &  137 &  17 & [ -165,    5] \\
{[}HB89{]} $1339+053$                & 0.265710 & 0.267077 &     324 &  255 &  50 & [  220,  410] \\
PB\,04055                        & 0.382744 & 0.382578 &     -36 &  157 &  40 & [ -197,   95] \\
Mrk\,279                         & 0.030451 & 0.029147 &    -379 &  996 &  24 & [ -690,  -52] \\
2XMM\,J$135315.8+634546$         & 0.088200 & 0.081687 &   -1794 & 1244 &  22 & [-2353,-1386] \\
2XMM\,J$135315.8+634546$         & 0.088200 & 0.084853 &    -922 & 1906 &  18 & [-1386, -477] \\
SDSS\,J$135625.55+251523.7$      & 0.164009 & 0.162546 &    -377 &  103 &  14 & [ -510, -318] \\
SDSS\,J$135625.55+251523.7$      & 0.164009 & 0.163033 &    -251 &  205 &  10 & [ -318, -190] \\
2XMM\,J$141348.3+440014$         & 0.089600 & 0.083608 &   -1649 & 5574 &  20 & [-2522, -900] \\
2XMM\,J$141348.3+440014$         & 0.089600 & 0.089244 &     -98 &  216 &  15 & [ -200,  -23] \\
2XMM\,J$141348.3+440014$         & 0.089600 & 0.089787 &      51 &  171 &  11 & [  -23,  159] \\
NGC\,5548                        & 0.017175 & 0.013337 &   -1131 &  551 &  14 & [-1396, -870] \\
NGC\,5548                        & 0.017175 & 0.015613 &    -460 & 1887 &   7 & [ -870, -111] \\
NGC\,5548                        & 0.017175 & 0.017252 &      23 &  620 &   4 & [ -111,  145] \\
QSO\,J$1435+3604$                & 0.429945 & 0.427258 &    -563 &  646 &  99 & [ -795, -270] \\
Mrk\,0478                        & 0.079055 & 0.069106 &   -2764 &   87 &  12 & [-2870,-2620] \\
Mrk\,0478                        & 0.079055 & 0.071107 &   -2208 &  381 &  14 & [-2440,-2044] \\
Mrk\,0478                        & 0.079055 & 0.078577 &    -133 &   78 &   8 & [ -223,  -37] \\
2MASX\,J$14510879+2709272$       & 0.065000 & 0.064179 &    -231 & 1600 &  26 & [ -600,  100] \\
2MASX\,J$15085291+6814074$       & 0.058637 & 0.055747 &    -818 &  414 &  18 & [ -900, -730] \\
2MASX\,J$15085291+6814074$       & 0.058637 & 0.056845 &    -508 & 1100 &  24 & [ -701, -335] \\
2MASX\,J$15085291+6814074$       & 0.058637 & 0.060397 &     498 &   10 &  34 & [ -190,  102] \\
2MASS\,J$1521396.7+033729.2$     & 0.126500 & 0.124562 &    -516 & 1489 &  30 & [ -780, -240] \\
Mrk\,290                         & 0.030227 & 0.028963 &    -368 &  430 &  10 & [ -675, -120] \\
2MASX\,J$18324966+5340219$       & 0.045000 & 0.043762 &    -355 &  145 &   7 & [ -420, -300] \\
2MASX\,J$18324966+5340219$       & 0.045000 & 0.044276 &    -208 &  152 &  10 & [ -300, -130] \\
Mrk\,509                         & 0.034397 & 0.033311 &    -315 &  357 &   2 & [ -450, -161] \\
Mrk\,509                         & 0.034397 & 0.034533 &      40 &  523 &   2 & [ -161,  320] \\
Mrk\,1513                        & 0.062977 & 0.057532 &   -1536 &  251 &   5 & [-1600,-1452] \\
Mrk\,1513                        & 0.062977 & 0.063143 &      47 &   47 &   6 & [ -132,   95] \\
Q\,$2135-147$                    & 0.200470 & 0.199582 &    -222 &  353 &   6 & [ -320, -109] \\
Q\,$2135-147$                    & 0.200470 & 0.200903 &     108 & 1141 &   8 & [ -109,  350] \\
Mrk\,304                         & 0.065762 & 0.054711 &   -3109 &  284 &   9 & [-3400,-2813] \\
Mrk\,304                         & 0.065762 & 0.061238 &   -1273 & 5506 &  16 & [-2530, -200] \\
UGC\,12163                       & 0.024684 & 0.024240 &    -130 & 1437 &  15 & [ -400,  175] \\
MR\,$2251-178$                   & 0.063980 & 0.062921 &    -298 &  300 &   7 & [ -550,   50] \\
6dF\,J$2301520-55083$            & 0.141000 & 0.138436 &    -674 &  395 &   8 & [ -800, -580] \\
6dF\,J$2301520-55083$            & 0.141000 & 0.139147 &    -487 &  421 &   8 & [ -580, -380] \\
NGC\,7469                        & 0.016317 & 0.009797 &   -1923 &  373 &   6 & [-2070,-1748] \\
NGC\,7469                        & 0.016317 & 0.013258 &    -902 &  181 &   7 & [-1060, -755] \\
NGC\,7469                        & 0.016317 & 0.014176 &    -631 &   26 &   9 & [ -755, -300] \\
2MASS\,J$23215113-7026441$       & 0.300000 & 0.301334 &     308 &  126 &  14 & [  175,  380] \\
\hline
\end{longtable}
\twocolumn

\noindent background quasar. The first was in PG\,0953$+$414 at $z_{abs}=0.0681$, $v_{offset}=43,018$~{\kms}.  The second was in SDSS\,J110406.94$+$314111.4 at $z_{abs}=0.2365$, $v_{offset}=44,447$~{\kms}.  The third was in 1ES\,1553+113 at $z_{abs}=0.1876$, $v_{offset}=47,834$~{\kms}. 

It should be noted that other studies have found intervening systems in some of the quasars listed in this survey, such as {\NV} absorption at $z=0.20701$ toward HE\,0226$-$4110 as reported in \citet{2011ApJ...743..180S}, and {\NV} absorption at $z=0.227$ toward HB\,0107$-$025 as reported in \citet{2014ApJ...784....5M}. However, these systems had strong absorption lines in lower ionization species, while {\NV} was barely detected.
With our focus on {\NV} absorbers, and the $5\sigma $ and $3\sigma $ detection limits on {\NV}$\lambda 1238.821$ and {\NV}$\lambda 1242.804$ respectively, these systems could not be included in our survey due to not meeting our minimum detection criteria. 

There are only two potentially intrinsic systems with central velocities found outside the associated region that can reasonably be assumed to be physically associated with the host quasar. The system at $-6262$\,\kms\ in the spectrum of 2MASX\,J$004818.99+394111.8$\ is a NAL that is apparently line-locked with two other systems at $-660$\,\kms\ and $+321$\,\kms. The system at $-6138$~{\kms} in the spectrum of quasar 2XMM\,J$100420.0+051300$\ is a BAL with a width of nearly $5000$~{\kms}.  As mentioned in \S\ref{sec:intro}, absorption features with such large widths are more readily explained through an intrinsic origin. Hereafter, we designate the associated {\NV} systems, the apparently line-locked systems, and the BAL system as \emph{intrinsic} systems, consistent with previous results regarding using {\NV} to trace intrinsic absorption \citep[e.g.,][]{2013MNRAS.435.1233G}. We list the intrinsic systems in our sample in Table~\ref{table:SystemProperties3}, along with the emission redshift of the quasar, the absorption redshift of the system, the offset velocity of the system, the equivalent width of the {\NV}~$\lambda$1238.821 line, the error in the equivalent width, and the velocity limits of integration for the system. 

Due to the dearth of intrinsic NAL systems containing {\NV} doublet lines with $|v_{offset}|>5000$~{\kms}, it is therefore assumed that intrinsic {\NV} NAL systems are not found outside of the adopted velocity cutoff defining the associated region.  Including quasars that do not cover this region in their spectra would therefore bias our sample.  As such, we created a subsample of quasars using only those quasars for which there is at least partial wavelength coverage, if not complete wavelength coverage, of the $|v_{offset}|<5000$~{\kms} associated region.

There are 16 quasars that only partially cover the adopted associated region in this subsample.  They are: PG\,0003+158, HE\,0153$-$4520, ESO\,031$-$G$-$008, 2MASSI\,J0803592+433258, QSO\,J0910+1014, SDSS\,J092554.70+400414.1, QSO\,J1009+0713, TON\,1187, SDSS\,J110406.94+314111.4, ESO\,265$-$G23, QSO\,B1309+3531, TON\,236, SDSS\,J113137.16+155645.3, ESO\,113$-$45, 2MASX\,J21362313$-$6224008, and 2MASS\,J23215113$-$7026441.  PG\,0003+158, SDSS\,J092554.70+400414.1, and QSO\,B1309+3531, are radio-loud quasars.  The other 13 are radio-quiet.  Despite only partially covering the associated region, one intrinsic {\NV} system was found in the spectra ESO\,265$-$G23, while two intrinsic {\NV} systems were found in the spectrum of 2MASS\,J23215113$-$7026441. 

Physically, we note that there is evidence that the velocity range that is deemed ``associated'' can and probably should be luminosity-dependent \citep[e.g.,][]{2008ApJ...672..102G,2002ApJ...569..641L}. For the average 3000\,\AA\ luminosity of this sample, an inspection of Figure~1 in \citet{2002ApJ...569..641L} implies that intrinsic absorbers lie within 10,000\,\kms\ (or less at lower luminosities). This is entirely consistent with the paucity of \NV\ absorbers found at larger velocity offsets.

In addition, we also required that this subsample contain only Type 1 quasars.  Type 2 quasar spectra do not have a bright, featureless continuum or broad emission lines.  Type 2 quasar spectra tend to be reddened and also include a large contribution from starlight in the host galaxy, making these objects physically distinct from Type 1 quasars.  This would both make NALs more difficult to detect and bias the sample.  Although the Type 2 quasars were searched, no {\NV} absorption lines were found. Including such objects in our survey would therefore bias our sample, reducing the number of systems found per quasar.

\begin{figure*}
\includegraphics[width=1.0\textwidth, trim = 100 30 130 80, clip=true]{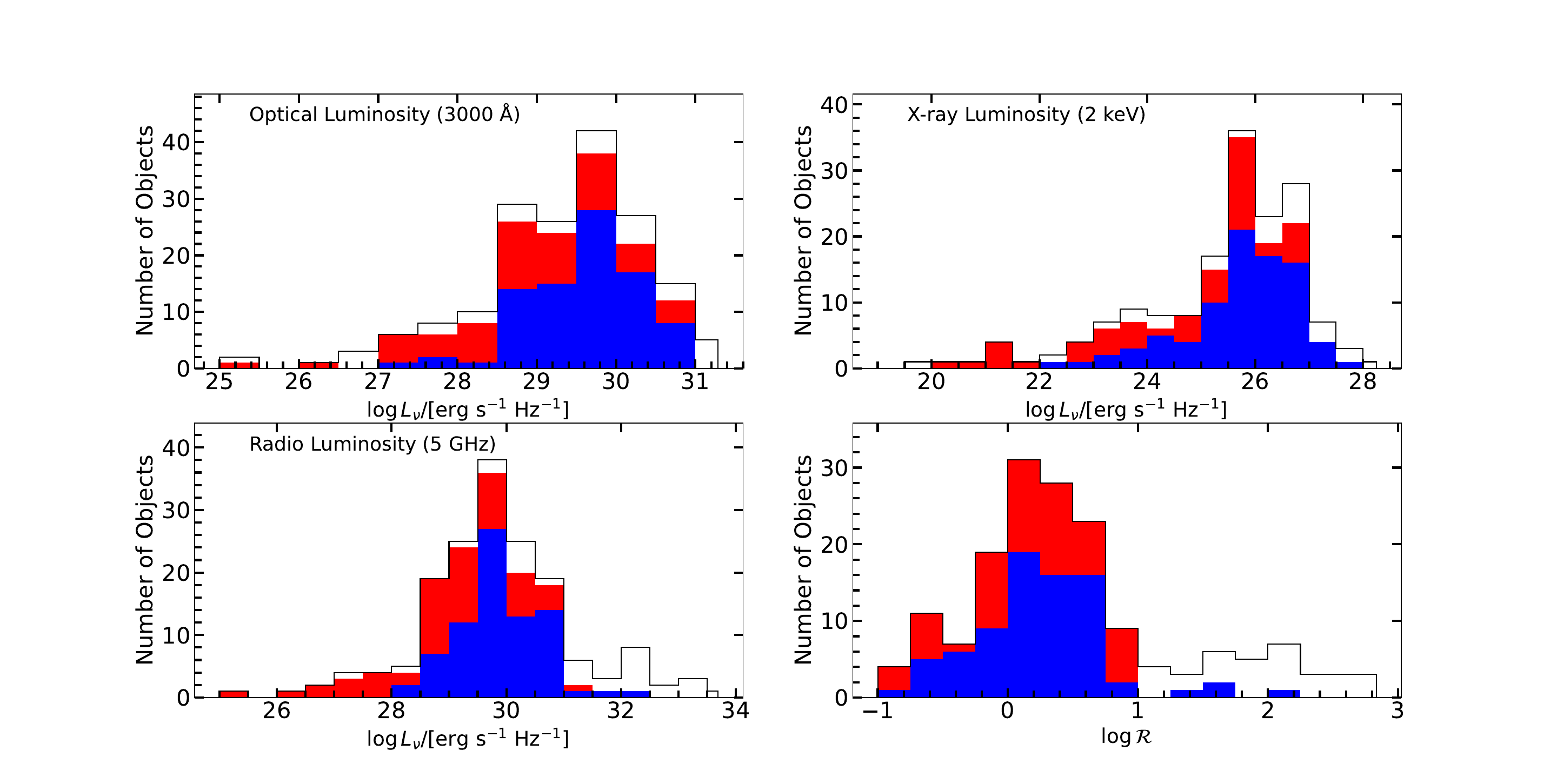}
\caption[Subsample Quasar Properties]{Distribution of optical ($3000$~$\mbox{\AA}$), radio (5~GHz), and X-ray (2~keV) luminosities, and the radio-loudness parameter for quasars whose spectra at least partially cover the {\NV} associated region.  These quasars form our subsample.  The unshaded regions represent radio-loud quasars.  The red shaded regions represent radio-quiet quasars.  The blue shaded regions represent quasars that did not have a radio detection, and so only have an upper limit on $L_{\nu }$(5 GHz).  The three regions are not cumulative.  The combination of the three regions represents all quasars. \label{fig:subsample}}
\end{figure*}

Finally, we also excluded BL Lac objects from the sample, as the continuum we see in their spectra comes from the jet, not the accretion disk.  Therefore, any intrinsic systems found would lie between us and the jet, rather than in the vicinity of the accretion disk, and so would arise in a different environment than the intrinsic {\NV} systems we are surveying.  It is of note that the number of BL Lac objects removed from the subsample (14) is nearly equivalent to the number of radio-loud quasars in the subsample (16).

The subsample created using these criteria includes 175 of the original 428 quasars.  Figure~\ref{fig:subsample} is the same as Figure~\ref{fig:QuasarProperties3}, but uses only the quasars in this subsample.  Comparing Figures~\ref{fig:QuasarProperties3} and \ref{fig:subsample}, we note that the original sample has higher optical luminosities than the \NV\ subsample.  These high luminosity quasars are at a higher redshift than is required for the {\NV} associated region to be covered in the spectra, due to the HST/COS gratings.  These high optical luminosity quasars that were removed often corresponded to high X-ray and radio luminosities, as well.  Additionally, due to the removal of BL Lac objects from this sample, some of the objects with the largest radio-loudness parameter were removed.

\subsection{Fraction of Quasars With an Intrinsic System}
In this new subsample, we found a $5\sigma $ detection of {\NV} was possible over at least 95{\%} of the associated regions of the quasars at an observed equivalent width of 83~m$\mbox{\AA}$.  We found 77 intrinsic NALs\footnote{The intrinsic label for a NAL is given to those lying within 5000\,\kms\ of the quasar redshift \emph{and} featuring NV absorption.} and 1 intrinsic BAL within the spectra of the 175 quasars in this subsample. Twenty-seven quasars contained only one intrinsic system, eleven quasars had two intrinsic systems within their spectra, eight quasars had three intrinsic systems, and one quasar had four intrinsic systems. There are 47 quasars ($27_{-5}^{+6}$\%, 90\%\ confidence) that had at least one intrinsic system.  This fraction of quasars containing an intrinsic {\NV} system is consistent with those found in variability studies, such as \citet{2004ApJ...601..715N} or \citet{2004ApJ...613..129W}, who found intrinsic systems in 25\% and 27\% of quasars, respectively.

We explore how often multiple intrinsic NALs appear towards the same quasar, compared to random expectations. With an average of $\frac{77}{175}=0.44$\,\NV\ NALs/quasar, a Poisson distribution predicts $112/175$ ($49/175$, $10/175$, $1/175$) quasars with 0 (1, 2, 3) \NV\ NALs. This is inconsistent with the observed distribution, suggesting that all sightlines do not have an equal likelihood of hosting \NV\ NALs in their spectra. 

The high-redshift survey from \citet{cc15} found no evidence that there are preferred sightlines to quasars with multiple intrinsic absorbers.  Additionally, although \citet{cc15} used different methodology to find intrinsic systems, the fraction of quasars with at least one intrinsic {\NV} system in that survey, $11-12${\%}, is less than half of what we found at low redshift.  However, if that study had simply used the presence of {\NV} absorption as an indicator of intrinsic absorption as is done in this paper, then that high-redshift sample would have exhibited a higher fraction of quasars with intrinsic absorption.  Of the 73 quasars in the high-redshift sample, 26 quasars had at least one associated system with {\NV} absorption present, resulting in a $36_{-7}^{+8}$\%\ (90\%\ confidence) occurrence rate for {\NV} absorption, which is consistent with the results of this low-redshift sample.

\subsection{System Properties}
\label{sec:sysprop}

\begin{figure*}
\includegraphics[width=1.0\textwidth, trim = 110 25 130 80, clip=true]{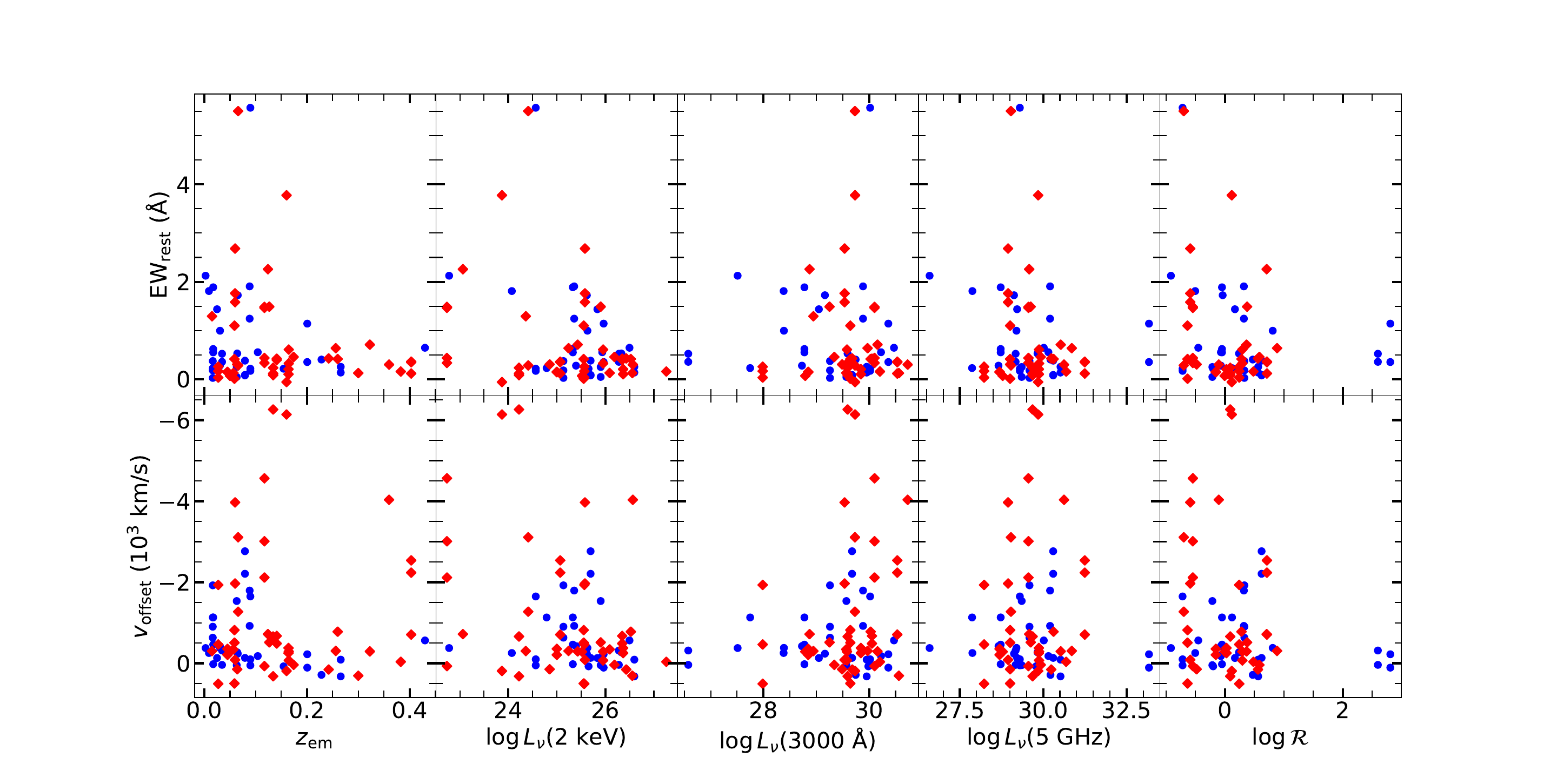}
\caption[System and Quasar Property Comparison]{Comparison of the properties of the intrinsic systems ($EW$ and $v_{offset}$) vs the properties of the host quasars ($z_{Em}$, $L(2$~keV), $L(3000$~$\mbox{\AA}$), $L(5$~GHz), $\mathcal R$).  Blue points represent quasars that have a radio detection, while red points represents quasars that did not, and so only have an upper limit for the radio luminosity. \label{fig:SystemProperties}}
\end{figure*}

Using this survey, we examined the statistical properties of the {\NV} absorption systems and their relationships to the quasars hosting them. The systems were described with their {\NV} equivalent widths and velocity offsets. From the quasars, we considered their optical, X-ray, and radio luminosities, as well as redshift and radio-loudness parameters. For quasars with intrinsic {\NV} absorbers, Figure~\ref{fig:SystemProperties} shows plots of the absorber properties against the quasar properties. We performed Spearman rank correlation tests on each combination of parameters, but found no significant trends according to those tests. We further compared the equivalent width and velocity offset distribution for two subsamples divided at the median value of each quasar and system property. These are reported in Table~\ref{table:statSysProp}. The systems were divided based on the median value of a given property so there were equal numbers in the ``low'' and ``high'' bins. The two groups were then compared using both a {\ks} test and an {\ad} test to see if the quasar property appears to relate to the given absorber property (equivalent width or velocity offset). 

\begin{table*}
\begin{threeparttable}
\caption[Statistical Tests for Correlations Between System Properties]{Statistical Tests for Correlations Between System Properties \label{table:statSysProp}}
\begin{tabular}{| c | c | c | }
\hline\hline
Quasar 									& $EW$\tnote{b}			& $v_{offset}$\tnote{b} \\
Property\tnote{a} 							& $(\mbox{\AA})$ 			& ({\kms}) \\
\hline
High $z$ vs Low $z$ &  0.14, 0.81 &  0.20, 0.34 \\
med $[z] =$ 0.0882 & -0.45, $>$0.25 &  0.40,  0.23 \\ \hline
High $L_\nu(3000$~$\mbox{\AA})$ vs Low $L_\nu(3000$~$\mbox{\AA})$ &  0.16, 0.74 &  0.21, 0.37 \\
$\log \text{med}[L_\nu(3000$~$\mbox{\AA})] =$ 29.64 & -0.60, $>$0.25 &  0.47,  0.21 \\ \hline
High $L_\nu(5$~$GHz)$ vs Low $L_\nu(5$~$GHz)$ &  0.19, 0.43 &  0.18, 0.52 \\
$\log \text{med}[L_\nu(5$~$GHz)] =$ 29.56 & -0.22, $>$0.25 & -0.54, $>$0.25 \\ \hline
High $L_\nu(2$~$keV)$ vs Low $L_\nu(2$~$keV)$ &  0.18, 0.55 &  0.29, 0.08 \\
$\log \text{med}[L_\nu(2$~$keV)] =$ 25.55 & -0.14, $>$0.25 &  1.89,  0.05 \\ \hline
High $\mathcal{R}$ vs Low $\mathcal{R}$ &  0.17, 0.53 &  0.10, 0.97 \\
$\log \text{med}[\mathcal{R}] =$ 0.09 & -0.39, $>$0.25 & -0.93, $>$0.25 \\ \hline
System property & & \\ \hline
High $EW$ vs Low $EW$ & N/A, N/A &  0.18, 0.56 \\
$\text{med}[EW] = 373$\,m\AA & N/A, N/A &  0.32, $>$0.25 \\ \hline
High $v_{offset}$ vs Low $v_{offset}$ &  0.21, 0.39 & N/A, N/A \\
$\text{med}[v_{offset}] = -376$\,km~s$^{-1}$ &  1.64,  0.07 & N/A, N/A \\ \hline
\end{tabular}
\begin{tablenotes}
\item [a] The quasar property that is used to divide the subsample into two groups based on the median of that property. The two groups are then compared using both the {\ks} and {\ad} tests to see if the quasar property affects the given property (equivalent width or velocity offset) of the system. See \S\ref{sec:sysprop} of the text for details.
\item [b] The quasar property being compared.  Results in each cell are in two lines.  The first line is the KS-statistic and the $p$-value of the KS-test.  The second line is the Anderson-Darling statistic and the $p$-value of the Anderson-Darling test.  Tests with a chance probability $p<0.01$ are in bold. 
\end{tablenotes}
\end{threeparttable}
\end{table*}

\begin{table}
\begin{threeparttable}
\caption[Statistical Tests for Correlations Between Quasar Properties]{Statistical Tests for Correlations Between Quasar Properties  \label{table:statDefVsNoDef}}
\begin{tabular}{| c | c | c | }
\hline\hline
Quasar & KS & AD \\
Property\tnote{a} &  &  \\
\hline
%
$z$     &  0.18, 0.21 &  0.07, $>$0.25 \\ \hline
$L(3000$~$\mbox{\AA})$ &  0.14, 0.48 &  0.36, 0.24 \\ \hline
$L(5$~GHz$)$ &  0.14, 0.47 &  0.22, $>$0.25 \\ \hline
$L(2$~keV$)$ &  0.19, 0.17 &  2.47, 0.03 \\ \hline
$\mathcal{R}$ &  0.20, 0.12 &  2.34, 0.04 \\ \hline

\end{tabular}
\begin{tablenotes}
\item Note: All tests are performed comparing those quasars with at least one intrinsic {\NV} system in their spectra to those quasars without any intrinsic {\NV} systems.
\item [a] The quasar property being compared.  Results in reported in two cells.  The first cell is the KS-statistic and the $p$-value of the KS-test.  The second cell is the Anderson-Darling statistic and the $p$-value of the Anderson-Darling test. 
See \S\ref{sec:sysprop} of the text for details. \\
\end{tablenotes}
\end{threeparttable}
\end{table}

In addition to comparing the properties of quasars that host {\NV} systems, it is also useful to compare the properties of quasars that host systems with those that do not in order to determine if {\NV} systems have a preference for a specific type of quasar. The statistical results of the {\ks} (KS) and {\ad} (AD) tests in which we compare the properties of quasars that host intrinsic {\NV} systems vs those without an intrinsic system can be seen in Table~\ref{table:statDefVsNoDef}. There are no statistically significant differences between the populations in any parameter. 

\begin{figure*}
\includegraphics[width=1.0\textwidth, trim = 80 30 130 80, clip=true]{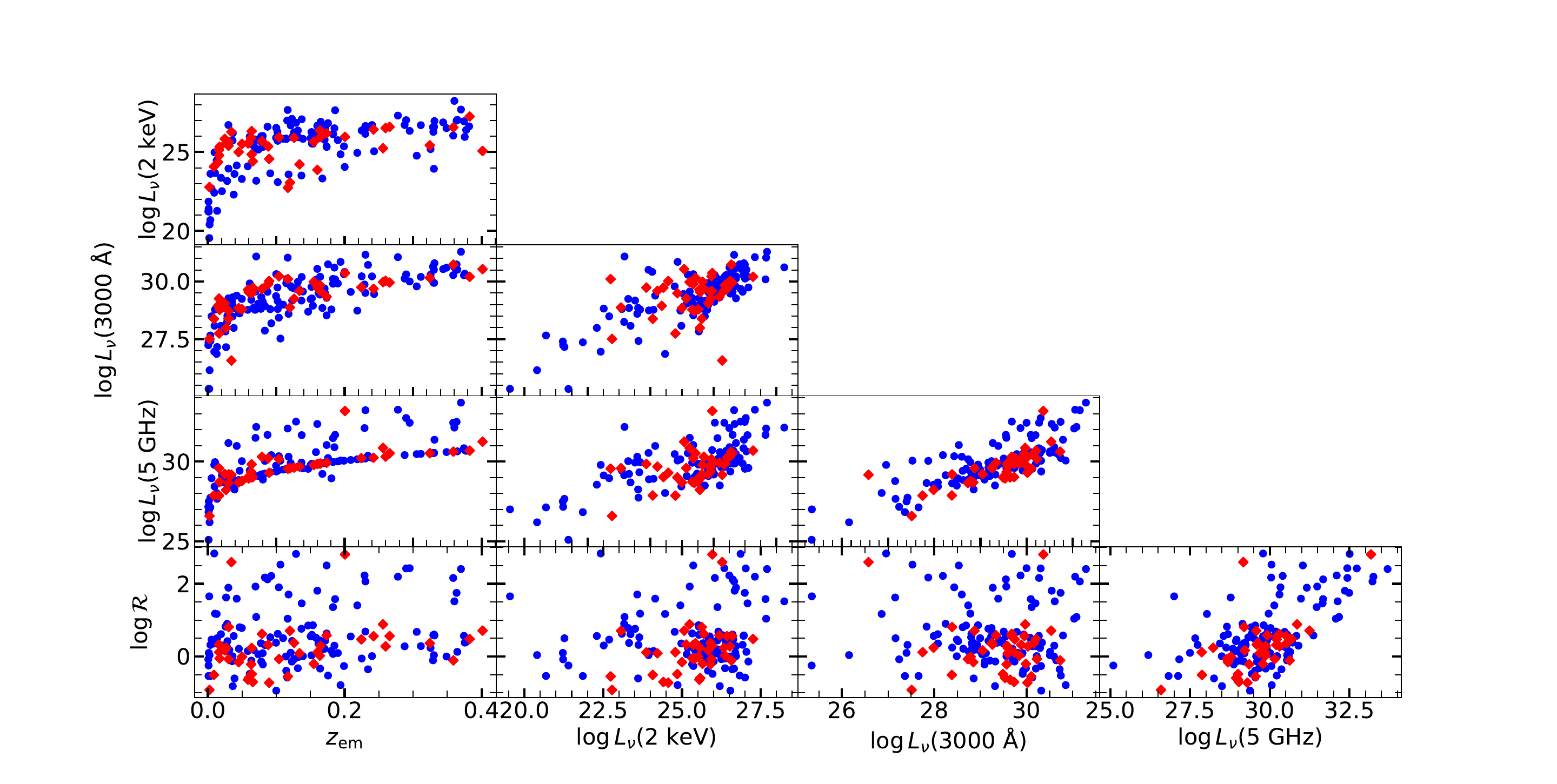}
\caption[Pair Plot of Quasar Properties]{A pair plot of quasar properties ($z_{Em}$, $L(X-ray)$, $L(optical)$, $L(radio)$, $\mathcal R$) in the survey. The blue points represent quasars without an intrinsic NAL, and the red points represent those quasars containing at least one intrinsic system.  All quasars within our sample that have an intrinsic NAL present in their spectra are radio-quiet quasars. \label{fig:AllQuasarProperties}}
\end{figure*}

\begin{figure}\begin{center}
\centering
\includegraphics[width=0.5\textwidth, trim = 400 30 415 80, clip=true]{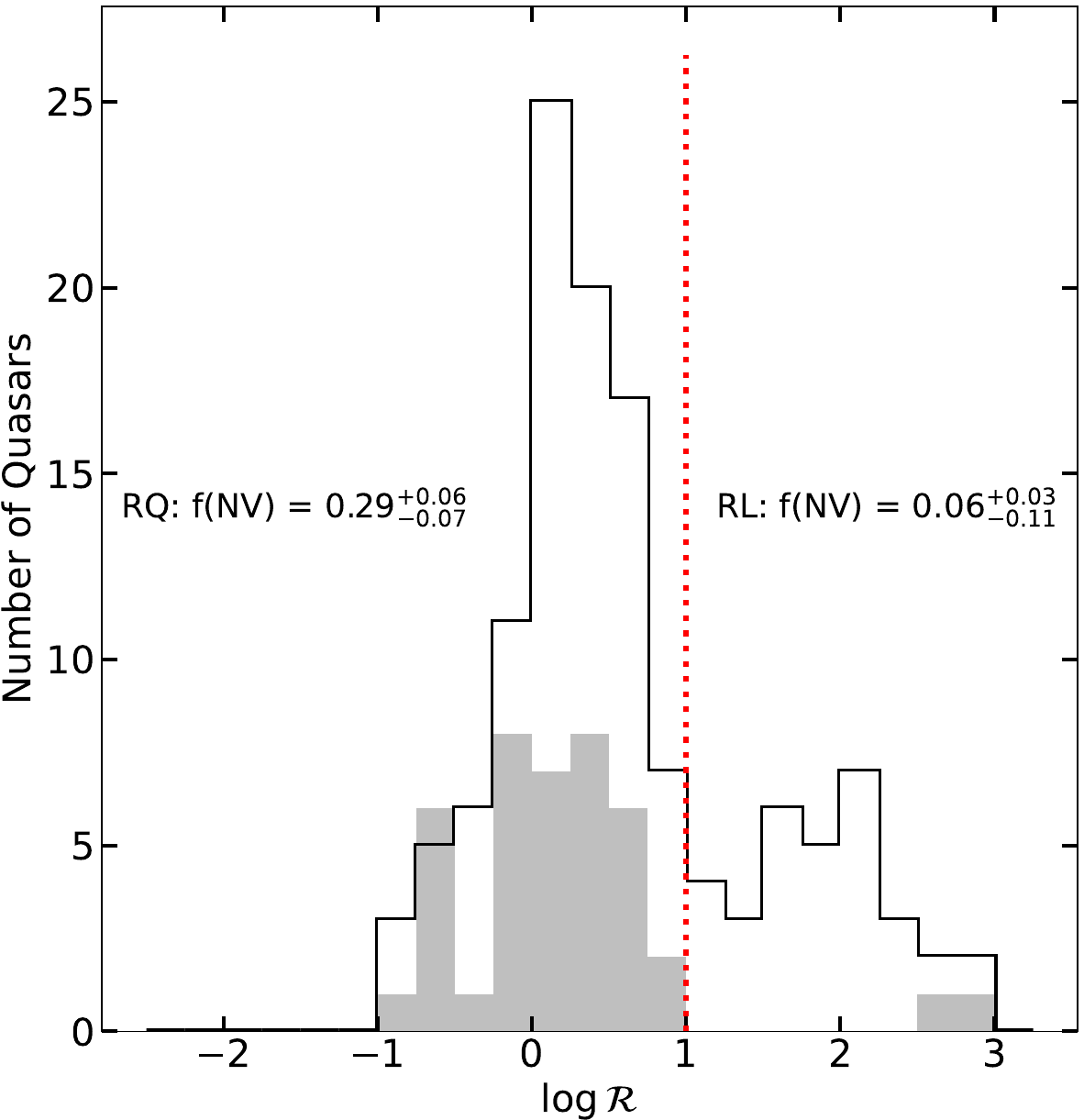}
\caption[Distribution of Radio-loudness Parameter]{Distribution of the radio-loudness parameter, $\mathcal{R}$, for the quasars in our subsample. The black shaded region corresponds to the quasars with one or more intrinsic {\NV} systems, while the non-shaded region corresponds to quasars without an intrinsic {\NV} system. The delimiter between radio-loud quasars and radio-quiet quasars is $\mathcal{R}=10$, denoted in the figure with a dashed red vertical line.  Only two radio-loud quasars feature an intrinsic {\NV} system. \label{fig:RValueDist}}
\end{center}\end{figure}

In Figure~\ref{fig:AllQuasarProperties}, we compare the properties of quasars that host intrinsic {\NV} systems (red) to those without an intrinsic system (blue). The five quasar properties plotted are redshift, X-ray luminosity, optical luminosity, radio luminosity, and radio-loudness parameter.  No obvious differences between the quasars that host an intrinsic {\NV} system, and those that do not, appear except for those containing radio luminosity and radio-loudness parameter.  In those plots, quasars containing an intrinsic system have lower radio luminosities and lower radio-loudness parameters.  This trend is highlighted in Figure~\ref{fig:RValueDist}, where we plot the distribution of quasars with respect to radio-loudness parameter. The shaded region represents those quasars with at least one intrinsic system, while the unshaded region represents those without intrinsic systems.  The red vertical line represents the criterion for radio-loudness adopted by \citet{1989AJ.....98.1195K,1994AJ....108.1163K}, $\mathcal R\geq 10$.

As can be seen, of the 175 quasars in our sample with the associated region at least partially covered, 34 of them are radio-loud, 133 are radio-quiet, and 8 are either insufficiently constrained or have no radio data available. Of the 167 definitely radio-constrained, 39/133 ($29^{+6}_{-7}$\%) radio-quiet quasars have intrinsic {\NV} absorption, while only 2/34 ($6^{+3}_{-11}$\%) radio-loud quasars have intrinsic {\NV} absorption. If intrinsic systems are equally likely to occur in radio-loud and radio-quiet quasars, the binomial probability states that there is less than a 0.1\% chance for us to find only two intrinsic systems toward the 34 radio-loud quasars. 

\subsection{Sample Comparison}

We compare the results from this sample with our companion survey of high-redshift quasars \citep{cc15}.  The \citet{cc15} sample has a higher quasar redshift range of $1.4$--$5$, while the redshifts of quasars in this final subsample were all $z<0.5$.  Another difference is that the quasars of the high-redshift sample were more optically luminous than the majority of our sample, without much overlap.  Thus, any differences between the two samples could potentially be due to either redshift evolution or to variations in optical luminosity.

\begin{figure}\begin{center}
\centering
\includegraphics[width=0.5\textwidth, trim=95 320 700 80, clip=True]{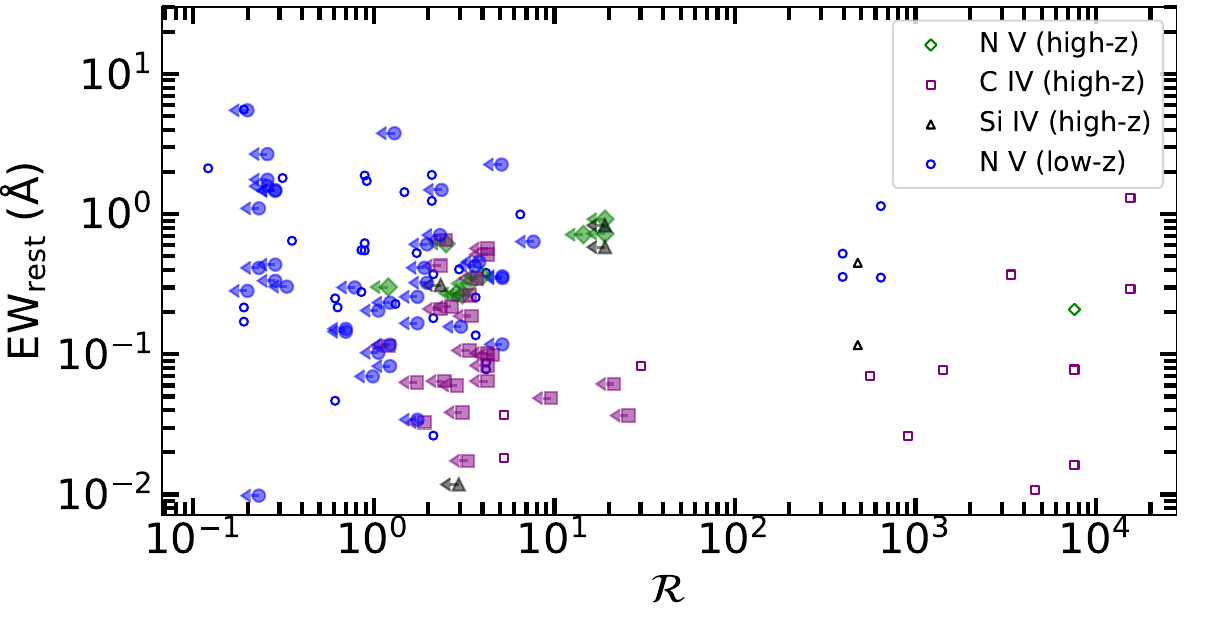}
\caption[Equivalent Width vs. Radio-loudness Parameter]{The rest-frame equivalent width, $EW_{rest}$, vs. the radio-loudness parameter, $\mathcal{R}$ of the current low-redshift sample and the high-redshift sample from \citet{cc15}.  Unfilled symbols represent those quasars that have a defined radio-loudness parameter, while filled symbols are those quasars without radio detections, and so whose radio-loudness parameter is only an upper limit.  The {\NV} systems found in this paper are plotted as blue circles.  The green diamonds, purple squares, and black triangles represent systems from \citet{cc15}, measuring their equivalent width using {\NV}, {\CIV}, and {\SiIV}, respectively.
\label{fig:EWvsRBoth}}
\end{center}\end{figure}

\begin{figure}\begin{center}
\centering
\includegraphics[width=0.5\textwidth, trim=110 320 700 80, clip=True]{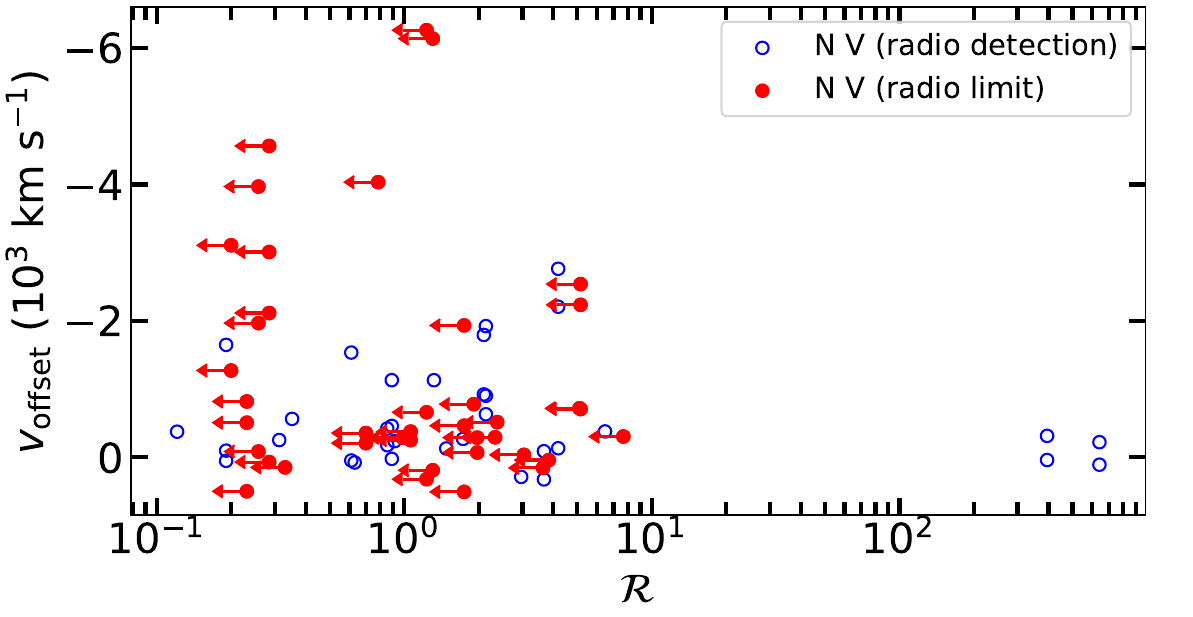}
\caption[Velocity Offset vs. Radio-loudness Parameter]{The velocity offset vs. radio-loudness parameter of the current sample.  Blue symbols represent those quasars that have a defined radio-loudness parameter, while red symbols are those quasars without radio detections, and so whose radio-loudness parameter is only an upper limit.  There appears to be a distinct lack of systems with large $\mathcal{R}$ and large $\mathrm{v_{offset}}$.  \label{fig:VvsRHST}}
\end{center}\end{figure}

\begin{figure}\begin{center}
\centering
\includegraphics[width=0.5\textwidth, trim=100 320 700 75, clip=True]{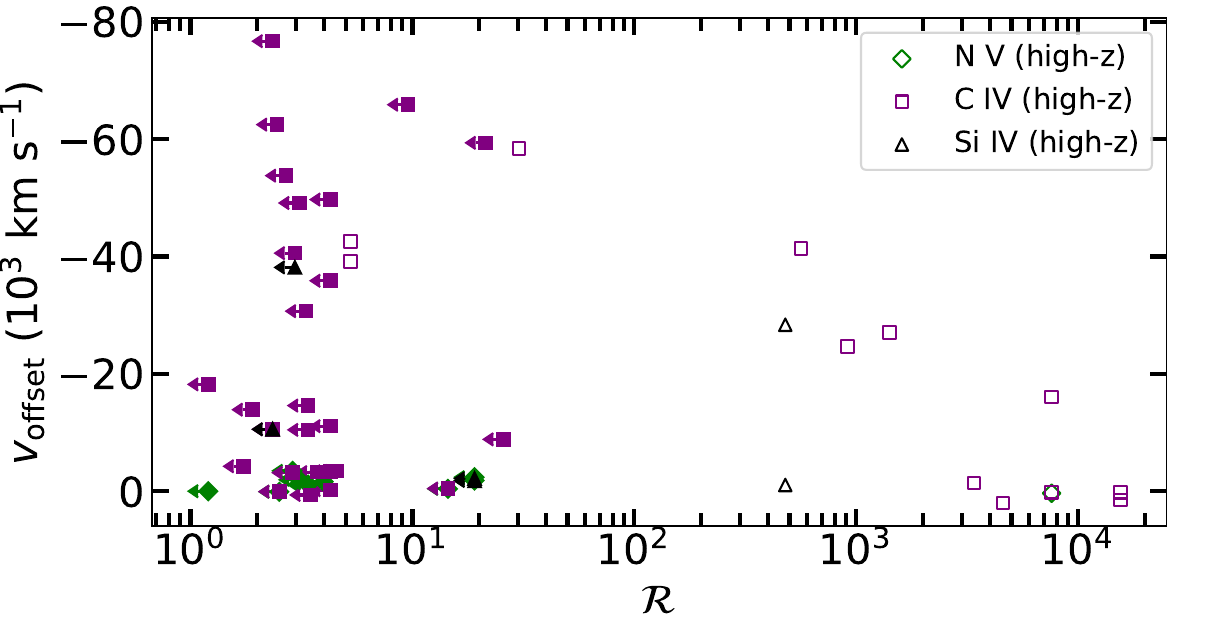}
\caption[VLT Sample Velocity Offset vs Radio-Loudness]{Same as Figure~\ref{fig:VvsRHST}, but for the high-redshift VLT sample, with symbols as in Figure~\ref{fig:EWvsRBoth}.  Although the ranges over which both axes are covered are much different, just as in Figure~\ref{fig:VvsRHST} there is a distinct lack of systems with large $\mathcal{R}$ and $\mathrm{v_{offset}}$.  \label{fig:VvsRVLT}}
\end{center}\end{figure}

The first trend that the high-redshift and low-redshift samples have in common is the relationship between rest-frame equivalent width of {\NV} \(\lambda1238.821\) and radio-loudness parameter (See Figure~\ref{fig:EWvsRBoth}).  In both the high-redshift and the low-redshift samples there is a statistical dearth of systems with high radio-loudness and equivalent width, regardless of the ion used to detect the system. Only four out of 78 {\NV} intrinsic systems lie at with $\mathcal{R}>100$ in this low-redshift sample.  However, there are 70 systems residing in the spectra of $\mathcal{R}<10$\ quasars. Of those 70, 26\% had $EW_{rest}>1\mbox{\AA}$. In the high-redshift sample, some systems were found in radio-loud quasars, though aside from one outlier, there are no systems with $EW_{rest}>0.3$~$\mbox{\AA}$ and $\mathcal{R}>20$.  Additionally, the outlier was found using {\CIV}, not {\NV}. Even though there are some systems from the high-redshift sample that were found in radio-loud quasars using the definition from \citet{1989AJ.....98.1195K, 1994AJ....108.1163K}, there is only one intrinsic {\NV} system found above $\mathcal{R}>20$, and that system had a small rest-frame equivalent width (0.21~$\mbox{\AA}$).  Additionally, those {\NV} systems found between $\mathcal{R}=10$ and $\mathcal{R}=20$ have only upper limits on the radio-loudness parameter.  
They could potentially reside within radio-quiet quasars.  Finally, although it does appear that the systems from the low-redshift sample generally have a greater equivalent width, those systems from the high-redshift sample that do have a relatively large equivalent width are almost all from radio-quiet quasars. 

A second similarity between the high-redshift and low-redshift samples is seen in velocity offset vs. radio-loudness parameter (See Figures~\ref{fig:VvsRHST} and \ref{fig:VvsRVLT}). Although they are on greatly different scales, both the high-redshift and low-redshift samples show a lack of systems at both high radio-loudness parameter and high velocity offset.  As mentioned in the previous paragraph, there is a dearth of {\NV} intrinsic systems within the spectra of low-redshift  $\mathcal{R}>10$ quasars. If we instead sub-divide the radio-quiet quasars (that have radio detections), there are 15 systems within the spectra of quasars $1<\mathcal{R}<10$, but only five of those systems have $|v_{offset}|>1000$~{\kms}, and only eight of those systems have $|v_{offset}|>500$~{\kms}. Only two have $|v_{offset}|>2000$~{\kms}. There are 32 systems within the spectra of quasars with $\mathcal{R}<1$\ (combining both radio-detected and those with sufficiently constrained radio non-detection limits).  Of those 32 systems, 44{\%} have $|v_{offset}|>500$~{\kms}, 34{\%} have $|v_{offset}|>1000$~{\kms}, and 19{\%} have $|v_{offset}|>2000$~{\kms}. Therefore, quasars with higher radio-loudness parameters do not appear to host systems with velocities as high as those with lower radio-loudness parameters.  When we compare velocity offset to optical luminosity instead (see Figure~\ref{fig:SystemProperties}), a third similarity arises, where the systems with the highest velocity offsets also have the highest optical luminosities, as has been noted previously in \citet{2008ApJ...672..102G}. This is consistent with velocity and radio-loudness parameter being anti-correlated and suggests that increased luminosity allows systems to be accelerated to higher velocities.  

\section{Discussion}
\label{sec:discussion}
Our primary result from searching the archive of HST/COS observations of quasars/AGN is that only two (6\%) of the 34 radio-loud quasars host four of the 78 intrinsic {\NV} systems, whereas their occurrence in radio-quiet quasars is 73 systems in 39/133 (29\%) quasars. [Note: Two of the eight quasars either without radio information or insufficiently constraining non-detection limits have an intrinsic system.] We compare the distributions of radio-loudness parameter for quasars with and without intrinsic {\NV} systems in Figure~\ref{fig:RValueDist}. Despite the obvious dearth of radio-loud quasars with intrinsic {\NV} systems, a KS test yields an 8\%\ probability that the two $\log \mathcal{R}$\ distributions are the same, which is insufficient to claim any difference. This might suggest that there is nothing peculiar about the radio properties of quasars with and devoid of intrinsic {\NV} systems.  However, if we take the incidence rate of \NV\ absorbers in radio-quiet quasars ($\frac{73}{133}=0.55$\,\NV/quasar) and apply it to the 34 radio-loud quasars, we would statistically expect 19 \NV\ absorption systems. The Poisson probability of expecting 19 and finding four or less is \(4 \times 10^{-5}\), suggesting something significant. There are two ways to interpret this: (1) there is a bias in the sample of radio-loud quasars; or (2) intrinsic {\NV} systems avoid low-redshift radio-loud quasars.

We first address the question of bias. In \S\ref{sec:first}, we explore whether a bias could be due to orientation effects in the sample of radio-loud quasars. In \S\ref{sec:cos} we consider if a bias could be introduced by the manner in which COS observers select objects. Finally, \S\ref{sec:answer}, we discuss possible physical reasons for differences between the radio-loud and radio-quiet subsamples.

\subsection{FIRST Radio Observations}
\label{sec:first}

\begin{figure*}\begin{center}
\centering
\includegraphics[width=0.8\textwidth]{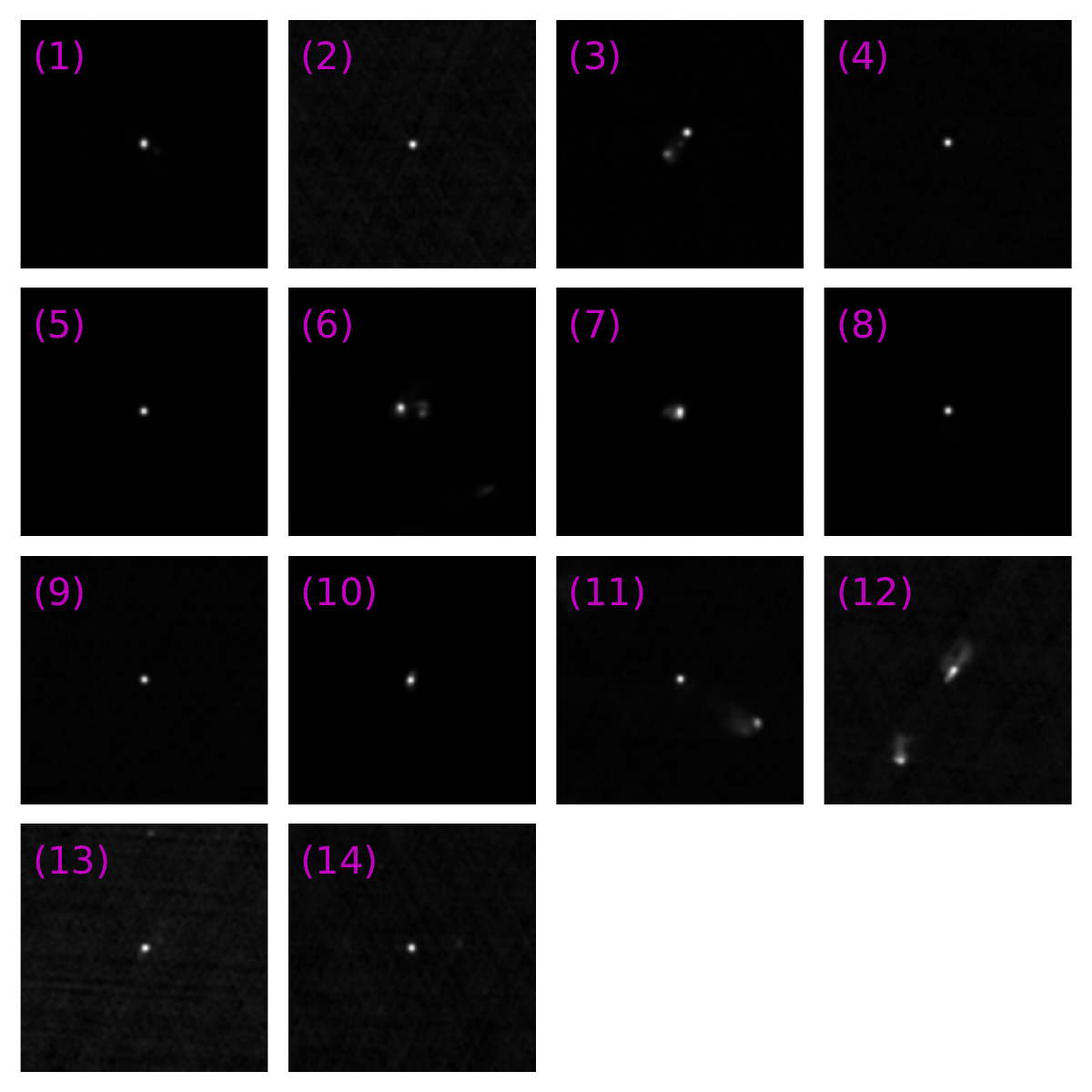}
\caption[Radio Image Mosaic]{Radio images of the 14 radio-loud quasars that were observed by FIRST.  The indices in the top left corners correspond to the following quasars: (1) MRK 1501, (2) WISEA\,J$080650.80+523506.9$, (3) SBS\,$0924+606$B, (4) WISEA\,J$103118.52+505335.8$, (5) MRK 0421, (6) NGC 3690, (7) TON 0580, (8) MRK 0231, (9) FBQS\,J$131217.7+351521$, (10) FBQS\,$J142700.4+234800$, (11) {[}HB89{]}\,$1425+267$, (12) SBS 1503+570, (13) SBS 1648+547, (14) 2MASX\,J$17034196+5813443$.
All radio images are $4' \times 4'$, centered on the quasar.  Of those 14 radio-loud quasars, all but three (3, 6, 12) exhibit compact radio morphologies, based on their core-to-lobe flux ratios, which implies that these quasars have low inclination angles, and are relatively face on.}
\label{fig:radioImages}
\end{center}\end{figure*}

In order to determine if our findings were consistent with orientation effects, we looked at FIRST radio images of the radio-loud quasars in our sample.  Of the 34 radio-loud quasars that cover the {\NV} emission line, 14 were covered by FIRST. As shown in Figure~\ref{fig:radioImages}, 11 of the 14 radio-loud quasars (quasars 1, 2, 4, 5, 7--11, 13, 14 in Figure~\ref{fig:radioImages}) have core-dominated morphologies, with (at most) minor flux contributions from jets.  Only three of the FIRST images (quasars 3, 6, 12) show sufficiently powerful radio lobes that would suggest an edge-on sightline to the quasar.  This means that the majority of radio-loud quasars in our sample of 34 likely have sightlines that reside relatively close to the poles. Though the number statistics are small, this would seem to be inconsistent with a randomly oriented population, in which we would expect more lobe-dominated than core-dominated objects. This apparent bias favoring core-dominated objects has remained even after the 14 BL Lac objects, which are known to be core-dominated, have been removed from the sample, as was described in {\S}~\ref{sec:Results}. 

While the radio-loud, FIRST-detected, subsample is relatively small, we can try to estimate the likelihood that the observed deficit of {\NV}-absorbing radio-loud objects is due to orientation effects. For this estimate, we stipulate that radio-loud and radio-quiet quasars have the same structure in terms of the geometric distribution of {\NV}-absorbing gas. So, we can use the observation that 39/133 ($\sim30$\%) of the radio-quiet quasars show {\NV} absorption as a starting point. If the 30\%\ fraction of quasars having {\NV} absorption applies equally to both core-dominated and lobe-dominated radio-loud subsamples, then we would expect that $\sim1$\ lobe-dominated quasar and $\sim4$\ core-dominated quasars would show {\NV} absorption lines. Of the FIRST-detected quasars shown, only FBQS\,J$101000.7+300321$\ (quasar 3 in Figure~\ref{fig:radioImages}), a lobe-dominated quasar showed intrinsic {\NV} absorption. The Poisson probabilities of \([1, 0]\) in the {[}lobe-, core-{]} dominated subsamples when \([1,4]\) are expected is \(\sim[40, 2]\)\%.

However, it still remains to be seen whether the 30\%\ incidence should apply equally to both lobe- and core-dominated objects. If this is the result of an orientation effect, we would expect there to be a higher incidence among lobe-dominated objects and a lower incidence among core-dominated objects.  With only $3$\ lobe-dominated quasars in the sample, it is not possible to determine if there is an orientation preference in the incidence of {\NV} absorbing gas that leads to the overall deficit in the radio-loud subsample.

\subsection{Possible Bias in HST/COS Archive}
\label{sec:cos}

A potential reason for the bias toward core-dominated objects is one introduced by sample selection due to the science goals of HST/COS observers. Those science goals would naturally select Type 1 objects that are less affected by reddening that would attenuate the UV spectrum. Type 1 objects likely lie within $60^\circ$\ of the accretion disk rotation axis or jet axis \citep[e.g.,][]{2016ApJ...830...82M,2015MNRAS.454.1202S,2005AA...436..817A,1989ApJ...336..606B} so that the sightlines do not pass through the dusty torus. The division between core-dominated and lobe-dominated objects occurs at a lower angle, around $42^\circ$\ \citep[see, for example, ][]{2016ApJ...830...82M}. Taken at face value, this implies that truly random orientations of Type 1 quasars resulting in a core-dominated morphology ($0$--$42^\circ$) are equally likely as those resulting in a lobe-dominated morphology ($42$--$60^\circ$). The Poisson probability of observing a 11-to-3 ratio of core-dominated to lobe-dominated objects when a 1-to-1 ratio is expected is about 2\%. Hence, the null hypothesis that the observed ratio is consistent with the expected ratio is not ruled out at high confidence. That is, it is possible, though seemingly unlikely, that there may not be a bias considering only Type 1 objects.

Another possible bias could arise if the radio-loud quasars were selected using radio-surveys that are cross-matched with optical surveys \citep[e.g., FIRST and SDSS,][]{2002AJ....124.2364I,2015ApJ...801...26H}. In such a case, a quasar with an optical component in the same location as the radio component would make for an easier identification than for a radio-loud quasar with radio lobes that are not positionally coincident with the optical component of the quasar.  

\subsection{Why is there a difference in the incidence of {\NV} absorption between radio-loud and radio-quiet quasars?}
\label{sec:answer}
In the absence of obvious observational biases, we turn to physical causes for the absence of intrinsic {\NV} absorption in low-redshift, radio-loud quasars.  This could point toward a physical difference between radio-loud and radio-quiet quasars in terms of the existence and location of {\NV}-bearing gas. Of course, radio-loud quasars must have \emph{some} {\NV}-bearing gas as evidenced by the presence of the {\NV}$\lambda$1239 emission line. So the dearth of {\NV} intrinsic \emph{absorption} points to either a structural/geometric difference or a temporal/evolutionary difference between radio-loud and radio-quiet quasars.

The most obvious structural/geometric difference between radio-loud and radio-quiet quasars is the power of the jet. The seminal explanation for the production of the jet came from \citet{1977MNRAS.179..433B}, who identified both the black hole spin and field strength of a magnetized accretion disc as important components. This scenario can be impacted by differences in black hole mass \citep[e.g.,][]{2011MNRAS.416..917C}, or accretion rate \citep[e.g.,][]{2017ApJ...849....4S}, but seems to be triggered by mergers \citep[e.g.,][]{2015ApJ...806..147C}.

Having both a rapidly spinning and massive black hole would affect the radiation field that is being emitted by the quasar. We know from observations that radio-loud quasars have harder X-ray spectra \citep[e.g.,][]{2011ApJ...726...20M, 2011ApJS..196....2S}. The spin both permits the inner edge of the magnetized disc to enter the ergosphere of the black hole and provides an additional rotational component that the field must cope with. Since the inner edge of the disc is closer than for a non-rotating black hole, the disc can be hotter, yielding a harder SED for radio-loud quasars. In turn, this could change the manner in which an outflow can be launched/driven from the disc.  A harder spectrum could have the effect of ionizing the gas to a higher degree. 

\begin{figure}\begin{center}
\centering
\includegraphics[width=0.5\textwidth]{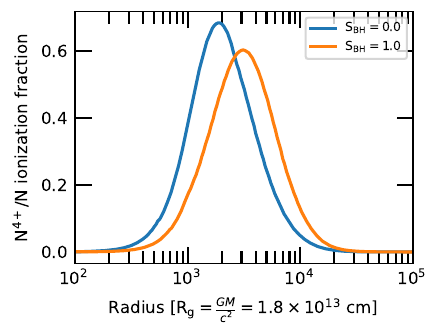}
\caption[{N\(^{4+}\)} ionization fraction]{{N\(^{4+}\)} ionization fraction on the surface of a standard thin-disk. Using Cloudy photoionization models, we have computed the ionization fraction on the surface of a standard thin-disk model. In these simulations, we assume a \(10^8\,\mathrm{M}_\odot\) black hole accreting at a rate of \(1.8\,\mathrm{M}_\odot\,\mathrm{yr}^{-1}\). The blue curve is for a non-rotating hole with the thin-disk of \citet{1973A&A....24..337S}. The orange curve is for a maximally spinning block hole with the thin-disk of \citet{1973blho.conf..343N}. The harder ionizing spectrum from the inner disk around a rotating-black hole results in {N\(^{4+}\)} appearing at larger radii.}
\label{fig:nvfrac}
\end{center}\end{figure}

If we stipulate that the gas launched from the disk around a spinning black hole is subject to a harder ionizing spectrum, we can speculate on the consequences of this for the solid angle subtended by the outflow. The solid angle is subject to the fraction of driving photons that come from the parts of the disk from below where the gas is launched. A larger fraction results in a larger vertical component of the radiative force per unit mass relative to the radial component. With a harder ionizing spectrum, but similar luminosity, the {\NV}-bearing gas would be launched from farther out in the disk. For example, using Cloudy photoionization simulations \citep{Chatzikos2023,Gunasekera2023} in a standard thin-disk model \citep{1973A&A....24..337S,1973blho.conf..343N}, the \(\mathrm{N^{4+}/N}\) ionization fraction peaks around \(\sim2000\,\mathrm{R}_g\) on the disk surface around a non-rotating \(10^8\,M_\odot\) black hole accreting at \(1.8\,M_\odot\,\mathrm{yr}^{-1}\) (Fig.~\ref{fig:nvfrac}). However, maximizing the black hole spin will increase that radius to \(\sim3000\,\mathrm{R}_g\). At larger radii, the gas will experience a smaller vertical driving force relative to radial driving potentially yielding a smaller solid angle. Subtending a smaller solid angle, then, would result in only observing the outflow in absorption for the most extreme inclination angles.

These orientation effects could be supported by differences seen in low- and high-redshift radio-loud quasars.  Although only a few intrinsic {\NV} systems were found in high-redshift, radio-loud quasars through the partial coverage method in \citet{cc15}, the fact remains that \emph{some} intrinsic systems were found. However, if we simply use {\NV} absorption as an indicator of a system being intrinsic, 9 of the 19 radio-loud quasars in that survey had associated {\NV} absorption.  If there really is a difference between the low- and high-redshift intrinsic system occurrence rate (that is not observational bias), then the answer is likely tied to the black hole mass (statistically higher in high-redshift samples), the mass accretion rate (statistically higher in high-redshift samples), and also possibly the mass fuelling rate (potentially peaking at $\zem \sim 2$\ as evidenced by the the peak of the quasar number density and the star-formation rate). Radio-loud objects at high-redshift still have massive spinning black holes, and {\NV}-bearing gas is still launched at larger radii. However, since the accretion rate is larger, so too are the luminosity and the local radiation pressure from where the {\NV}-bearing gas is launched. Hence, it is conceivable that the gas is launched with a larger vertical component of velocity and {\NV} systems subsequently subtend a larger solid angle relative to the inner disk/black hole. A larger solid angle at high-redshift would then imply that the {\NV}-bearing gas can be observed with a higher incidence in absorption.

Unfortunately, it is difficult to determine the primary effect responsible for the distinct lack of {\NV} systems within radio-loud quasars.  Each of the above structural or evolutionary differences seem to imply orientation effects in which radio-loud quasars with relatively low-inclination angle sightlines are less likely to exhibit {\NV} absorption than those with high-inclination angle sightlines.

\section{Summary}
\label{sec:Summary}

In this survey, we found that 175 of the 428 low-redshift quasars in the HST/COS archive (as of August 2015) were Type 1 quasars with sufficient wavelength coverage that their associated regions are at least partially covered by HST/COS spectra.  Within those 175 quasars, we found 78 intrinsic {\NV} systems, including 77 NALs and 1 BAL.  We described their properties, demographics, and relation of their properties to those of the quasars hosting them.  Our findings and conclusions are as follows:
\begin{enumerate}[1.]
\item Within our sample, $\sim 24\%$\ $(42/175)$\ of quasars contain at least one intrinsic {\NV} NAL in their spectrum.
\item Comparisons with the high-redshift VLT sample from \citet{cc15} show that both samples simultaneously have a lack of {\NV} systems at large equivalent width and high radio-loudness parameter.  Similarly, there is a lack of systems with both high velocity offset and high radio-loudness parameter in both samples.  Both samples appear to support the idea that larger optical luminosities allow for larger velocity offsets. 
\item The intrinsic {\NV} systems were preferentially discovered within the spectra of $39/133$ radio-quiet quasars.  A statistically significant dearth of systems was found in the spectra of the radio-loud quasars (\(6_{-11}^{+3}\)\% vs. \(29_{-7}^{+6}\)\% for radio-quiet quasars).
\item Core-dominated morphologies were discovered for 11 of the 14 radio-loud quasars observed by FIRST, which suggests that our sample is biased toward sightlines close to the poles.  It is possible, though unlikely, that the dearth of {\NV} absorption in radio-loud quasars could be from a sample selection bias stemming from using only Type 1 objects. This potentially suggests that the polar regions in radio-loud quasars may be physically different than those of radio-quiet quasars with regard to the presence of {\NV}-bearing gas.
\end{enumerate}

\section*{Acknowledgements}

Support for Program number HST-AR-12846.002-A was provided by NASA through a grant from the Space Telescope Science Institute, which is operated by the Association of Universities for Research in Astronomy, Incorporated, under NASA contract NAS5-26555. Further support was provided by NSF grant AST-0807993. TM acknowledges support from JSPS KAKENHI Grant Number 21H01126. This research has made use of the NASA/IPAC Extragalactic Database, which is funded by the National Aeronautics and Space Administration and operated by the California Institute of Technology. This work made use of Astropy:\footnote{http://www.astropy.org} a community-developed core Python package and an ecosystem of tools and resources for astronomy \citep{astropy:2013, astropy:2018, astropy:2022}. We thank the anonymous referee for thoughtful and speedy reviews.

\section*{Data Availability}
The data underlying this article were accessed from the Hubble Space Telescope archive (\url{https://mast.stsci.edu/search/ui/\#/hst}, datasets listed in Table~\ref{table:QuasarObservations}). The derived data generated in this research will be shared on reasonable request to the corresponding author.



\bibliographystyle{mnras}
\bibliography{refs} 



\appendix
\section{Quasar Observations and Properties}

Here, we present two tables relating to this work as referenced in \S\ref{sec:Sample}. In Table~\ref{table:QuasarObservations}, we list datasets used. The table presents the QSO name (column 1), J2000 coordinates (column 2), QSO emission redshift (\(\zem\), column 3), the COS Proposal ID (column 4), the proposal principal investigator (PI, column 5), the minimum (\(\lambda_\mathrm{down}\)) and maximum (\(\lambda_\mathrm{up}\) wavelengths of the spectral coverage (column 6), the corresponding absorber redshifts (\(z_\mathrm{down,\NV}\), \(z_\mathrm{up,\NV}\), column 7) and offset velocities (\(v_\mathrm{down,\NV}\), \(v_\mathrm{up,\NV}\), column 8).

In Table~\ref{table:QuasarProperties3}, we list the observed properties of the objects. The table presents the QSO name (column 1), QSO emission redshift (\(\zem\), column 2), the limiting fraction (3\(\sigma\)-confidence) of the 3000\,\AA\ monochromatic luminosity attributable to the host galaxy (column 3), the monochromatic optical luminosity (\(L_\nu(3000\,\mathrm{\AA})\), column 4), the observed radio frequency (\(\nu\), column 5) and adopted radio spectral index (\(\alpha_r = d \log L_\nu / d \log \nu\), column 6) from which the monochromatic radio luminosity (\(L_\nu(5\,\mathrm{GHz})\), column 7) was derived, the radio-loudness parameters (\(\mathcal{R}\), column 8), the adopted X-ray spectral index (\(\alpha_r = d \log L_\nu / d \log \nu\), column 9), the monochromatic X-ray luminosity (\(L_\nu(2\,\mathrm{keV})\), column 10),and references for the adopted photometery (column 11).

\onecolumn
\begin{landscape}
{\small 
\begin{longtable}{ | c | c | c | c | c | c | c | c | }
\caption{Quasar Observations\label{table:QuasarObservations}}\\
\hline \hline 
    & Coordinates  &                 & COS     & PI   & {[}$\lambda _{\text{down}},\lambda _{\text{up}}${]} & & {[}$v_{\text{down, N V}},v_{\text{up, N V}}${]} \\
QSO & (RA DEC, J2000) & $z_{\text{em}}$ & Prop ID & Name & ($\mbox{\AA}$) & {[}$z_{\text{down, N V}},z_{\text{up, N V}}${]} & ({\kms}) \\ \hline \hline
\endfirsthead

\caption{[}{]}{(continued)}\\
\hline \hline
    & Coordinates &                 & COS     & PI   & {[}$\lambda _{\text{down}},\lambda _{\text{up}}${]} & & {[}$v_{\text{down, N V}},v_{\text{up, N V}}${]} \\
QSO & (RA DEC, J2000) & $z_{\text{em}}$ & Prop ID & Name & ($\mbox{\AA}$) & {[}$z_{\text{down, N V}},z_{\text{up, N V}}${]} & ({\kms}) \\ \hline \hline
\endhead

\hline
\endlastfoot
WISEA J000024.41-124547.9 & 00h00m24.41s -12d45m47.91s & 0.198988 & 12604 & Fox & $[1134,1274]$ & $[-0.0846, 0.0284]$ & $[-79001.,-45653.]$ \\
 & & &  &  & $[1300,1571]$ & $[0.0494,0.2681]$ & $[-39720., 16793.]$ \\
 & & &  &  & $[1589,1762]$ & $[0.2827,0.4223]$ & $[20195.,50716.]$ \\
\hline
WISEA J000543.07-500654.8 & 00h05m43.07s -50d06m54.84s & 0.033446 & 12936 & Jenkins & $[1154,1294]$ & $[-0.0685, 0.0445]$ & $[-31015.,  3201.]$ \\
 & & &  &  & $[1309,1571]$ & $[0.0566,0.2681]$ & $[ 6656.,60511.]$ \\
 & & &  &  & $[1589,1762]$ & $[0.2827,0.4223]$ & $[63780.,92623.]$ \\
\hline
PG 0003+158 & 00h05m59.24s +16d09m49.00s & 0.450900 & 12038 & Green & $[1134,1798]$ & $[-0.0846, 0.4514]$ & $[-1.e+05, 1.e+02]$ \\
\hline
MRK 0335 & 00h06m19.52s +20d12m10.50s & 0.025785 & 15439 & Schartel & $[1134,1798]$ & $[-0.0846, 0.4514]$ & $[-33990.,100060.]$ \\
\hline

\end{longtable}}
\twocolumn
\end{landscape}

\onecolumn
\begin{landscape}
{\small 
\begin{ThreePartTable}
\begin{TableNotes}
\footnotesize
\item[] References -- (0) \citet{2009AJ....138..845O}, (1) \citet{1998AJ....115.1693C}, (2) \citet{2018MNRAS.473.4937S}, (3) \citet{2012GASC..C...0000S}, (4) \citet{2024A&A...682A..34M}, (5) \citet{2007SDSS6.C...0000:}, (6) \citet{1990PKS...C......0W}, (7) \citet{1986ApJS...61....1B}, (8) \citet{2020A&A...641A.136W}, (9) \citet{2011ApJ...739...57K}, (10) \citet{2006A&A...455..161L}, (11) \citet{1987ApJ...313..651E}, (12) \citet{2001ApJS..133....1U}, (13) \citet{1994IAUC.6100....1W}, (14) \citet{1994ApJS...95....1E}, (15) \citet{2007ApJS..171..376C}, (16) \citet{1991ApJS...75....1B}, (17) \citet{1987ApJ...323..243W}, (18) \citet{2005ApJS..161..185U}, (19) \citet{2004SDSS2.C...0000:}, (20) \citet{2012GASC..C...0000S}, (21) \citet{1979ApJ...230...79N}, (22) \citet{1991rc3..book.....D}, (23) \citet{1996ApJS..103...81C}, (24) \citet{2006ApJS..164...52S}, (25) \citet{2024ApJS..274...22E}, (26) \citet{2003ASPC..295..465P}, (27) \citet{2007ApJ...664...53A}, (28) \citet{1991ApJS...75.1011G}, (29) \citet{1996AJ....111.1431B}, (30) \citet{2006AJ....131.2826S}, (31) \citet{2007ApJS..173..457B}, (32) \citet{2024ApJS..274...22E}, (33) \citet{2010MNRAS.403.2088L}, (34) \citet{2010ApJ...713L..11Z}, (35) \citet{2005A&A...435..857M}, (36) \citet{1990MNRAS.243..692M}, (37) \citet{2000AJ....120..604S}, (38) \citet{2011AJ....142..105P}, (39) \citet{2010MNRAS.401.1151M}, (40) \citet{2009ApJ...703.1597A}, (41) \citet{2011ApJS..193...28P}, (42) \citet{2017ApJS..233...17R}, (43) \citet{2006AJ....131.1163S}, (44) \citet{1981A&AS...45..367K}, (45) \citet{2003PASJ...55..351T}, (46) \citet{2009A&A...495..691M}, (47) \citet{2016MNRAS.460.4433H}, (48) \citet{1995ApJS...98..369B}, (49) \citet{1982MNRAS.199..943H}, (50) \citet{1976AuJPA..39....1W}, (51) \citet{1994ApJS...90..179G}, (52) \citet{2004ApJ...615..135S}, (53) \citet{2008ApJ...678...22H}, (54) \citet{2016A&A...592A...5F}, (55) \citet{1975AJ.....80..771S}, (56) \citet{2015ApJ...810...14A}, (57) \citet{2015MNRAS.450.2658M}, (58) \citet{2011ApJ...726...43A}, (59) \citet{2007A&A...476..759B}, (60) \citet{2009A&A...495..421B}, (61) \citet{2006MNRAS.368..479G}, (62) \citet{1983MNRAS.205..793W}, (63) \citet{2013AJ....146....5P}, (64) \citet{2006ApJ...639...37K}, (65) \citet{2010A&A...524A..64C}, (66) \citet{2011A&A...533A..79M}, (67) \citet{2010A&A...520A..47M}, (68) \citet{2004MNRAS.349.1397C}, (69) \citet{2017MNRAS.464.1306C}, (70) \citet{1970ApJ...162..743A}, (71) \citet{1978AJ.....83..451P}, (72) \citet{2004ApJ...605..670P}, (73) \citet{2010MNRAS.402.2403M}, (74) \citet{1994ApJS...90..173G}, (75) \citet{1994ApJS...91..111W}, (76) \citet{2010ApJ...717.1243G}, (77) \citet{2003MNRAS.342.1117M}, (78) 
\citet{2008MNRAS.384..775M}, (79) \citet{1972AuJPA..23....1S}, (80) \citet{2011A&A...526A.102B}, (81) \citet{2007ApJS..170...33P}, (82) \citet{2010ApJ...713..180Z}, (83) \citet{1981AJ.....86..854K}, (84) \citet{2014A&A...572A..59M}, (85) \citet{2005A&A...440..855G}, (86) 
\citet{2004SDSS3.C...0000:}, (87) \citet{1995ApJS...97..347G}, (88) \citet{2008ApJ...677...37O}, (89) \citet{2009ApJS..183...17Y}, (90) \citet{2006ApJS..162...38A}, (91) \citet{2001AJ....122.2791S}, (92) \citet{2009ApJS..182..378W}, (93) \citet{2011MNRAS.412..161G}, (94) \citet{2006ApJS..165..229F}, (95) \citet{2005MNRAS.356.1029B}, (96) \citet{2015MNRAS.446.2985V}, (97) \citet{2014ApJS..212...18B}, (98) \citet{2005A&A...436..837H}, (99) \citet{2005ApJ...621L..21C}, (100) \citet{2004ApJ...606..213T}, (101) \citet{2007MNRAS.381..228P}, (102) \citet{2006A&A...449..425M}, (103) \citet{2009AJ....137...42R}, (104) \citet{2015ApJ...801...26H}, (105) \citet{1990ApJS...72..621L}, (106) \citet{2017ApJ...851...33P}, (107) \citet{2015MNRAS.454.3864B}, (108) \citet{2007AJ....134..648M}, (109) \citet{2005A&A...435..521N}, (110) \citet{2006AJ....132..546G}, (111) \citet{2008MNRAS.384..316L}, (112) \citet{2010A&A...520A.113B}, (113) \citet{2011A&A...536A..84V}, (114) \citet{2009ApJ...705.1196T}, (115) \citet{2018ApJS..234...18B}, (116) \citet{2006A&A...449..559B}, (117) \citet{2007ApJ...662..909K}, (118) \citet{2016MNRAS.459..366G}, (119) \citet{2005A&A...432...15P}, (120) \citet{1978ApJS...36..317W}, (121) \citet{1981MNRAS.194..795G}, (122) \citet{1996ApJS..103..145W}, (123) \citet{2005A&A...434..385G}, (124) \citet{1997A&AS..122..235L}, (125) \citet{2018ApJS..238...32K}, (126) \citet{2009ApJS..183...17Y}, (127) \citet{2011ApJ...738...96M}, (128) \citet{2016ApJS..224...40W}, (129) \citet{2008MNRAS.383..581D}, (130) \citet{2004ApJS..153..119M}, (131) \citet{2011A&A...536A..49G}, (132) \citet{1992ApJS...79..331W}, (133) \citet{1983ApJS...52..341M}, (134) \citet{1989MNRAS.240..833T}, (135) \citet{2007ApJ...661..719U}, (136) \citet{1975AuJPA..38....1W}, (137) \citet{1982ApJ...256....1C}, (138) \citet{2011ApJ...736..139K}, (139) \citet{2005MNRAS.358.1423O}, (140) \citet{2006MNRAS.366..953B}, (141) \citet{2005ApJ...621..123U}, (142) \citet{2011ApJ...742...49T}, (143) \citet{1978ApJ...224...22O}, (144) \citet{1994A&AS..106..119L}, (145) \citet{2007AJ....133.1236K}, (146) \citet{2015ApJS..218....8D}, (147) \citet{2007AJ....133..313A}, (148) \citet{2007MNRAS.378...23J}, (149) \citet{1975AuJPA..34....1B}, (150) \citet{1982ApJ...260..437R}, (151) \citet{2004AJ....127.1405C}, (152) \citet{1970ApL.....5...29W}, (153) \citet{2004MNRAS.351....1S}, (154) \citet{2005ApJ...629...61K}, (155) \citet{2016A&A...588A..70B}, (156) \citet{2005ApJS..156...13M}, (157) \citet{2011ApJS..193...15M}, (158) \citet{2009AJ....138..991R}, (159) \citet{2007MNRAS.382..194N}, (160) \citet{2007MNRAS.379.1359M}, (161) \citet{2012A&A...538A..26R}, (162) \citet{2010ApJ...719...45C}, (163) \citet{2005ApJS..161....9K}, (164) \citet{2016MNRAS.460.2385W}, (165) \citet{2011ApJ...728...28W}, (166) \citet{1969ApJ...157....1K}, (167) \citet{2009A&A...502...61M}, (168) \citet{2006MNRAS.366..339B}, (169) \citet{2005ApJ...624..189W}, (170) \citet{jarrett2000}, (171) \citet{2010ApJ...716...30A}, (172) \citet{1968cgcg.book.....Z}, (173) \citet{1997A&AS..124..259R}, (174) \citet{2009ApJ...690.1181S}, (175) \citet{1989ApJS...70..257B}, (176) \citet{2006A&A...448L..49R}, (177) \citet{1997MNRAS.285..511B}, (178) \citet{2002AJ....124..675C}, (179) \citet{2004ApJ...601..723Y}, (180) \citet{2011A&A...534A..36K}, (181) \citet{2003ApJ...582..105Y}, (182) \citet{2009A&A...505..541C}, (183) \citet{2007ApJ...663..103L}, (184) \citet{2008MNRAS.385.1656S}, (185) \citet{2007ApJ...670..992F}, (186) \citet{2004MNRAS.351..161P}, (187) \citet{2004MNRAS.347..854V}, (188) \citet{2014MNRAS.437.3929S}, (189) \citet{1969AuJPA...7....3E}, (190) \citet{2009ApJ...690.1322W}, (191) \citet{2013wise.rept....1C}, (192) \citet{2011MNRAS.411.2353P}, (193) \citet{2017ApJ...838..139S}, (194) \citet{2015ApJS..218...23A}, (195) \citet{2006AJ....131.1872P}, (196) \citet{2010A&A...516A..56H}, (197) \citet{2004ApJ...613..752G}
\end{TableNotes}
\begin{longtable}{ | c | l | l | c | r | l | r | r | c | r | c | }
\caption{Sample Quasars and Their Properties\label{table:QuasarProperties3}}\\
\hline \hline 
    \multicolumn{2}{c}{~~} &             &                               & \multicolumn{4}{c}{Radio}                                                      &  \multicolumn{2}{c}{Soft X-ray} & \\ \hline
    &                      & $\log f_\mathrm{gal}$ & $L_{\nu}(3000$\,\AA$)$        & $\nu $ &                  & $L_{\nu}(5\,\mathrm{GHz})$ &                       &                  & $L_{\nu}(2\,\mathrm{keV})$ & \\
QSO & $z_{\text{em}}$     &                   & (erg s$^{-1}$ Hz$^{-1}$)      & (GHz)  & $\alpha_{r}$ & (erg s$^{-1}$ Hz$^{-1}$)   & $\log ({\mathcal R})$ & $\alpha_{x}$ & (erg s$^{-1}$ Hz$^{-1}$)  & References\\ \hline\hline
\endfirsthead

\caption[]{(continued)}\\
\hline \hline
    \multicolumn{2}{c}{~~} &            &                               & \multicolumn{4}{c}{Radio}                                                      &  \multicolumn{2}{c}{Soft X-ray} & \\ \hline
    &                      & $\log f_\mathrm{gal}$ & $L_{\nu}(3000$\,\AA$)$        & $\nu $ &                  & $L_{\nu}(5\,\mathrm{GHz})$ &                       &                  & $L_{\nu}(2\,\mathrm{keV})$ & \\
QSO & $z_{\text{em}}$     &                   & (erg s$^{-1}$ Hz$^{-1}$)      & (GHz)  & $\alpha_{r}$ & (erg s$^{-1}$ Hz$^{-1}$)   & $\log ({\mathcal R})$ & $\alpha_{x}$ & (erg s$^{-1}$ Hz$^{-1}$)  & References\\ \hline\hline
\endhead

\hline
\bottomrule
\insertTableNotes
\endlastfoot
WISEA J000024.41-124547.9        & 0.198988 &        & $2.06 \times 10^{30}$ &  1.400 & -0.7 & $<1.13 \times 10^{30}$ & $<-0.26$ & -1.4 & $2.30 \times 10^{25}$ & 0,1,2 \\
WISEA J000543.07-500654.8        & 0.033446 &        & $2.14 \times 10^{28}$ &        &      &                        &          & -1.0 & $1.25 \times 10^{23}$ & 3,4 \\
PG 0003+158                      & 0.450900 &        & $1.28 \times 10^{31}$ &  3.917 & -0.9 & $ 2.73 \times 10^{33}$ & $  2.33$ & -0.7 & $2.51 \times 10^{25}$ & 5,6,7,8 \\
MRK 0335                         & 0.025785 &        & $6.82 \times 10^{27}$ &  4.924 & -0.5 & $ 4.54 \times 10^{28}$ & $  0.82$ & -0.7 & $3.43 \times 10^{25}$ &  9,10,11,12,13 \\
MRK 1501                         & 0.087227 &        & $3.56 \times 10^{29}$ &  1.522 &  0.5 & $ 4.72 \times 10^{31}$ & $  2.12$ & -0.6 & $4.04 \times 10^{26}$ & 14,15,16,17,18 \\
FBQS J0012-1022                  & 0.224388 & <-5.48 & $5.55 \times 10^{29}$ &  2.099 & -0.7 & $ 1.65 \times 10^{30}$ & $  0.47$ &      &                       &  1,19 \\
WISEA J002854.15+430815.4        & 0.092725 &        & $1.54 \times 10^{28}$ &  1.672 & -0.7 & $ 2.54 \times 10^{30}$ & $  2.22$ &      &                       &  1,20 \\
PG 0026+129                      & 0.142000 &        & $8.10 \times 10^{29}$ &  1.599 & -1.2 & $ 1.73 \times 10^{30}$ & $  0.33$ & -0.7 & $8.70 \times 10^{26}$ &  1,10,17,18,21 \\
\end{longtable}
\end{ThreePartTable}
}
\twocolumn
\end{landscape}

\section{Notes on Individual Objects with Intrinsic {\NV} Absorption}
\label{sec:notes}
\subsection{QSO\,J$0012-1022$ \(\zem = 0.228191\) PID: 12248}
\begin{figure}
\includegraphics[width=0.5\textwidth]{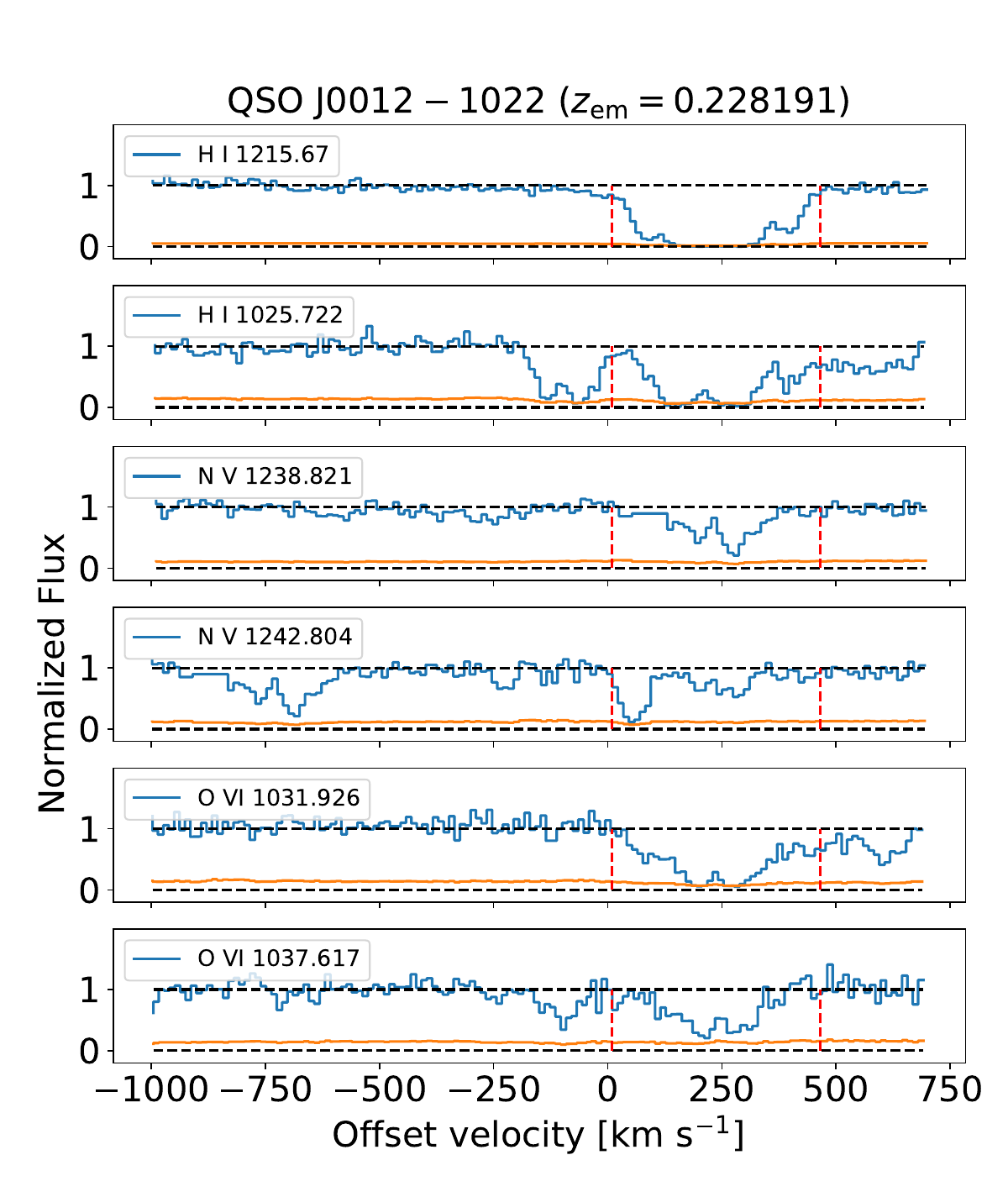}
\caption[QSO\,J$0012-1022$ System Plot]{Intrinsic Absorption in QSO\,J$0012-1022$. We show a velocity-stacked plot of the detected transitions for the intrinsic absorber in the archived HST/COS spectra. In each panel, the flux density is shown as a blue histogram, and the uncertainty is shown as an orange histogram. For clarity, spectra are shown with a sampling of 5 pixels per bin. A horizontal dashed black line mark the zero-flux level. Vertical red dashed lines encompass the absorption system at \(+282\)\,\kms\ relative to \(\zem = 0.228191\). The {\NV} \(\lambda1242\) intrinsic absorption is blended with Galactic {\SiII} \(\lambda1526\) absorption.}
\label{fig:q0012}
\end{figure}

COS observations for PID 12248 including this target have primarily focused on intervening absorbers. In Figure~\ref{fig:q0012}, we present velocity-stack spectra of the {\HI} \(\lambda\lambda1215, 1025\) lines, the {\NV} \(\lambda\lambda1238, 1242\) doublet, and the {\OVI} \(\lambda\lambda1031, 1037\) doublet. Two blended components are at an equivalent-width-weighted average velocity of \(+282\)\,\kms\  relative to \(\zem = 0.228191\). [Note: In all cases, we use the {\NV} \(\lambda1238\) transition for equivalent width weighting.]

\subsection{2MASS\,J$0036230.0+431640.2$ \(\zem = 0.12387\) PID: 11632}
\begin{figure}
\includegraphics[width=0.5\textwidth]{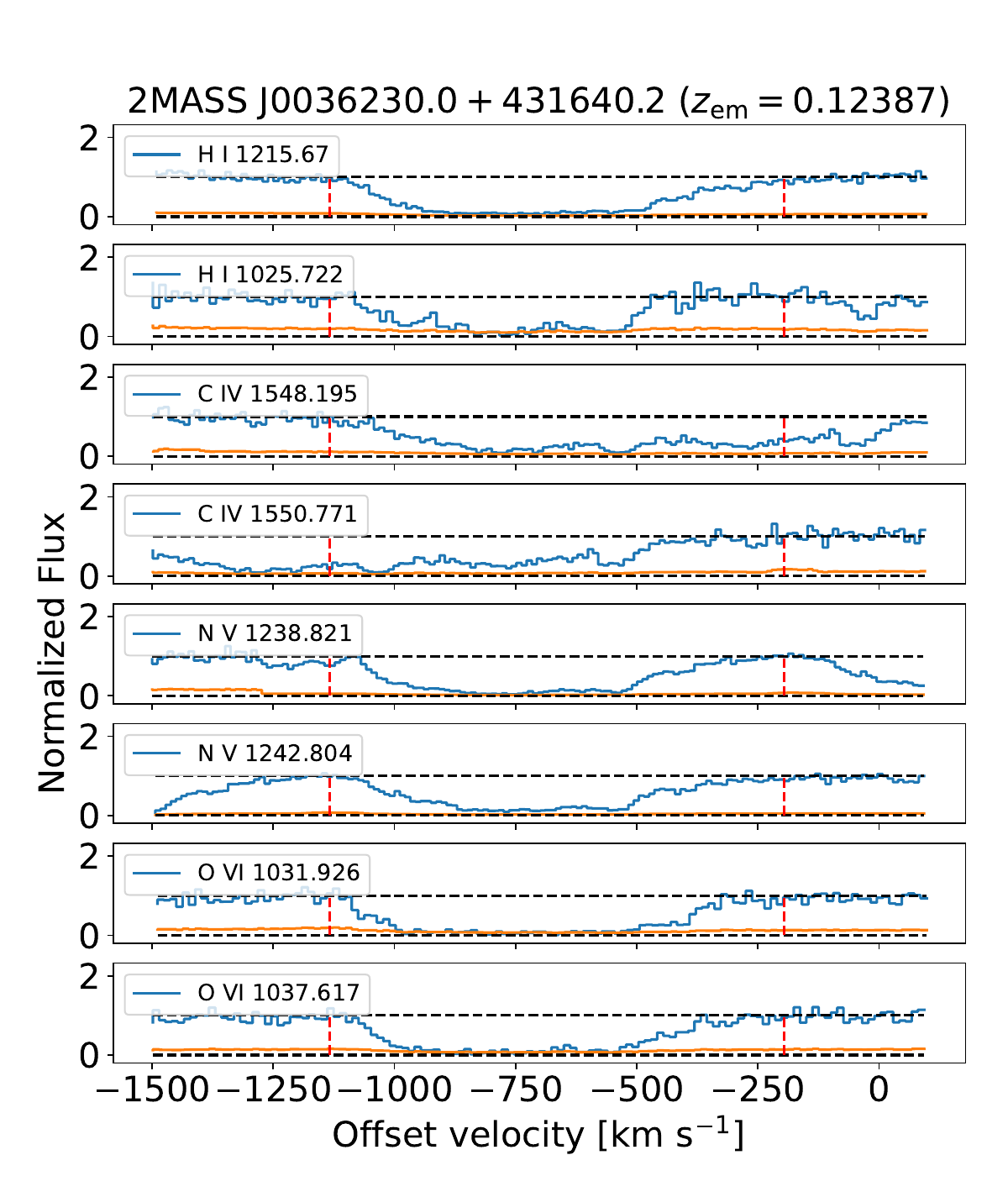}
\caption[2MASS\,J$0036230.0+431640.2$ System Plot]{Intrinsic Absorption in 2MASS\,J$0036230.0+431640.2$. We show a velocity-stacked plot of the detection transitions for the intrinsic absorber in the archived HST/COS spectra. In each panel, the flux density is shown as a blue histogram, and the uncertainty is shown as an orange histogram. For clarity, spectra are shown with a sampling of 5 pixels per bin. A horizontal dashed black line mark the zero-flux level. Vertical red dashed lines encompass the absorption system at \(-713\)\,\kms\ relative \(\zem = 0.12387\).}
\label{fig:hs0033}
\end{figure}

This target was originally observed to study M31 through quasar absorption sightlines. The reported redshift of \(\zem = 0.12\)\ from the Hamburg-Schmidt Survey \citep[][HS\,\(0033+4300\)]{1999AAS..134..483H} is not sufficiently precise for our purposes. From the peak of the {\HI} \(\lambda1215\) broad emission line in the COS spectrum, we adopt a redshift of \(\zem = 0.12387\). In Figure~\ref{fig:hs0033}, we present velocity-stack spectra of the {\HI} \(\lambda\lambda1215, 1025\) lines, the {\CIV} \(\lambda\lambda1548,1550\) doublet, the {\NV} \(\lambda\lambda1238, 1242\) doublet, and the {\OVI} \(\lambda\lambda1031, 1037\) doublet. The absorption spans the velocity range \([-1140,-190]\)\,\kms\ with a smooth trough indicative of an outflow. We adopt an equivalent-width-weighted average velocity of \(-713\)\,\kms\ relative \(\zem = 0.12387\)\ for the system.

\subsection{2MASX\,J$004818.99+394111.8$  \(\zem = 0.134\) PID: 11686}
\begin{figure}
\includegraphics[width=0.5\textwidth]{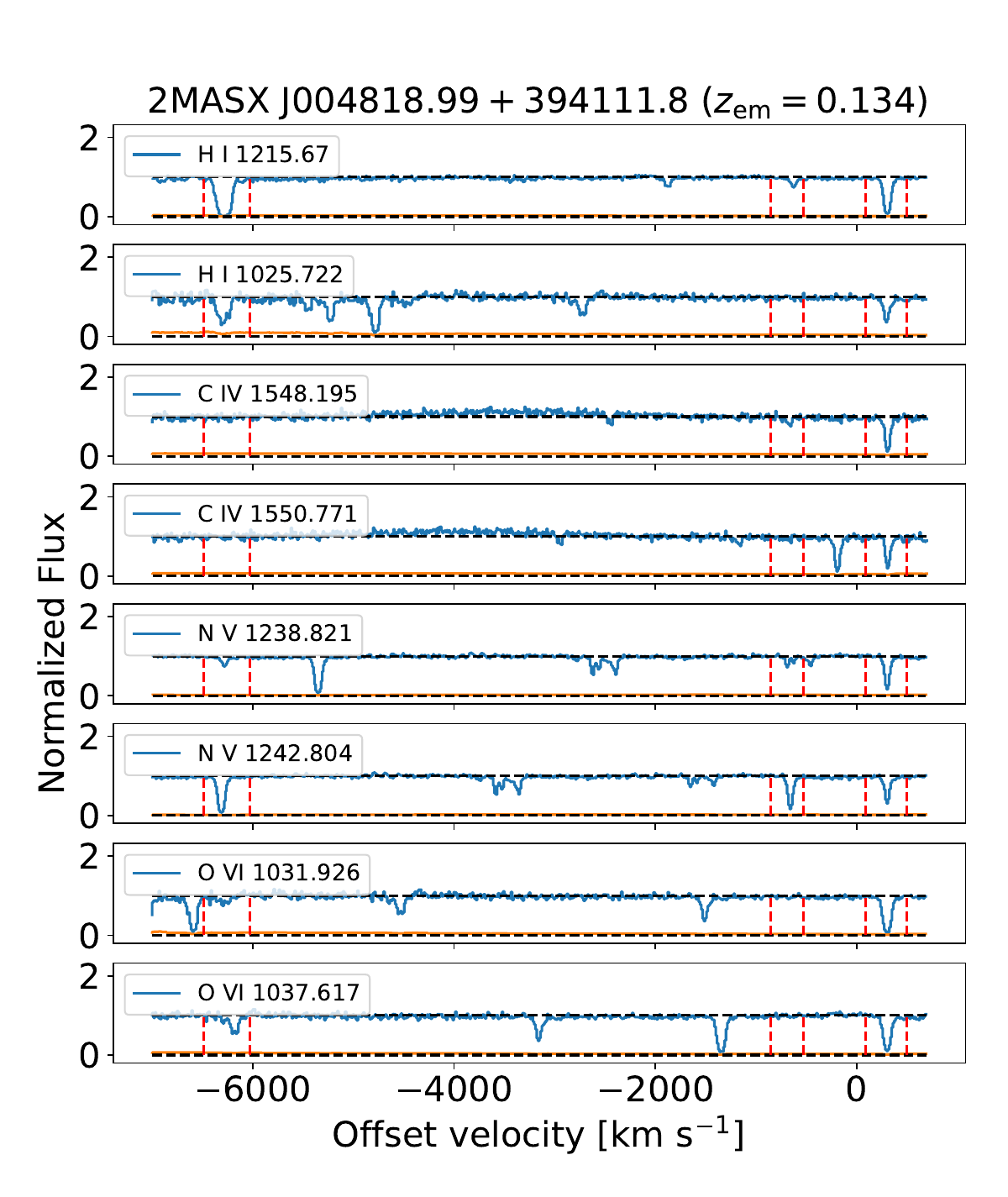}
\caption[2MASX\,J$004818.99+394111.8$ System Plot]{Intrinsic Absorption in 2MASX\,J$004818.99+394111.8$. We show a velocity-stacked plot of the detection transitions for the intrinsic absorber in the archived HST/COS spectra. In each panel, the flux density is shown as a blue histogram, and the uncertainty is shown as an orange histogram. For clarity, spectra are shown with a sampling of 5 pixels per bin. A horizontal dashed black line mark the zero-flux level. Vertical red dashed lines encompass the apparently line-locked absorption systems at \(-6262\)\,\kms, \(-660\)\,\kms, and \(+321\)\,\kms.}
\label{fig:ioand}
\end{figure}

The COS spectra of this object (Figure~\ref{fig:ioand}) features three components that appear to be line-locked. Two of the components are separated by the velocity displacement between {\HI} \(\lambda1215\)\ and {\NV} \(\lambda1238\) while the other pair is spaced by the separation of the {\NV} \(\lambda\lambda1238, 1242\) doublet. As these components are well-separated, we treat them as separate systems with velocities \(-6266\)\,\kms, \(-656\)\,\kms, and \(+311\)\,\kms\ relative to \(\zem = 0.134\).

\subsection{6dFGS\,gJ$015530.0-085704$  \(\zem = 0.164427\) PID: 12248}
\begin{figure}
\includegraphics[width=0.5\textwidth]{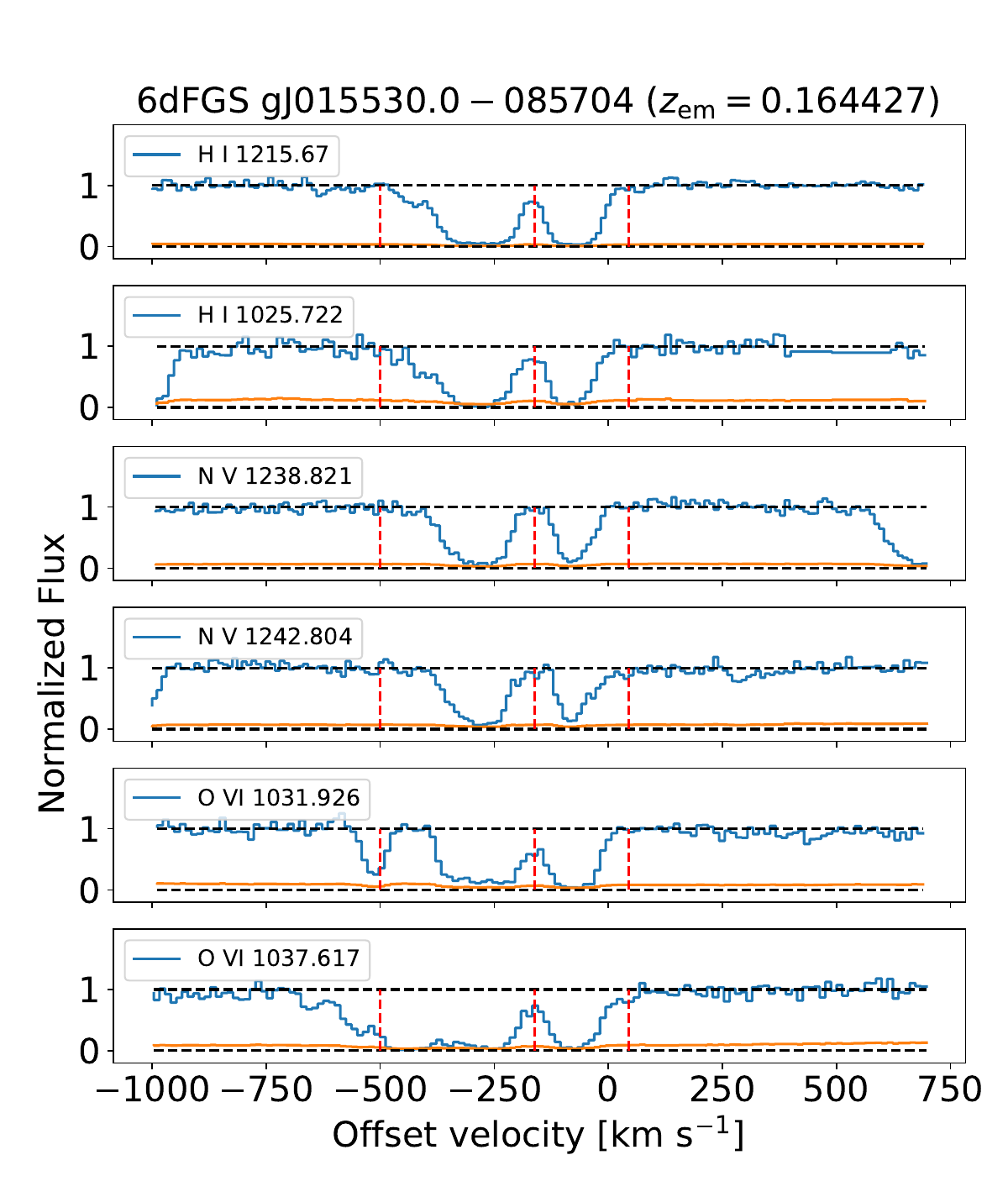}
\caption[6dFGS\,gJ$015530.0-085704$ System Plot]{Intrinsic Absorption in 6dFGS\,gJ$015530.0-085704$. We show a velocity-stacked plot of the detection transitions for the intrinsic absorber in the archived HST/COS spectra. In each panel, the flux density is shown as a blue histogram, and the uncertainty is shown as an orange histogram. For clarity, spectra are shown with a sampling of 5 pixels per bin. A horizontal dashed black line mark the zero-flux level. Vertical red dashed lines encompass the absorption systems at \(-284\)\,\kms\ and \(-69\)\,\kms\ relative to \(\zem = 0.164427\).}
\label{fig:q0155}
\end{figure}

COS observations for PID 12248 including this target have primarily focused on intervening absorbers. In Figure~\ref{fig:q0155}, we present velocity-stack spectra of the {\HI} \(\lambda\lambda1215, 1025\) lines, the {\NV} \(\lambda\lambda1238, 1242\) doublet, and the {\OVI} \(\lambda\lambda1031, 1037\) doublet.
In {\NV}, two components are cleanly separated at \(-284\)\,\kms\ and \(-69\)\,\kms\ relative to \(\zem = 0.164427\).

\subsection{2MASS\,J$02121832-0737198$  \(\zem = 0.17392\) PID: 12248}
\begin{figure}
\includegraphics[width=0.5\textwidth]{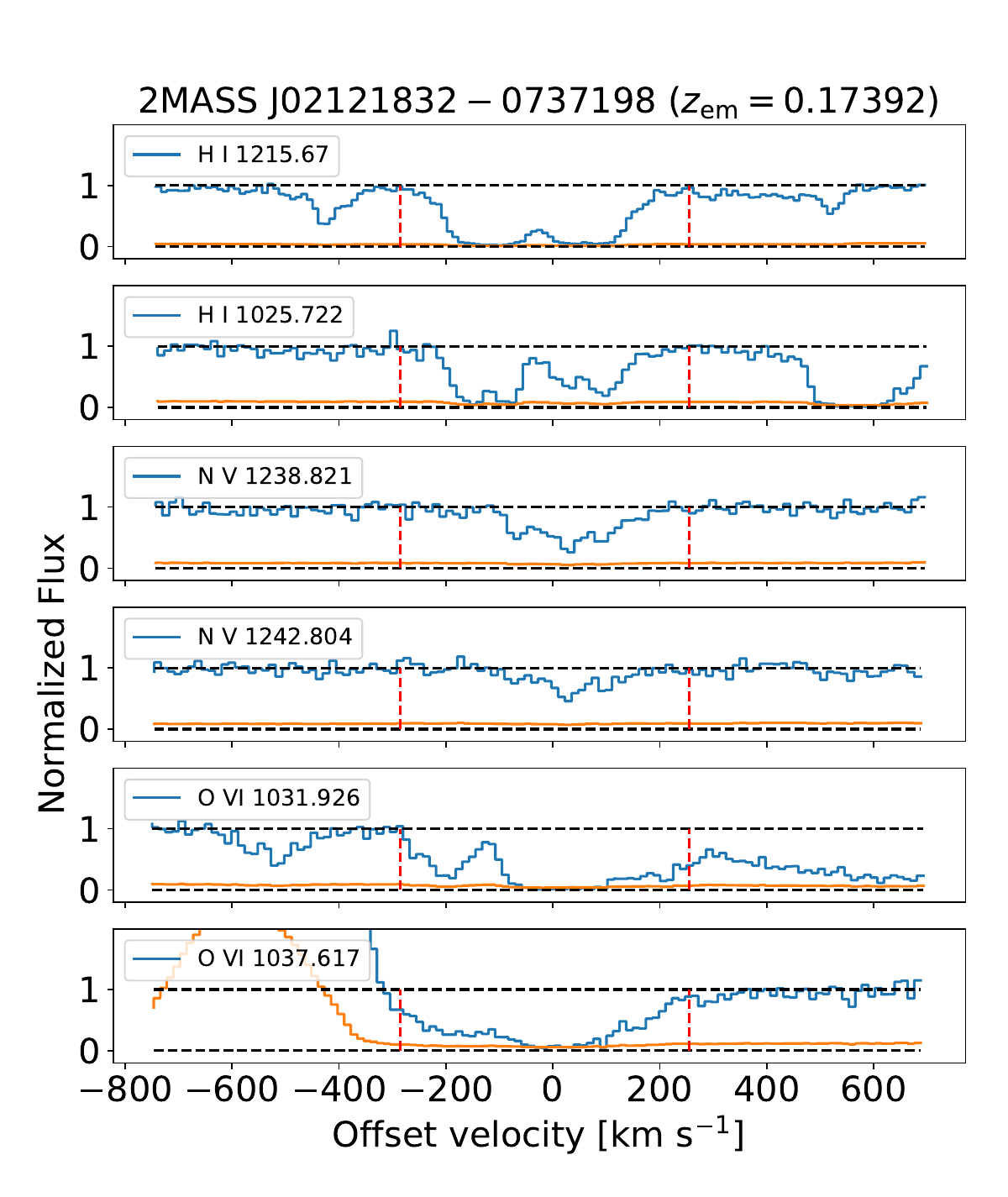}
\caption[2MASS\,J$02121832-0737198$ System Plot]{Intrinsic Absorption in 2MASS\,J$02121832-0737198$. We show a velocity-stacked plot of the detection transitions for the intrinsic absorber in the archived HST/COS spectra. In each panel, the flux density is shown as a blue histogram, and the uncertainty is shown as an orange histogram. For clarity, spectra are shown with a sampling of 5 pixels per bin. A horizontal dashed black line mark the zero-flux level. Vertical red dashed lines encompass the absorption systems at \(+48\)\,\kms\ relative to \(\zem = 0.17392\).}
\label{fig:q0212}
\end{figure}

COS observations for PID 12248 including this target have primarily focused on intervening absorbers. In Figure~\ref{fig:q0212}, we present velocity-stack spectra of the {\HI} \(\lambda\lambda1025, 1215\) lines, the {\NV} \(\lambda\lambda1238, 1242\) doublet, and the {\OVI} \(\lambda\lambda1031, 1037\) doublet.Two blended components are at an equivalent-width-weighted average velocity of \(+48\)\,\kms\ relative to \(\zem = 0.17392\).

\subsection{Mrk\,1044 \(\zem = 0.0165\) PID: 12212}
\begin{figure}
\includegraphics[width=0.5\textwidth]{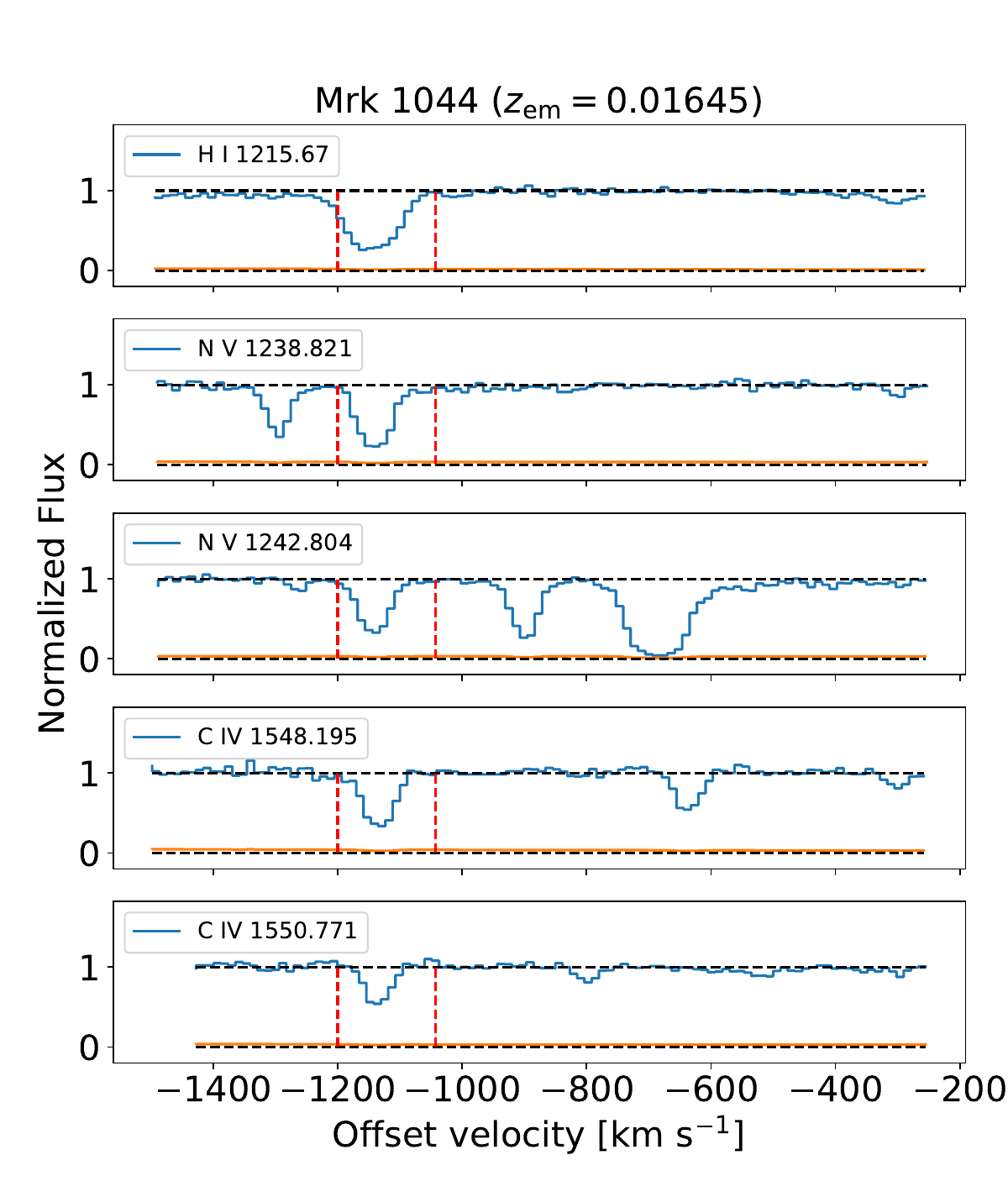}
\caption[MRK 1044 System Plot]{Intrinsic Absorption in MRK\,1044. We show a velocity-stacked plot of the detection transitions for the intrinsic absorber in the archived HST/COS spectra. In each panel, the flux density is shown as a blue histogram, and the uncertainty is shown as an orange histogram. For clarity, spectra are shown with a sampling of 5 pixels per bin. A horizontal dashed black line mark the zero-flux level. Vertical red dashed lines mark the single absorption component at \(-1131\)\,\kms.}
\label{fig:mrk1044}
\end{figure}

The COS spectrum of this object has been analyzed for absorption by intervening \citep[e.g.,][]{2016A&A...590A..68R} and Milky Way/HVC \citep[e.g.,][]{2014ApJ...787..147F} structures. From previous observations using STIS/G140M spectra, \citet{2005ApJ...620..183F} found two blueshifted absorption components in {\NV} at \(-1145\) and \(-295\)\,\kms. In our analysis of the COS spectrum (Fig.~\ref{fig:mrk1044}), we find an absorption component at an equivalent-width weighted average velocity of \(-1131\)\,\kms, coincident with the \(-1145\)\,\kms\ absorber reported by \citet{2005ApJ...620..183F}, but no absorption at \(-295\)\,\kms.

\subsection{Mrk\,595 \(\zem = 0.026982\) PID: 12275}
\begin{figure}
\includegraphics[width=0.5\textwidth]{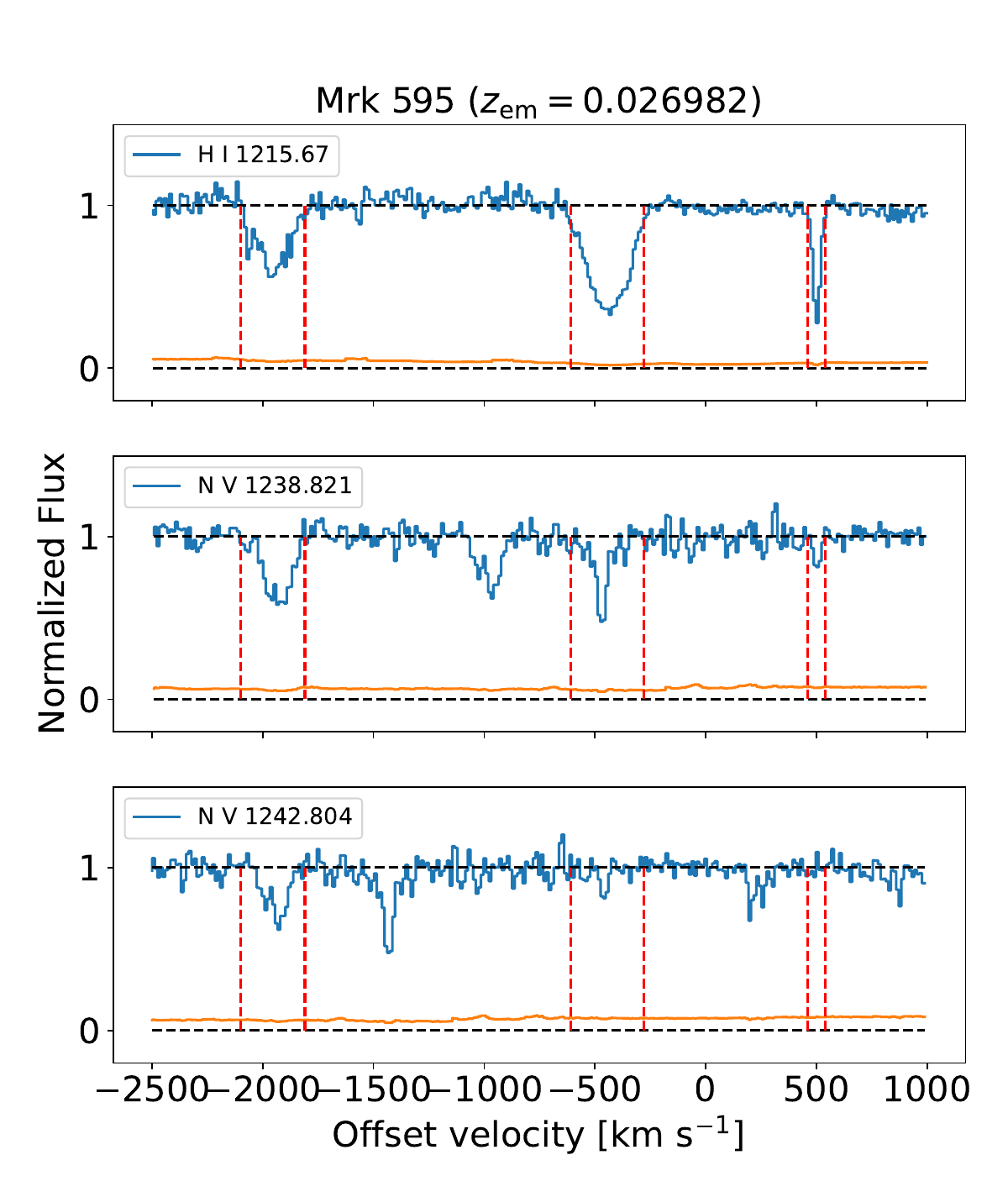}
\caption[Mrk 595 System Plot]{Intrinsic Absorption in Mrk\,595. We show a velocity-stacked plot of the detection transitions for the intrinsic absorber in the archived HST/COS spectra. In each panel, the flux density is shown as a blue histogram, and the uncertainty is shown as an orange histogram. For clarity, spectra are shown with a sampling of 5 pixels per bin. A horizontal dashed black line mark the zero-flux level. Vertical red dashed lines encompass the absorption components at \(-1927\)\,\kms, and \(-462\)\,\kms.}
\label{fig:mrk595}
\end{figure}

COS observations for PID 12275 including this target have primarily focussed on intervening absorbers. In Figure~\ref{fig:mrk595}, we present velocity-stack spectra of the {\HI} \(\lambda1215\) line and the {\NV} \(\lambda\lambda1238, 1242\) doublet. Two components are detected at \(-1927\)\,\kms, and \(-462\)\,\kms. These are clearly separated and we treat them as kinematically-isolated systems. A possible third component/system may be at \(\sim+500\)\,\kms, but it is not detected in {\NV} \(\lambda1242\).

\subsection{LB\,1727 \(\zem = 0.104\) PID: 11686}
The COS spectra and in-depth analysis of this target was presented in \citet[][see their Figure~3]{2011ApJ...739....7E}. The {\NV} doublet features absorption in three blended components in the velocity range \([-300,0]\)\,\kms\ relative to \(\zem = 0.104\). We adopt a equivalent-width weighted average velocity of \(-178\)\,\kms

\subsection{Ton\,951 \(\zem = 0.064\) PID: 12569}
This target was part of the QUEST sample and the COS spectra are presented in \citet[][see their Figure~14c]{2022ApJ...926...60V}. They report two blended components at \(+140\)\,\kms\ and \(+190\)\,\kms\ relative to the systemic redshift of \(\zem = 0.064\). Since they are blended, we combine them into a single equivalent-width-weighted system velocity of \(-149\)\,\kms.

\subsection{2XMM\,J$100420.0+051300$ \(\zem = 0.160116\) PID: 13423, 15227, 13347}
This target was part of the QUEST sample and the COS spectra are presented in \citet[][see their Figure~14f]{2022ApJ...926...60V}. As noted by those authors, the reproduction of the unabsorbed spectrum is challenging given the breath and depth of the broad absorption lines. The {\NV} absorption lies in the velocity range \([-8800,-3800]\)\,\kms\ relative to \(\zem = 0.160116\) with a equivalent-width weighted-average velocity of \(-6138\)\,\kms.

\subsection{Ton\,0488 \(\zem = 0.256074\) PID: 12025}

\begin{figure}
\includegraphics[width=0.5\textwidth]{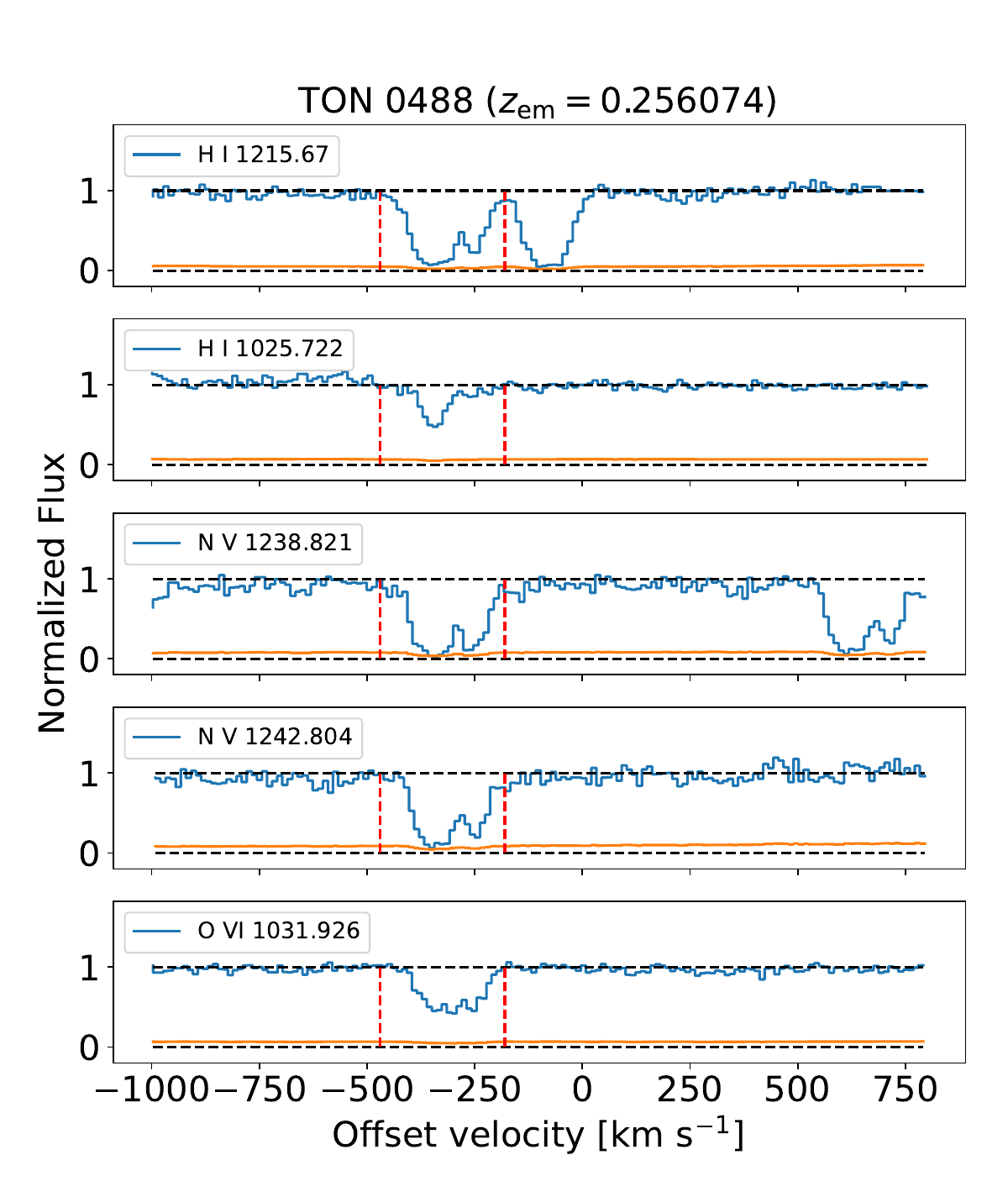}
\caption[Ton\,0488 System Plot]{Intrinsic Absorption in Ton\,0488. We show a velocity-stacked plot of the detection transitions for the intrinsic absorber in the archived HST/COS spectra. In each panel, the flux density is shown as a blue histogram, and the uncertainty is shown as an orange histogram. For clarity, spectra are shown with a sampling of 5 pixels per bin. A horizontal dashed black line mark the zero-flux level. Vertical red dashed lines encompass the absorption system at \(-305\)\,\kms\ relative to \(\zem = 0.256074\).}
\label{fig:ton0488}
\end{figure}

This object was part of the original GTO set of observations for PID 12025 and this dataset has primarily been used for studying intervening absorbers. 
In Figure~\ref{fig:ton0488}, we present velocity-stack spectra of the {\HI} \(\lambda\lambda1025,1215\) lines, the {\NV} \(\lambda\lambda1238, 1242\) doublet, and the {\OVI} \(\lambda1031.928\) line. The {\OVI} \(\lambda1037\) line is blended with geochoronal emission. The profiles are clearly a blend of complex kinematics in the velocity range \([-470,-180]\)\,\kms. Note that the {\HI} \(\lambda1215\) profile shows additional absorption in the velocity range \([-180,+51]\)\,\kms, but this seem unrelated to the intrinsic absorption. As there is no clear unabsorbed flux separating the blend of components, we adopt an equivalent-width-weighted average velocity of \(-305\)\,\kms\ relative to \(\zem = 0.256074\)\ for the system.

\subsection{2MASS\,J$10155924-2748289$  \(\zem = 0.2433\) PID: 13347}

\begin{figure}
\includegraphics[width=0.5\textwidth]{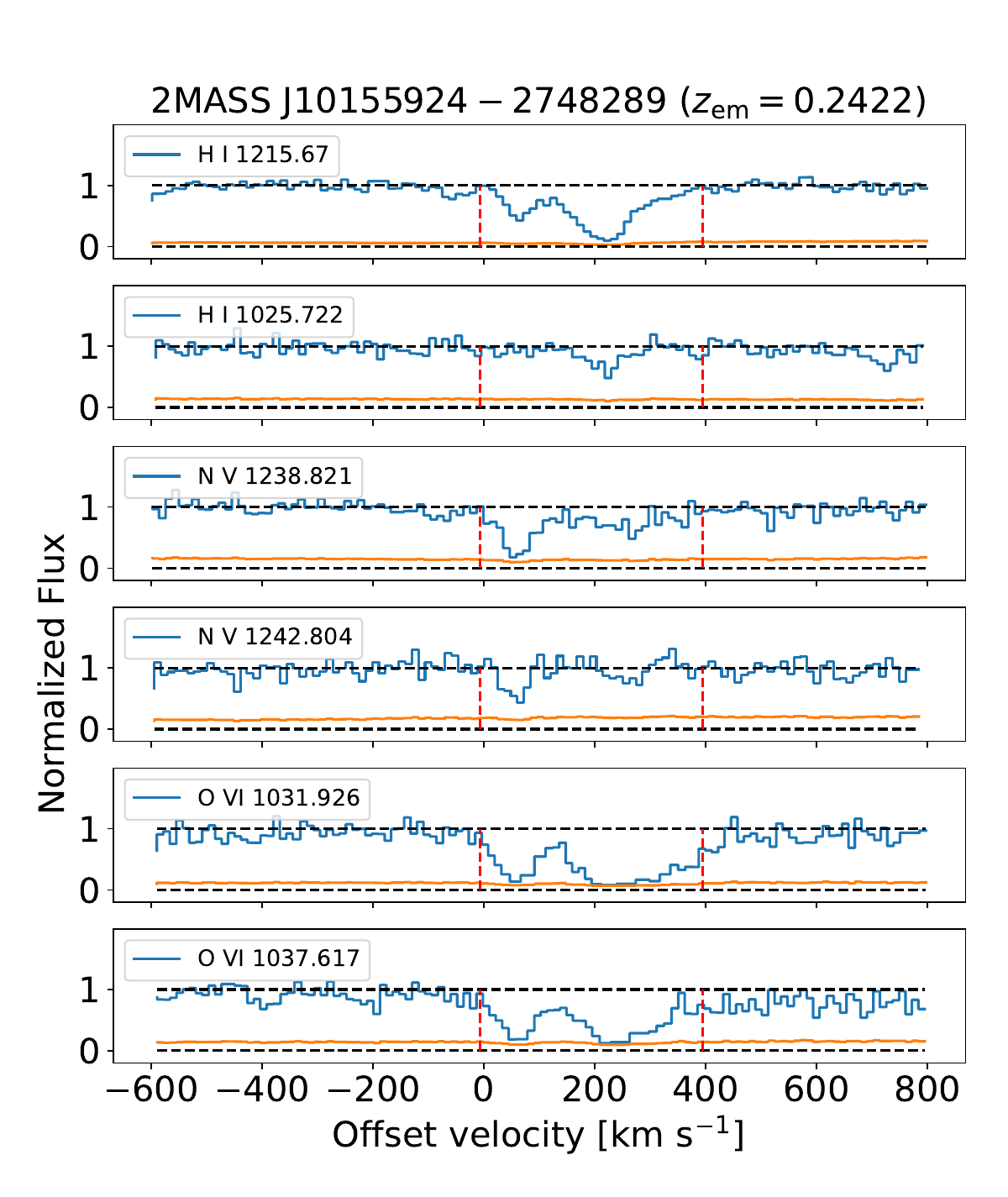}
\caption[2MASS\,J$10155924-2748289$ System Plot]{Intrinsic Absorption in 2MASS\,J$10155924-2748289$. We show a velocity-stacked plot of the detection transitions for the intrinsic absorber in the archived HST/COS spectra. In each panel, the flux density is shown as a blue histogram, and the uncertainty is shown as an orange histogram. For clarity, spectra are shown with a sampling of 5 pixels per bin. A horizontal dashed black line mark the zero-flux level. Vertical red dashed lines encompass the absorption component at \(+149\)\,\kms\ relative to \(\zem = 0.2422\).}
\label{fig:q1015}
\end{figure}

The COS observations for this object originally targeted intervening absorbers around dwarf galaxies in the outskirts of the Local Group. As there is no reported redshift for this object, we use the peak of the {\HI} \(\lambda1215\) line to adopt \(\zem = 0.2422\). In Figure~\ref{fig:q1015}, we present velocity-stacked spectra of the {\HI} \(\lambda1025, 1215\) lines, the {\NV} \(\lambda\lambda1238, 1242\) doublet, and the {\OVI} \(\lambda\lambda1031, 1037\) doublet. The profiles are clearly a blend of complex kinematics in the velocity range \([-10,+400]\)\,\kms. As there is no clear unabsorbed flux separating any components, we adopt an equivalent-width-weighted average velocity of \(+149\)\,\kms\ relative to \(\zem = 0.2422\)\ for the system.

\subsection{Mrk\,1253 \(\zem = 0.049986 \) PID: 13347}

\begin{figure}
\includegraphics[width=0.5\textwidth]{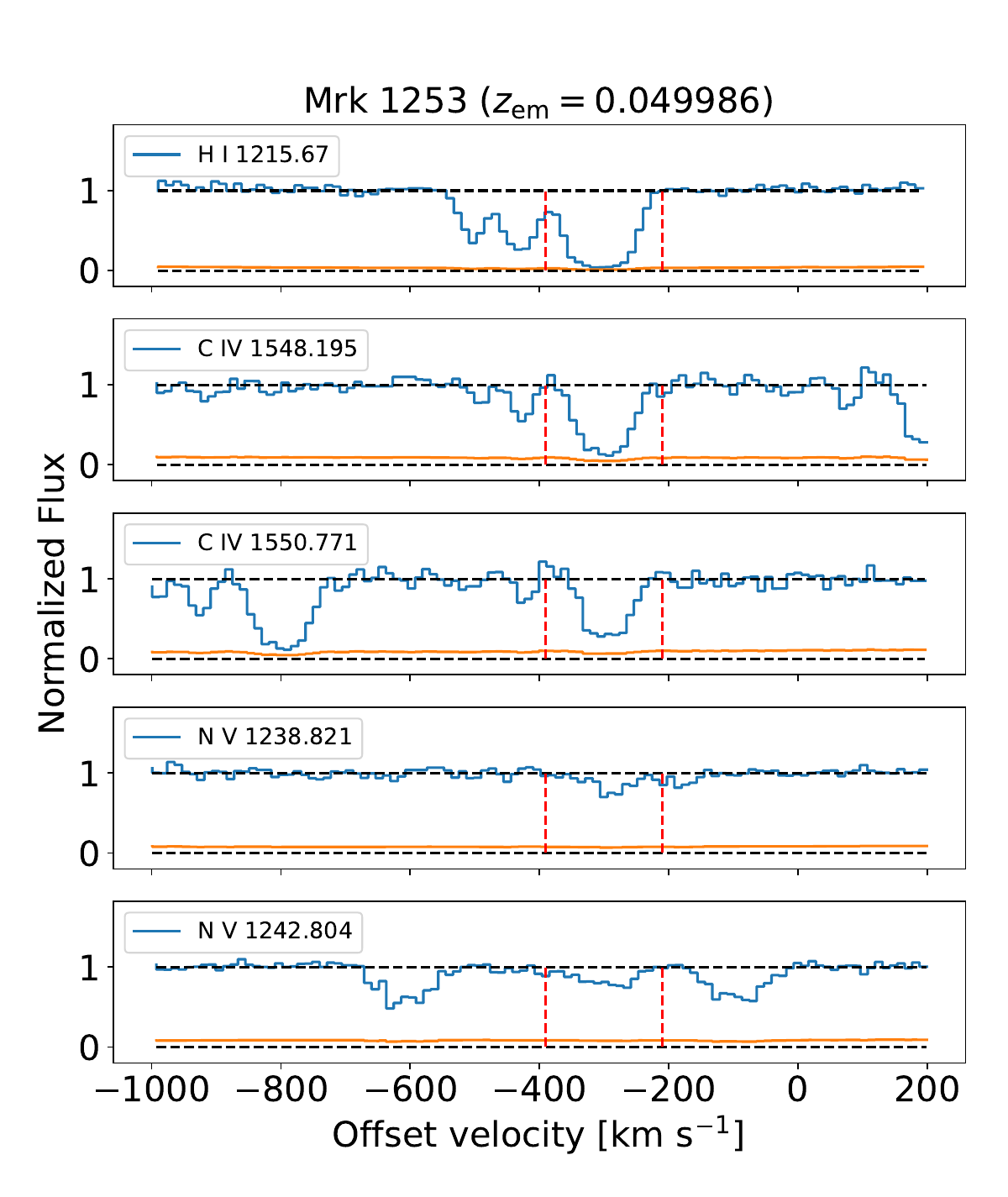}
\caption[Mrk 1253 System Plot]{Intrinsic Absorption in Mrk\,1253. We show a velocity-stacked plot of the detection transitions for the intrinsic absorber in the archived HST/COS spectra. In each panel, the flux density is shown as a blue histogram, and the uncertainty is shown as an orange histogram. For clarity, spectra are shown with a sampling of 5 pixels per bin. A horizontal dashed black line mark the zero-flux level. Vertical red dashed lines encompass the absorption component at \(-285\)\,\kms.}
\label{fig:mrk1253}
\end{figure}

The COS observations for this object originally targeted intervening absorbers around dwarf galaxies in the outskirts of the Local Group. In Figure~\ref{fig:mrk1253}, we present velocity-stacked spectra of the {\HI} \(\lambda1215\) line, the {\CIV} \(\lambda\lambda1548, 1550\) doublet, and the {\NV} \(\lambda\lambda1238, 1242\) doublet. The {\HI} \(\lambda1215\) absorption profile seems to contain three kinematic components. However, absorption in the {\NV} doublet only appears in one component at \(-285\)\,\kms.

\subsection{PG\,$1049-005$ \(\zem = 0.3599\) PID: 12248}
\begin{figure}
\includegraphics[width=0.5\textwidth]{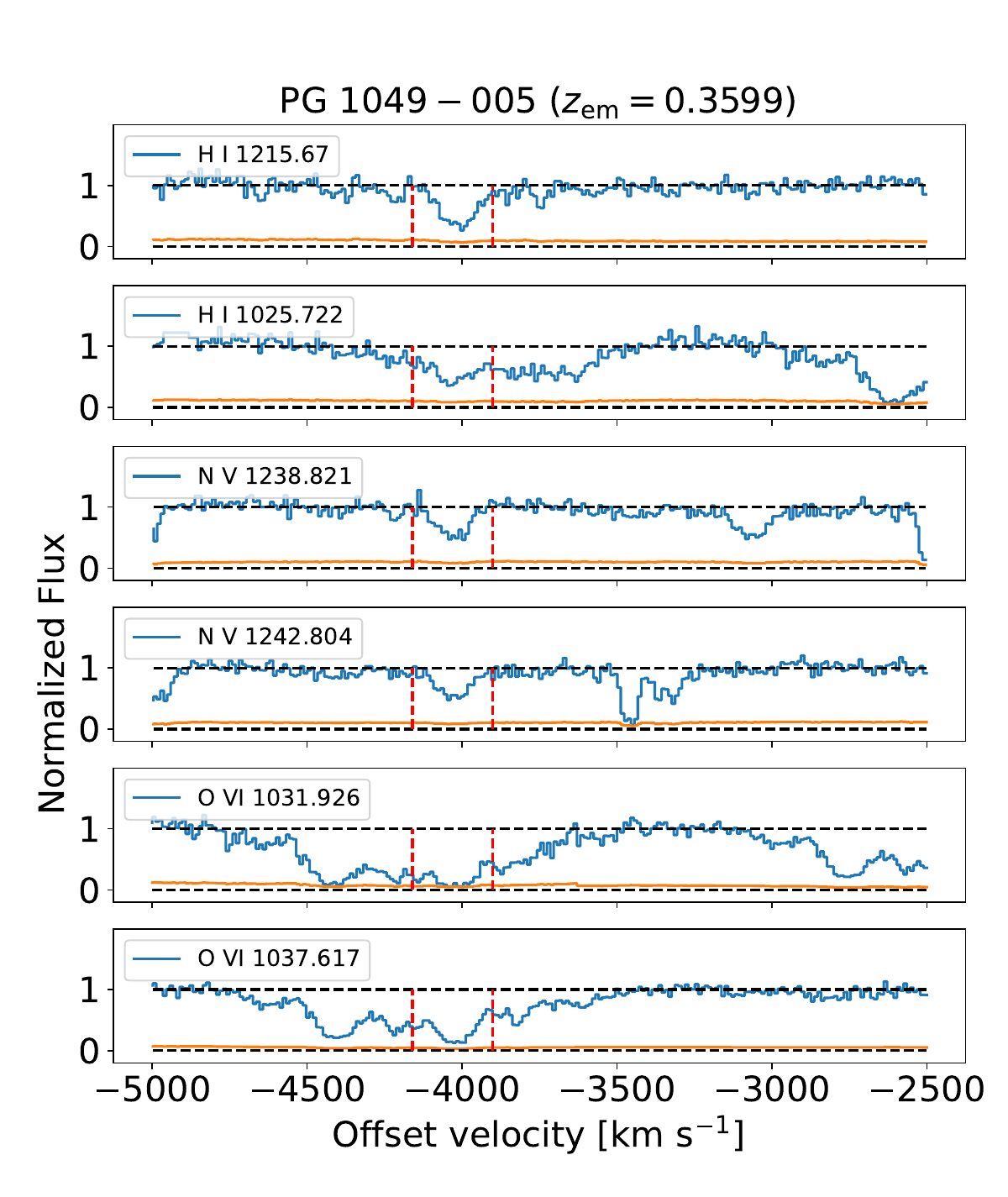}
\caption[PG\,$1049-005$ System Plot]{Intrinsic Absorption in PG\,$1049-005$. We show a velocity-stacked plot of the detection transitions for the intrinsic absorber in the archived HST/COS spectra. In each panel, the flux density is shown as a blue histogram, and the uncertainty is shown as an orange histogram. For clarity, spectra are shown with a sampling of 5 pixels per bin. A horizontal dashed black line mark the zero-flux level. Vertical red dashed lines encompass the absorption system at \(-4024\)\,\kms\ relative to \(\zem = 0.3599\). }
\label{fig:pg1049}
\end{figure}

COS observations for PID 12248 including this target have primarily focused on intervening absorbers. In Figure~\ref{fig:pg1049}, we present velocity-stacked spectra of the {\HI} \(\lambda1025, 1215\) lines, the {\NV} \(\lambda\lambda1238, 1242\) doublet, and the {\OVI} \(\lambda\lambda1031, 1037\) doublet. There is a single component lying at an equivalent-width-weighted average velocity of \(-4024\)\,\kms\ relative to \(\zem = 0.3599\).

\subsection{{[}VV2006{]}\,J$110313.0+414154$ \(\zem = 0.403095\) PID: 12248}
\begin{figure}
\includegraphics[width=0.5\textwidth]{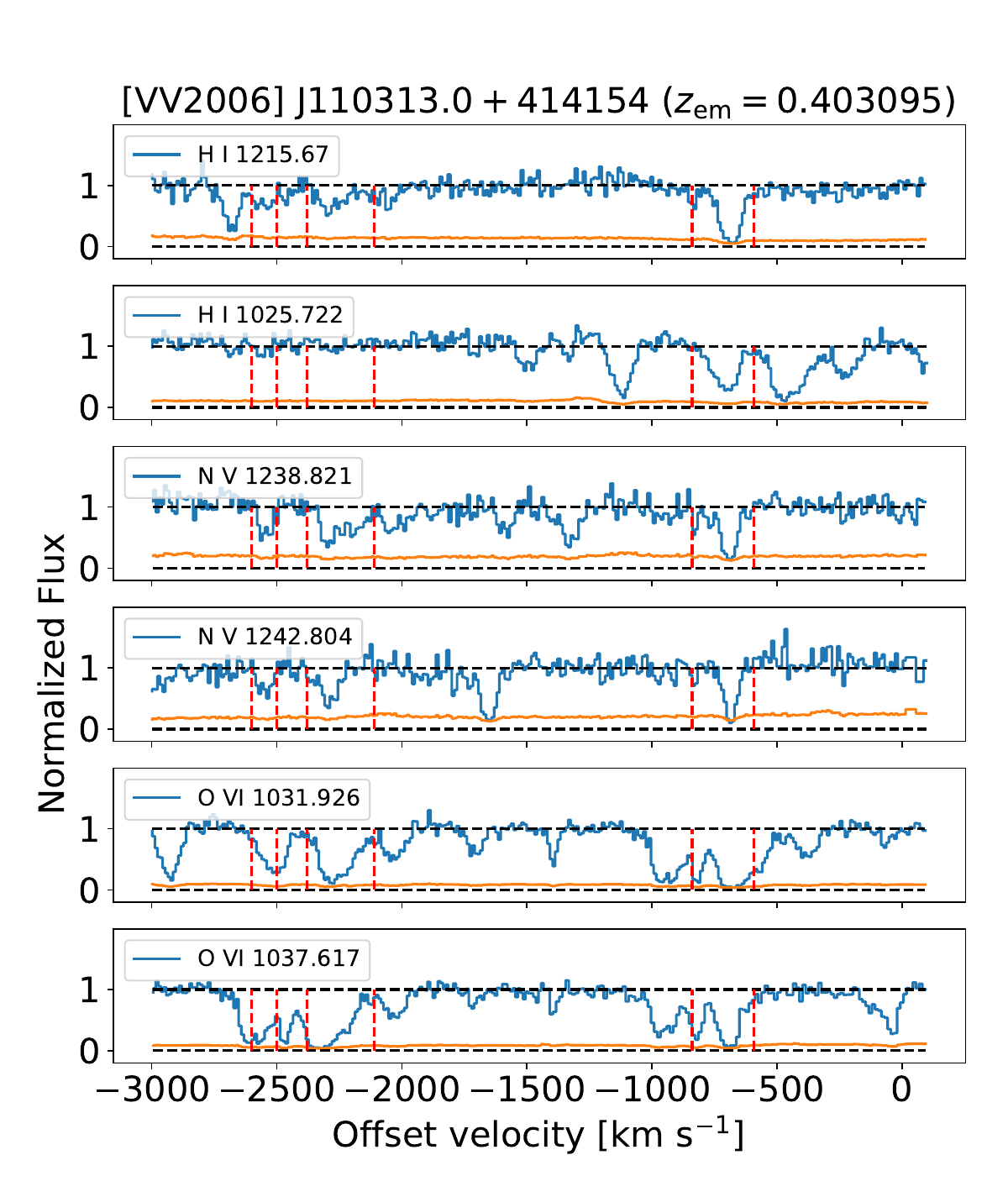}
\caption[{[}VV2006{]}\,J$110313.0+414154$ System Plot]{Intrinsic Absorption in {[}VV2006{]}\,J$110313.0+414154$. We show a velocity-stacked plot of the detection transitions for the intrinsic absorber in the archived HST/COS spectra. In each panel, the flux density is shown as a blue histogram, and the uncertainty is shown as an orange histogram. For clarity, spectra are shown with a sampling of 5 pixels per bin. A horizontal dashed black line mark the zero-flux level. Vertical red dashed lines encompass the absorption systems at \(-2542\)\,\kms, \(-2249\)\,\kms, and \(-707\)\,\kms\ relative to \(\zem = 0.403095\). }
\label{fig:q1103}
\end{figure}

COS observations for PID 12248 including this target have primarily focused on intervening absorbers. In Figure~\ref{fig:q1103}, we present velocity-stack spectra of the {\HI} \(\lambda1025, 1215\) lines, the {\NV} \(\lambda\lambda1238, 1242\) doublet, and the {\OVI} \(\lambda\lambda1031, 1037\) doublet. There are three clearly separated component in the velocity range \([-2600,-590]\)\,\kms\ lying at an equivalent-width-weighted average velocities of \(-2542\)\,\kms, \(-2249\)\,\kms, and \(-707\)\,\kms\ relative to \(\zem = 0.403095\).

\subsection{NGC\,3516 \(\zem = 0.00884\) PID: 12212}
\citet{2018ApJ...854..166D} analyzed the intrinsic absorption in the COS spectrum of this object (see their Figure 3). They report absorption components in the velocity range \(-1700\)\,\kms\ to \(0\)\,\kms. Most of the \NV\ absorption lies in the range \([-600,+30]\)\,\kms and we adopt an absorption \emph{system} velocity of \(-252\)\,\kms.

\subsection{2MASSi\,J$1118302+402553$ \(\zem = 0.154338\) PID: 11519}

\begin{figure}
\includegraphics[width=0.5\textwidth]{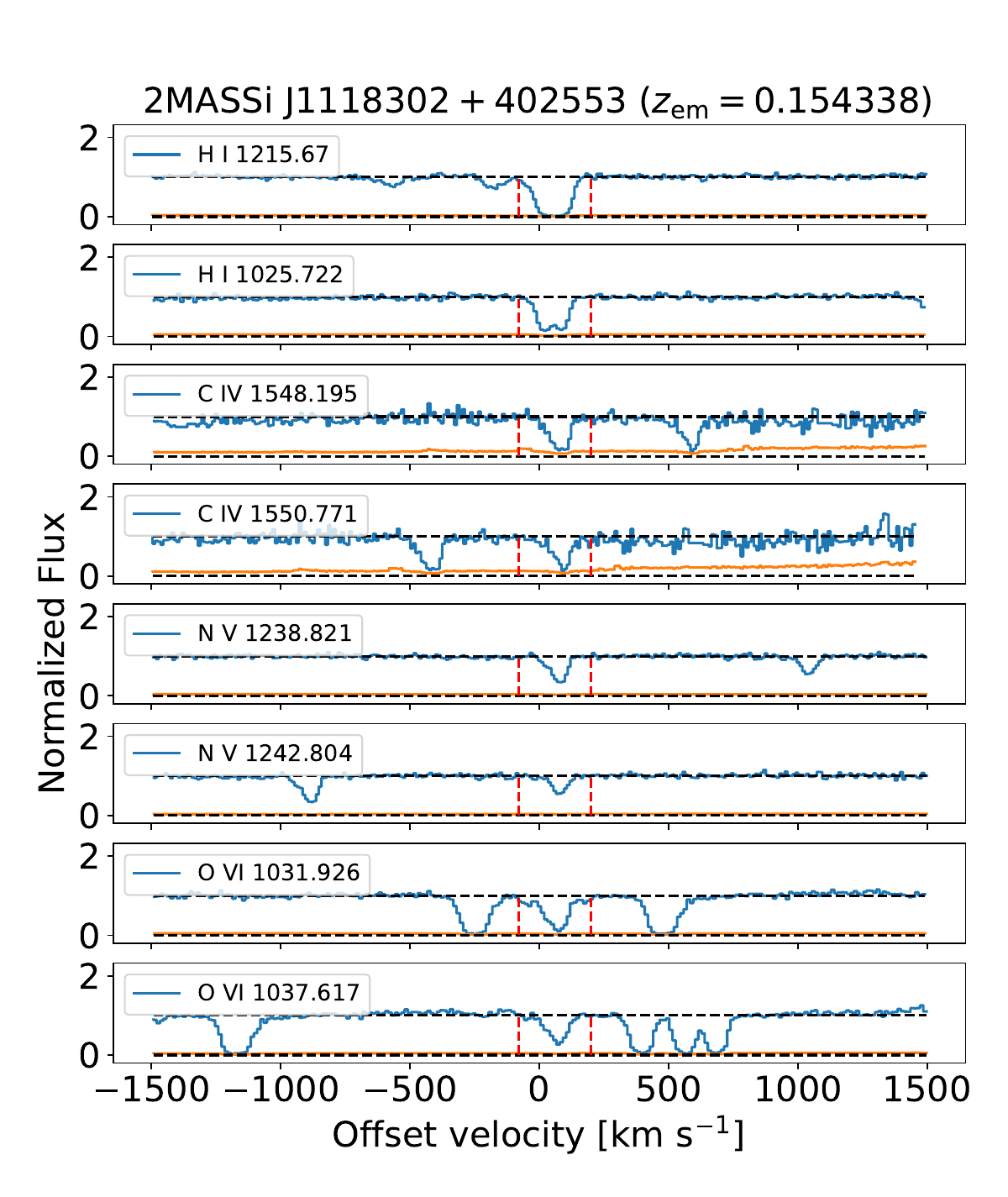}
\caption[2MASSi\,J$1118302+402553$ System Plot]{Intrinsic Absorption in 2MASSi\,J$1118302+402553$. We show a velocity-stacked plot of the detection transitions for the intrinsic absorber in the archived HST/COS spectra. In each panel, the flux density is shown as a blue histogram, and the uncertainty is shown as an orange histogram. For clarity, spectra are shown with a sampling of 5 pixels per bin. A horizontal dashed black line mark the zero-flux level. Vertical red dashed lines encompass the absorption system at \(-84\)\,\kms\ relative to \(\zem = 0.154338\).}
\label{fig:pg1115}
\end{figure}

This object, also known as PG\,\(1115+407\), was part of the original GTO set of observations for PID 11519. An apparently-single component is observed in the intrinsic absorption from {\HI} \(\lambda\lambda1025, 1215\) lines, the {\CIV} \(\lambda\lambda1548, 1550\) doublet, the {\NV} \(\lambda\lambda1238, 1242\) doublet, and the {\OVI} \(\lambda\lambda1031, 1037\) doublet (Fig.~\ref{fig:pg1115}) in the velocity range \([-80,+200]\)\,\kms. We adopt equivalent-width-weighted velocity of \(-84\)\,\kms\ relative to \(\zem = 0.154338\).

\subsection{ESO\,$265- \mathrm{G}023$ \(\zem = 0.056576\) PID: 12275}
\begin{figure}
\includegraphics[width=0.5\textwidth]{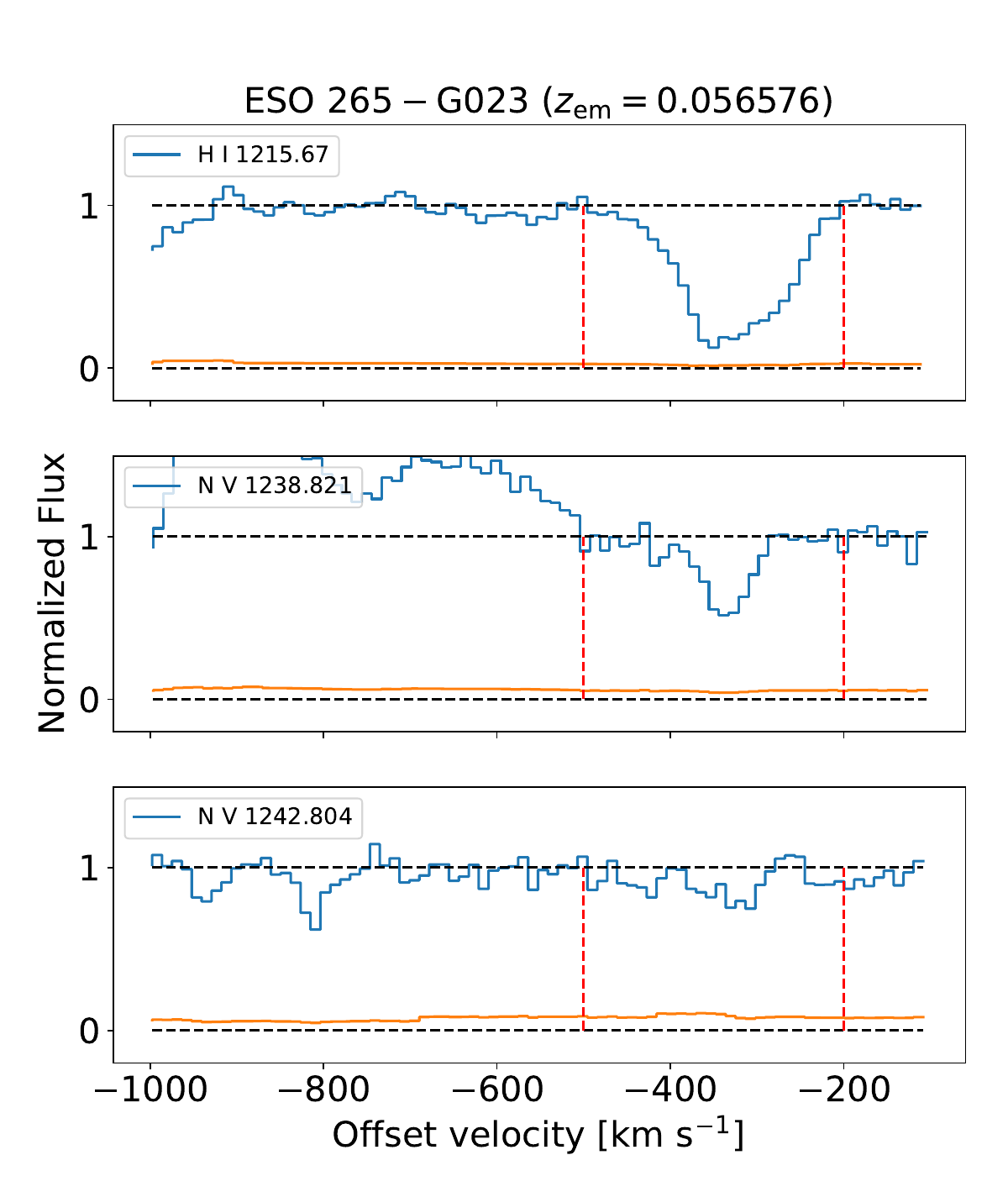}
\caption[ESO 265-G023 System Plot]{Intrinsic Absorption in ESO\,$265- \mathrm{G}023$. We show a velocity-stacked plot of the detection transitions for the intrinsic absorber in the archived HST/COS spectra. In each panel, the flux density is shown as a blue histogram, and the uncertainty is shown as an orange histogram. For clarity, spectra are shown with a sampling of 5 pixels per bin. The excess flux to the left of the {\NV} \(\lambda1238\) absorption is geochoronal emission from {\OI} \(\lambda1302.168\) and {\SiII} \(1304.370\). A horizontal dashed black line mark the zero-flux level. Vertical red dashed lines encompass the absorption component at \(-342\)\,\kms.}
\label{fig:eso265}
\end{figure}

COS observations for PID 12275 including this target have primarily focussed on intervening absorbers. In Figure~\ref{fig:eso265}, we present velocity-stacked spectra of the {\HI} \(\lambda1215\) line and the {\NV} \(\lambda\lambda1238, 1242\) doublet. One component is detected at \(-342\)\,\kms.

\subsection{Mrk\,1298 \(\zem = 0.059801\) PID: 12569}
This target was part of the QUEST sample and the COS spectra are presented in \citet[][see their Figure~3]{2022ApJ...926...60V}. The COS spectrum of Mrk\,1298 features complex absorption with many components \citep[\(\sim13\)\ as reported by ][]{2022ApJ...926...60V} spanning the velocity range \(-5000\)\,\kms\ to \(+400\)\,\kms\ relative to the systemic redshift of \(\zem = 0.059801\). [We note that the systemic redshift of Mrk\,1298 has been reported in range \(0.0598 < \zem < 0.062284\). \citet{2022ApJ...926...60V} used a systemic redshift of \(\zem = 0.060\). We adopt the most recent report from \citet{2023MNRAS.520.2351G}.] In keeping with our approach of grouping blended components into systems, we identify four systems at \(-3967\)\,\kms, \(-2720 \)\,\kms, \(-1967 \)\,\kms, 
\(-991\)\,\kms, and \(-87\)\,\kms.

\subsection{6dF\,J$115758.8-002221$  \(\zem = 0.259812\) PID: 11598}

\begin{figure}
\includegraphics[width=0.5\textwidth]{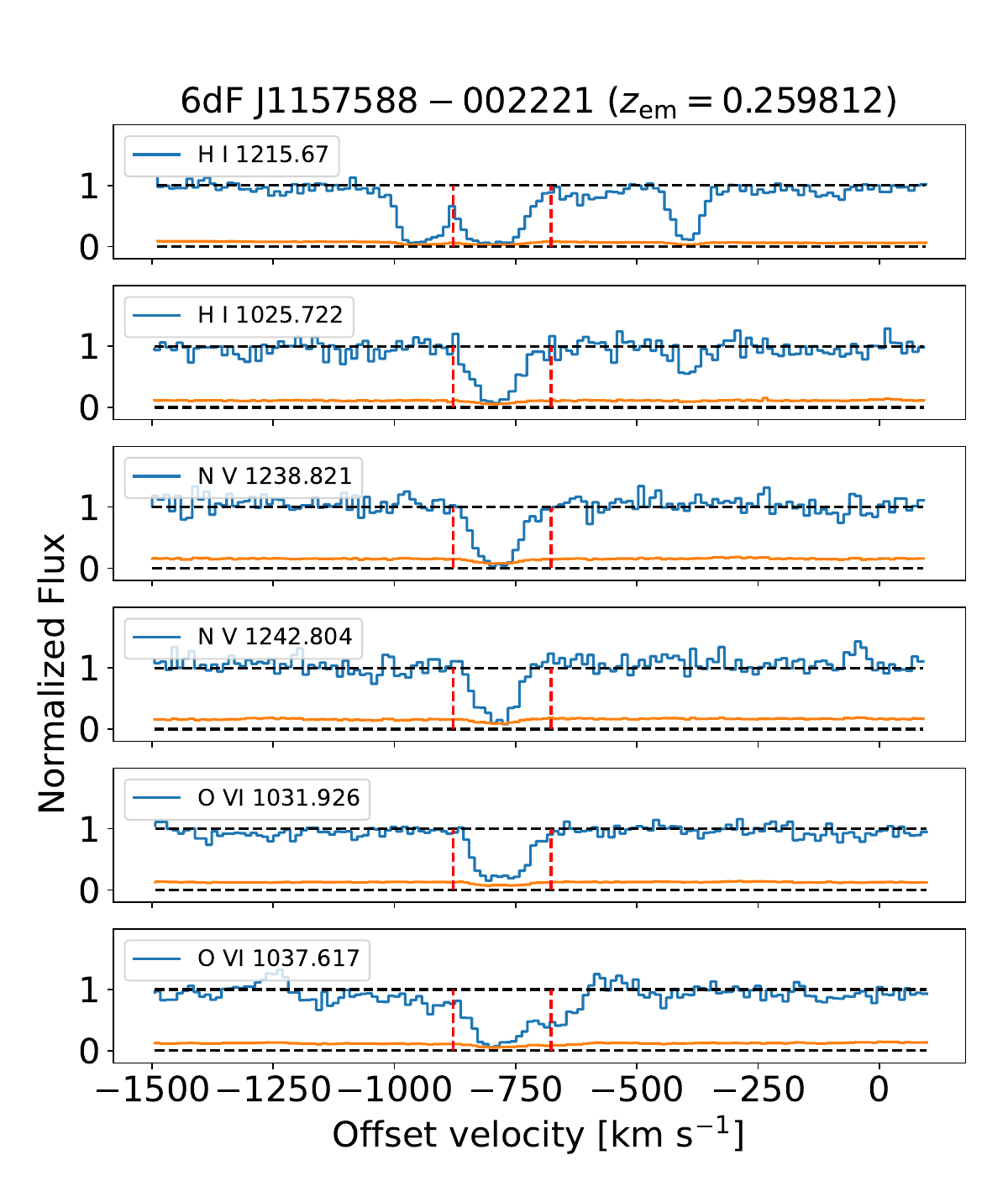}
\caption[6dF\,J$115758.8-002221$ System Plot]{Intrinsic Absorption in 6dF\,J$115758.8-002221$. We show a velocity-stacked plot of the detection transitions for the intrinsic absorber in the archived HST/COS spectra. In each panel, the flux density is shown as a blue histogram, and the uncertainty is shown as an orange histogram. For clarity, spectra are shown with a sampling of 5 pixels per bin. A horizontal dashed black line mark the zero-flux level. The {\HI} \(\lambda1215\) profile shows additional absorption as it is blended with the Galactic {\SiII} \(\lambda1526\) line. Vertical red dashed lines encompass the absorption system at \(-781\)\,\kms\ relative to \(\zem = 0.259812\).}
\label{fig:q1157}
\end{figure}

COS observations for this object were part of the COS-Halos project \citep{2013ApJ...777...59T,2013ApJS..204...17W} and have been used extensively in studies of intervening absorbers. In Figure~\ref{fig:q1157}, we present velocity-stack spectra of the {\HI} \(\lambda\lambda1025, 1215\) lines, the {\NV} \(\lambda\lambda1238, 1242\) doublet, and the {\OVI} \(\lambda\lambda1031, 1037\). The profiles appear to be a single component in the velocity range \([-880,-670]\)\,\kms. Note that the {\HI} \(\lambda1215\) profile shows additional absorption outside this range. This is a blend with the Galactic {\SiII} \(\lambda1526\) line. We adopt an equivalent-width-weighted average velocity of \(-781\)\,\kms\ relative to \(\zem = 0.259812\)\ for the system.

\subsection{NGC\,4051 \(\zem = 0.00234\) PID: 11834}
The COS spectrum for this object has been previously analyzed by \citet{2012ApJ...751...84K} (following up on an analysis of an HST/STIS observation by \citet{2001ApJ...557....2C}) who reported \NV\ absorption in five blended kinematics components in the velocity range \(-505\)\,\kms\ to \(-48\)\,\kms\ (see their Figures 3 and 7). Since these are all blended together, we adopt a single absorption \emph{system} velocity of \(-374\)\,\kms.

\subsection{ESO\,$267-13$ \(\zem = 0.01488\) PID: 12275}
\begin{figure}
\includegraphics[width=0.5\textwidth]{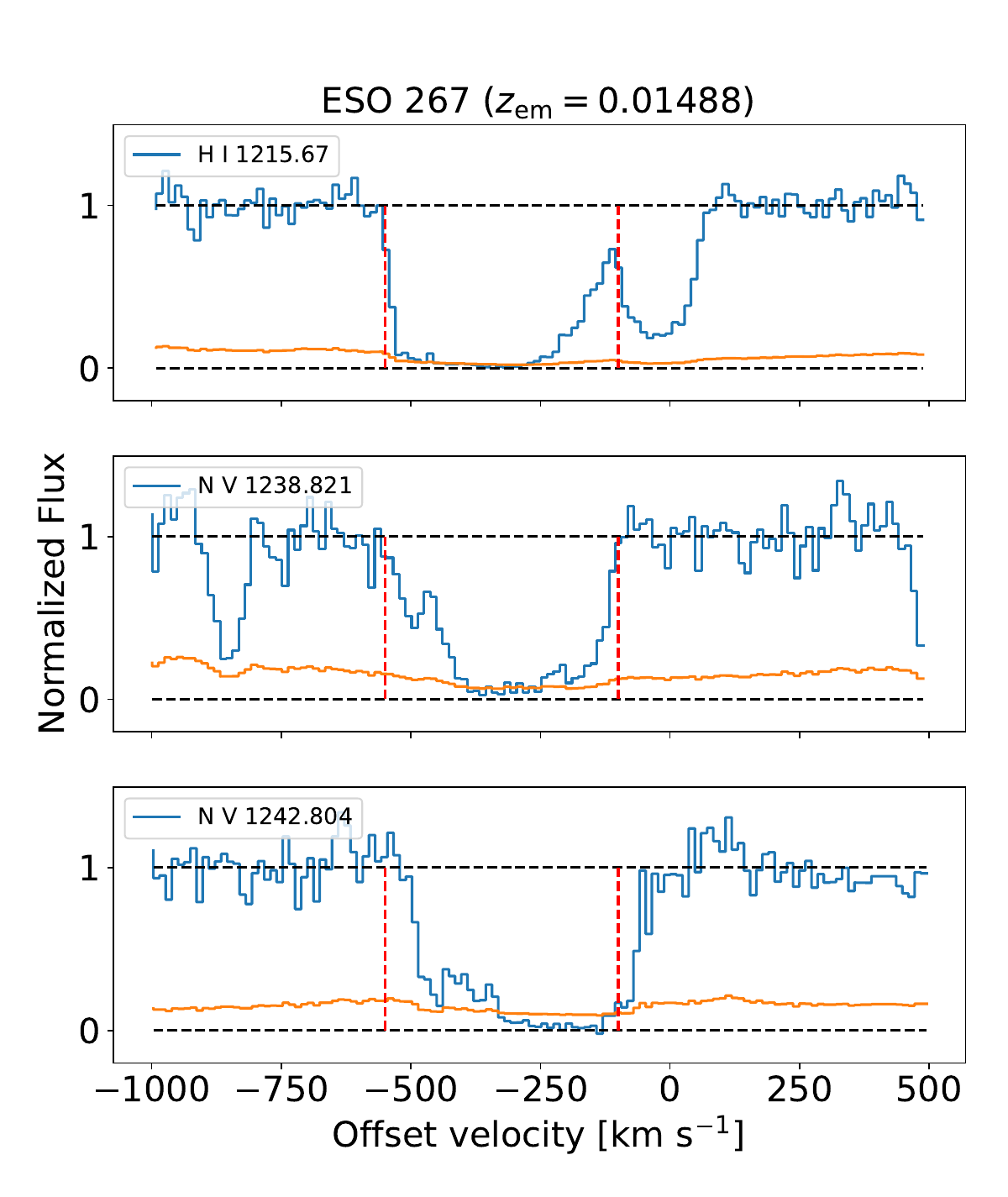}
\caption[ESO 267 System Plot]{Intrinsic Absorption in ESO\,267-G013. We show a velocity-stacked plot of the detection transitions for the intrinsic absorber in the archived HST/COS spectra. In each panel, the flux density is shown as a blue histogram, and the uncertainty is shown as an orange histogram. For clarity, spectra are shown with a sampling of 5 pixels per bin. A horizontal dashed black line mark the zero-flux level. Vertical red dashed lines mark the upper and lower velocity limits of the detected absorption at \(-550\)\,\kms\ and \(-100\)\,\kms, respectively.}
\label{fig:eso267}
\end{figure}

The COS spectrum of this object was originally taken for the purpose of ``Measuring gas flow rates in the Milky Way.'' The spectra have previously been analyzed for Milky Way and high-velocity cloud absorption \citep[e.g.,][]{2014ApJ...787..147F} and low-redshift IGM absorption \citep[e.g.,][]{2018MNRAS.476.4965M,2024MNRAS.530.3827S}. However, there has not been an analysis of this spectrum for intrinsic \NV\ absorption. We detect an \NV\ absorption system (Fig.~\ref{fig:eso267}) spanning the velocity range \(-500\)\,\kms -- \(-100\)\,\kms and we adopt a system velocity at the equivalent-width weighted average of \(-304\)\,\kms.

\subsection{2MASS\,J$12072101+2624292$  \(\zem = 0.322492\) PID: 12248}

\begin{figure}
\includegraphics[width=0.5\textwidth]{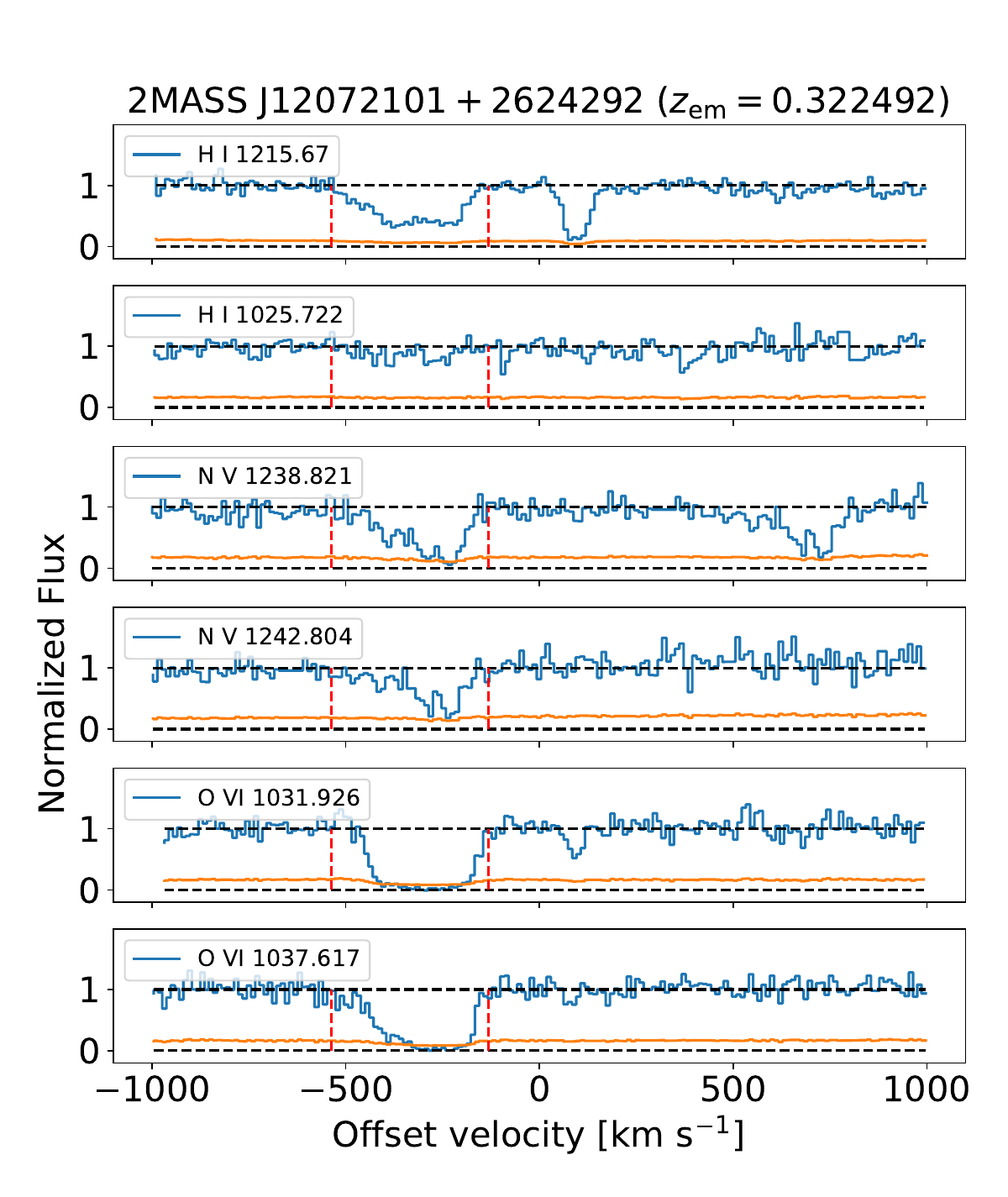}
\caption[2MASS\,J$12072101+2624292$ System Plot]{Intrinsic Absorption in 2MASS\,J$12072101+2624292$. We show a velocity-stacked plot of the detection transitions for the intrinsic absorber in the archived HST/COS spectra. In each panel, the flux density is shown as a blue histogram, and the uncertainty is shown as an orange histogram. For clarity, spectra are shown with a sampling of 5 pixels per bin. A horizontal dashed black line mark the zero-flux level. Vertical red dashed lines encompass the absorption system at \(-286\)\,\kms\ relative to \(\zem = 0.322492\). }
\label{fig:q1207}
\end{figure}

COS observations for PID 12248 including this target have primarily focused on intervening absorbers. In Figure~\ref{fig:q1207}, we present velocity-stack spectra of the {\HI} \(\lambda\lambda1215,1025\) lines, the {\NV} \(\lambda\lambda1238, 1242\) doublet, and the {\OVI} \(\lambda\lambda1031, 1037\) doublet. The blended components lying in the velocity range \([-540,-130]\)\,\kms\ are at an equivalent-width-weighted average velocity of \(-286\)\,\kms\ relative to \(\zem = 0.322492\).

\subsection{2MASS\,J$12305003+0115226$ \(\zem = 0.117\) PID: 11686}

\begin{figure}
\includegraphics[width=0.5\textwidth]{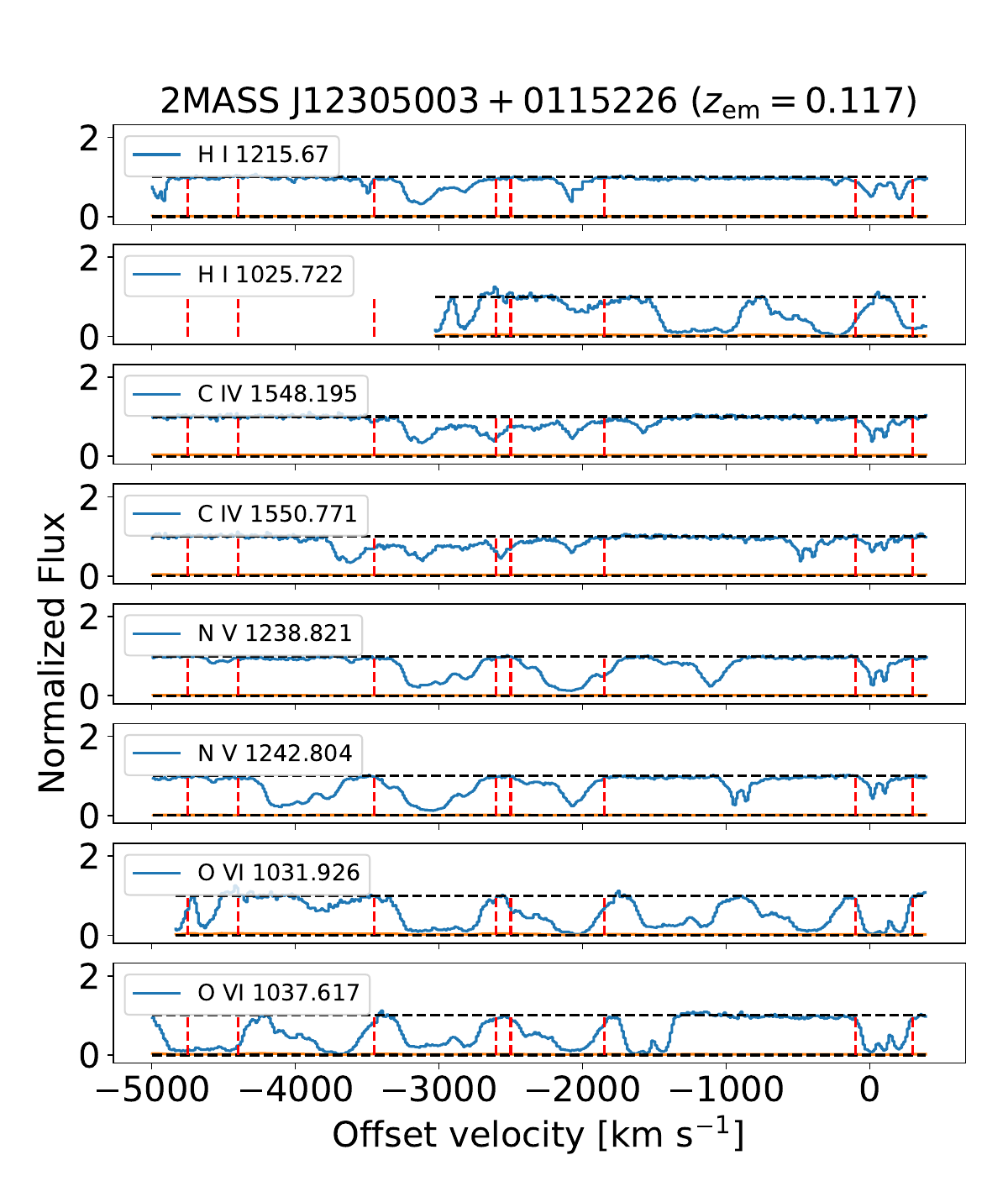}
\caption[2MASS\,J$12305003+0115226$ System Plot]{Intrinsic Absorption in 2MASS\,J$12305003+0115226$. We show a velocity-stacked plot of the detection transitions for the intrinsic absorber in the archived HST/COS spectra. In each panel, the flux density is shown as a blue histogram, and the uncertainty is shown as an orange histogram. For clarity, spectra are shown with a sampling of 5 pixels per bin. A horizontal dashed black line mark the zero-flux level. Vertical red dashed lines encompass the apparently line-locked absorption systems at \(-4563\)\,\kms, \(-3009\)\,\kms, \(-2112\)\,\kms, and \(+66\)\,\kms.}
\label{fig:rxj1230}
\end{figure}

STIS/E140M and FUSE spectra of this target were studied in detail by \citet[][see their Figure~3]{2003ApJ...598..922G}.  Here, we present the COS spectra covering the same wavelength range, taken 10.3 years later in the quasar frame in Fig~\ref{fig:rxj1230}. Several absorption systems reside in the velocity range \([-5000,+300]\)\,\kms. In keeping with our approach of grouping blended components, we adopt the following equivalent-width-weighted system velocities: \(-4563\)\,\kms, \(-3009\)\,\kms, \(-2112\)\,\kms, and \(+66\)\,\kms.

\subsection{[HB89]\,1339+053 \(\zem = 0.265710\)  PID: 12248}

\begin{figure}
\includegraphics[width=0.5\textwidth]{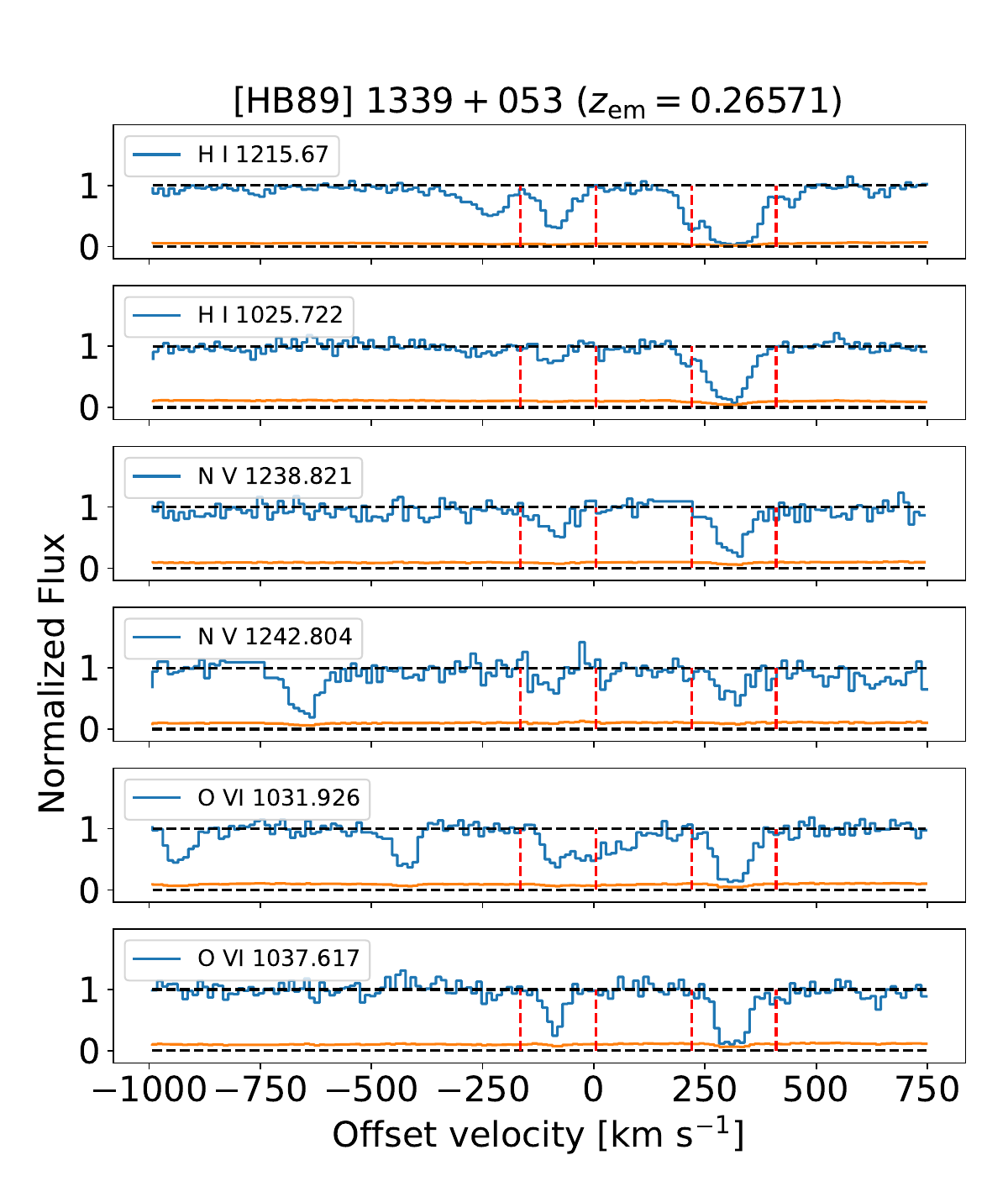}
\caption[{[}HB89{]}\,1339+053 System Plot]{Intrinsic Absorption in {[}HB89{]}\,1339+053. We show a velocity-stacked plot of the detection transitions for the intrinsic absorber in the archived HST/COS spectra. In each panel, the flux density is shown as a blue histogram, and the uncertainty is shown as an orange histogram. For clarity, spectra are shown with a sampling of 5 pixels per bin. A horizontal dashed black line mark the zero-flux level. Vertical red dashed lines encompass the absorption systems at \(-94\)\,\kms\ and \(+324\)\,\kms\ relative to \(\zem = 0.265710\).}
\label{fig:q1339}
\end{figure}

COS observations for PID 12248 including this target have primarily focused on intervening absorbers. In Figure~\ref{fig:q1339}, we present velocity-stack spectra of the {\HI} \(\lambda\lambda1025, 1215\) lines the {\NV} \(\lambda\lambda1238, 1242\) doublet, and the {\OVI} \(\lambda\lambda1031, 1037\) doublet. Two kinematic structures are observed in {\NV} absorption in the  velocity ranges \([-165,5]\)\,\kms\ and \([  220,  410]\)\,\kms\ relative to \(\zem = 0.265710\) with {\NV} \(\lambda1238\) equivalent-width-weighted average velocities of \(-94\)\,\kms\ and \(+324\)\,\kms, respectively.

\subsection{PB\,04055 \(\zem = 0.382744\) PID: 12248}

\begin{figure}
\includegraphics[width=0.5\textwidth]{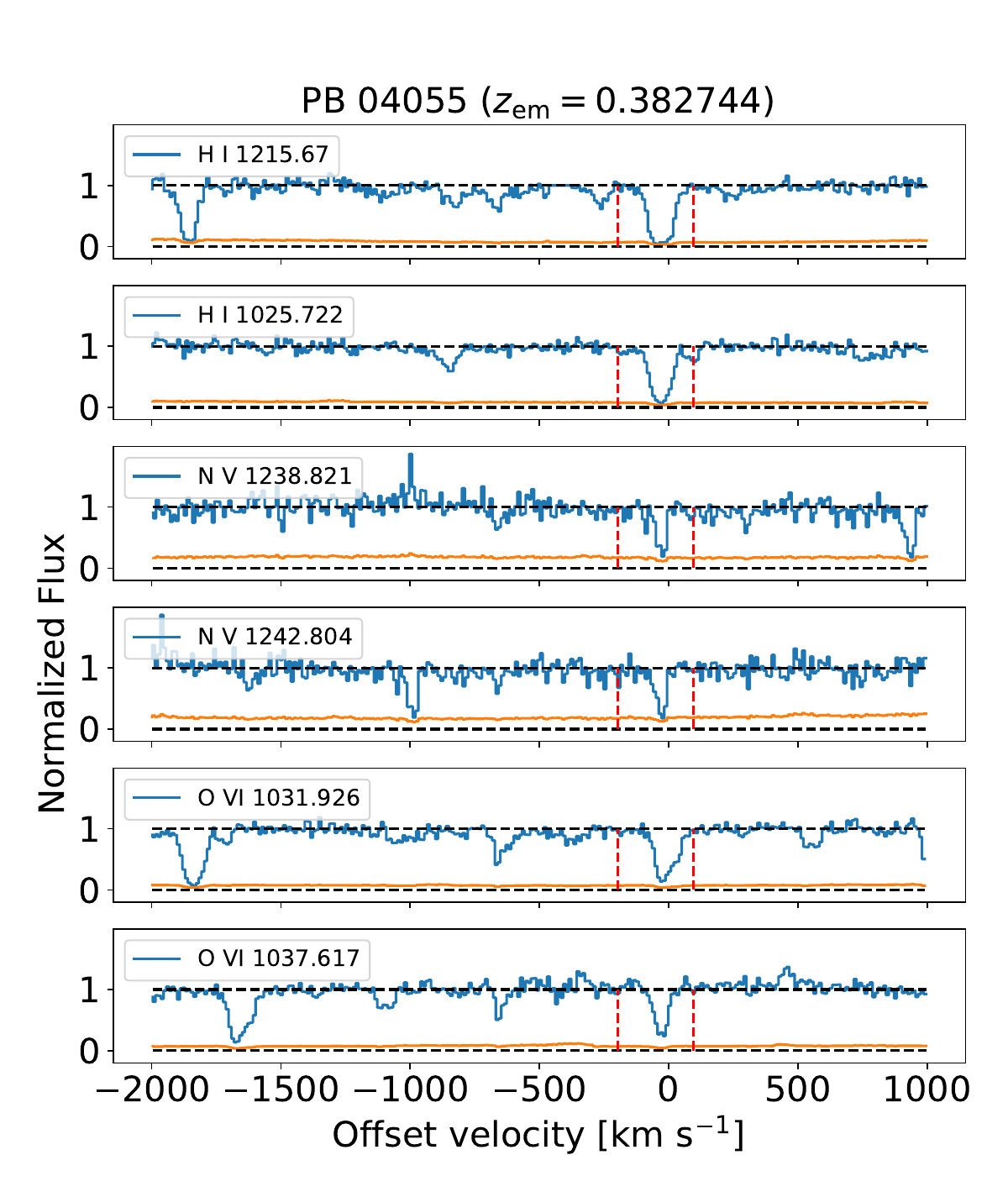}
\caption[PB\,04055 System Plot]{Intrinsic Absorption in PB\,04055. We show a velocity-stacked plot of the detection transitions for the intrinsic absorber in the archived HST/COS spectra. In each panel, the flux density is shown as a blue histogram, and the uncertainty is shown as an orange histogram. For clarity, spectra are shown with a sampling of 5 pixels per bin. A horizontal dashed black line mark the zero-flux level. Vertical red dashed lines encompass the absorption system at \(-29\)\,\kms\ relative to \(\zem = 0.382744\). }
\label{fig:pb04055}
\end{figure}

COS observations for PID 12248 including this target have primarily focused on intervening absorbers. In Figure~\ref{fig:pb04055}, we present velocity-stack spectra of the {\HI} \(\lambda\lambda1025, 1215\) lines, the {\NV} \(\lambda\lambda1238, 1242\) doublet, and the {\OVI} \(\lambda\lambda1031, 1037\) doublet. There is a single component lying at an equivalent-width-weighted average velocity of \(-29\)\,\kms\ relative to \(\zem = 0.382744\).

\subsection{Mrk\,279 \(\zem = 0.0305\) PID: 12212 (6593?)}
This target is one of a handful that have been subject to major UV/X-ray observing campaigns. Analysis of UV spectra using FUSE and HST/STIS from these campaigns are summarized in \citet{2004ApJS..152....1S}, \citet{2005ApJ...623...85G}, \citet{2007A&A...461..121C}, and \citet{2009ApJ...694..438S}. Intrinsic absorption in the HST/COS spectrum were presented by \citet{2015AAS...22543203S}.

Two primary absorption components are seen at \(-450\)\,\kms\ and \(+90\)\,\kms\ relative to the systemic redshift at \(\zem = 0.0305\), but with additional absorption in the \(-600\)\,\kms\ to \(+100\)\,\kms. In this velocity range, we adopt a absorption \emph{system} velocity of \(-369\)\,\kms.

\subsection{2XMM\,J$135315.8+634546$ \(\zem = 0.0882\) PID: 12569}

This target was part of the QUEST sample and the COS spectra are presented in \citet[][see their Figure~14l]{2022ApJ...926...60V}. They report two broad absorption systems in the velocity ranges \([-2200, -1500]\)\,\kms\ and \([-1400, -600]\)\,\kms\ relative to \(\zem = 0.088\). The velocity separation matches that of the {\NV} suggesting possible line-locking. These are separated by unabsorbed regions so we adopt two velocities at \(-1795\)\kms\ and \(-992\)\,\kms.

\subsection{SDSS\,J$135625.55+251523.7$ \(\zem = 0.164009\) PID: 12248}

\begin{figure}
\includegraphics[width=0.5\textwidth]{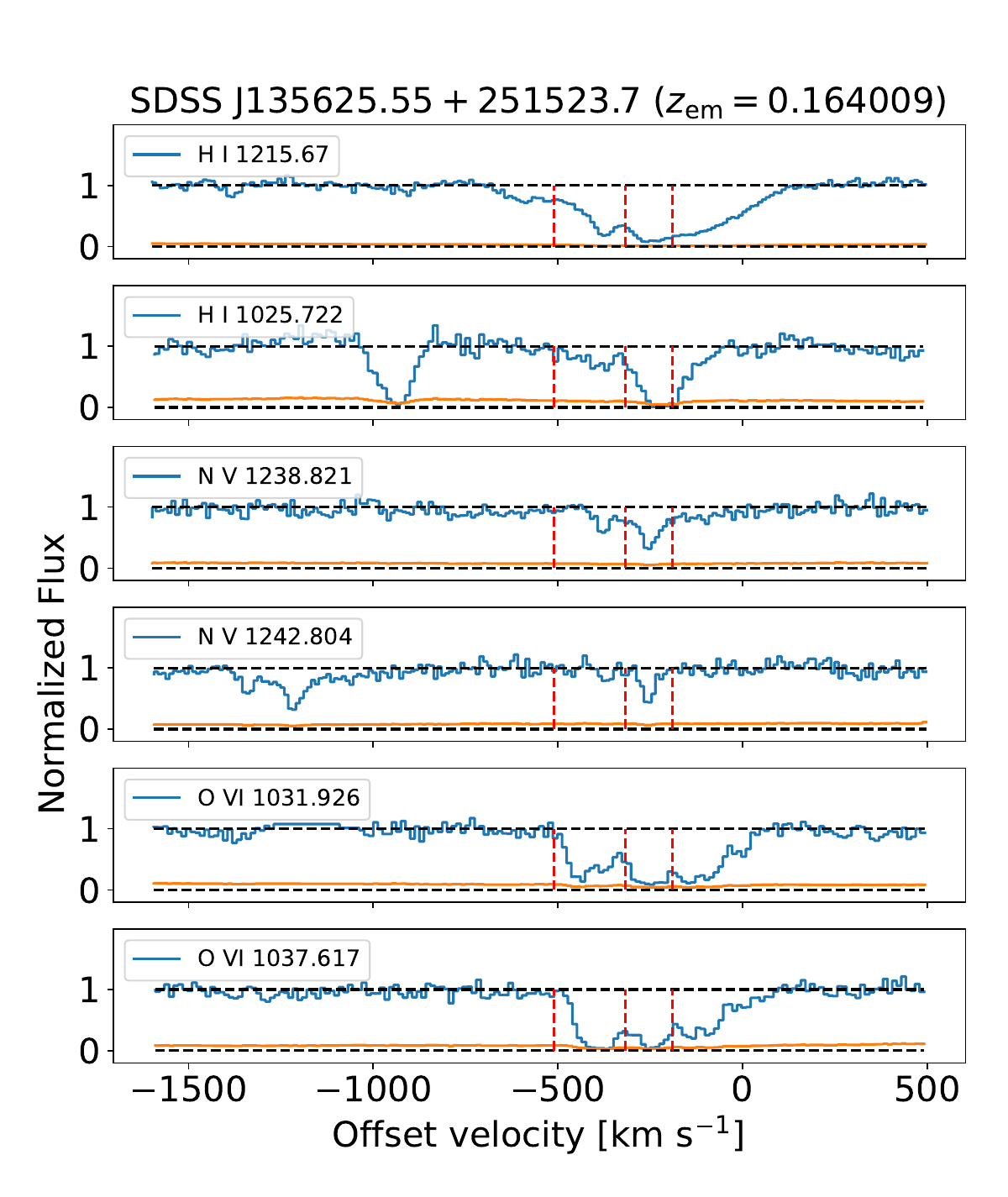}
\caption[SDSS\,J$135625.55+251523.7$ System Plot]{Intrinsic Absorption in SDSS\,J$135625.55+251523.7$. We show a velocity-stacked plot of the detection transitions for the intrinsic absorber in the archived HST/COS spectra. In each panel, the flux density is shown as a blue histogram, and the uncertainty is shown as an orange histogram. For clarity, spectra are shown with a sampling of 5 pixels per bin. A horizontal dashed black line mark the zero-flux level. Vertical red dashed lines encompass the absorption systems at \(-340\)\,\kms\ and \(-250\)\,\kms\ relative to \(\zem = 0.1265\). Additional absorption in the velocity range \([-745,+120]\)\,\kms, outside the displayed boundaries, is also evident in {\HI} \(\lambda1215\), as well as {\HI} \(\lambda1025\) and the {\OVI} \(\lambda\lambda1031,1037\) doublet (not shown).}
\label{fig:q1356}
\end{figure}

COS observations for PID 12248 including this target have primarily focused on intervening absorbers. In Figure~\ref{fig:q1356}, we present velocity-stack spectra of the {\HI} \(\lambda\lambda1025, 1215\) lines, the {\NV} \(\lambda\lambda1238, 1242\) doublet, and the {\OVI} \(\lambda\lambda1031, 1037\) doublet. In both {\HI} and {\OVI}, the system appears as a single blend of components. However, in {\NV}, two components are cleanly separated at \(-340\)\,\kms\ and \(-250\)\,\kms\ relative to \(\zem = 0.164009\).

\subsection{2XMM\,J$141348.3+440014$ \(\zem = 0.0896\) PID: 12569}
This target was part of the QUEST sample and the COS spectra are presented in \citet[][see their Figure~14m]{2022ApJ...926...60V} with a more in-depth analysis presented in \citet{2019MNRAS.487.5041H}. They report a mini-BAL in the velocity range \([-2800, -900]\)\,\kms, as well as a blended, apparently two-component, associated absorber at \(\sim0\)\,\kms. For this absorber, we measure an equivalent-width weighted average velocity of \(+18\)\,\kms\ relative to \(\zem = 0.0896\).

\subsection{NGC\,5548 \(\zem = 0.0172\) PID: 12212}
NGC\,5548 was part of a deep multiwavelength campaign \citep{2014Sci...345...64K,2015A&A...575A..22M}. \citet{2015A&A...577A..37A} present the analysis of the COS spectrum from that campaign. Originally, \citet{2003ApJ...594..116C} analyzed a high resolution STIS/E140M spectrum, inferring the presence of six absorption components at \(-1041\), \(-667\), \(-530\), \(-336\), \(-166\), and \(+78\)\,\kms. \citet{2015A&A...577A..37A} also report six absorbing components with different velocities: \(-1160\), \(-720\), \(-640\), \(-475\), \(-300\), and \(-40\)\,\kms\ (see their Figure~1). We identify the following velocity ranges that group blends of components: \([-1396,-870]\)\,\kms, \([-870,-111]\)\,\kms, and \([-111,145]\)\,\kms. These yield equivalent-width-weighted system velocities of \(-1131\)\,\kms, \(-460\)\,\kms, and \(+23\)\,\kms, respectively.

\subsection{QSO\,J$1435+3604$  \(\zem = 0.429945\) PID: 11598}

\begin{figure}
\includegraphics[width=0.5\textwidth]{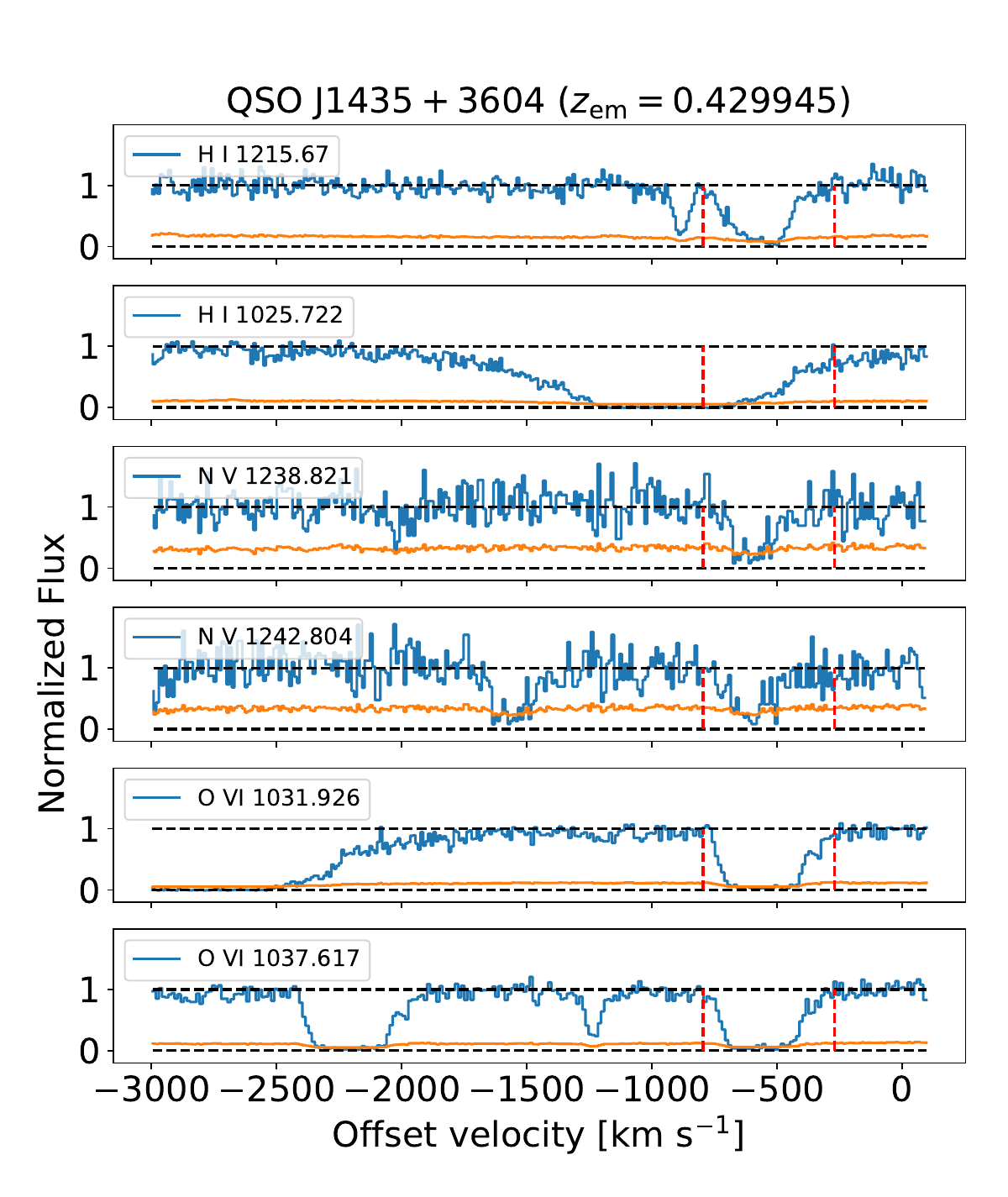}
\caption[QSO\,J$1435+3604$ System Plot]{Intrinsic Absorption in QSO\,J$1435+3604$. We show a velocity-stacked plot of the detection transitions for the intrinsic absorber in the archived HST/COS spectra. In each panel, the flux density is shown as a blue histogram, and the uncertainty is shown as an orange histogram. For clarity, spectra are shown with a sampling of 10 pixels per bin. A horizontal dashed black line mark the zero-flux level. Vertical red dashed lines encompass the absorption system at \(-550\)\,\kms\ relative to \(\zem = 0.429945\).}
\label{fig:q1435}
\end{figure}

COS observations for this object were part of the COS-Halos project \citep{2013ApJ...777...59T,2013ApJS..204...17W} and have been used extensively in studies of intervening absorbers. In Figure~\ref{fig:q1435}, we present velocity-stack spectra of the {\HI} \(\lambda\lambda1025, 1215\) lines, the {\NV} \(\lambda\lambda1238, 1242\) doublet, and the {\OVI} \(\lambda\lambda1031, 1037\) doublet. The profiles appear to be a single component in the velocity range \([-795,-270]\)\,\kms. We adopt an equivalent-width-weighted average velocity of \(-550\)\,\kms\ relative to \(\zem = 0.429945\)\ for the system.

\subsection{Mrk\,478 \(\zem = 0.079055\) PID: 12569}

This target was part of the QUEST sample and the COS spectra are presented in \citet[][see their Figure~14n]{2022ApJ...926...60V}. They report three kinematically distinct (unblended) components in the velocity range \(-2200\)\,\kms\ to \(+450\)\,\kms\ relative to the systemic redshift of \(\zem = 0.077\). While this is larger than the separation of the {\NV} doublet, the components are sufficiently narrow as to not blend the two transitions. Consequently, we maintain their separate velocities, but relative to \(\zem = 0.079055\). The new velocities are \(-2762\)\,\kms, \(-2211\)\,\kms, and \(-132\)\,\kms.

\subsection{2MASX\,J$14510879+2709272$ \(\zem = 0.065\) PID: 12248}

This target was part of the QUEST sample and the COS spectra are presented in \citet[][see their Figure~14o]{2022ApJ...926...60V}. They report four blended components in the range \([-600,+100]\)\,\kms\ relative to the systemic redshift of \(\zem = 0.065\). Since they are blended, we combine them into a single system velocity of \(-221\)\,\kms.

\subsection{2MASX\,J$15085291+6814074$ \(\zem = 0.058637\) PID: 12276}

\begin{figure}
\includegraphics[width=0.5\textwidth]{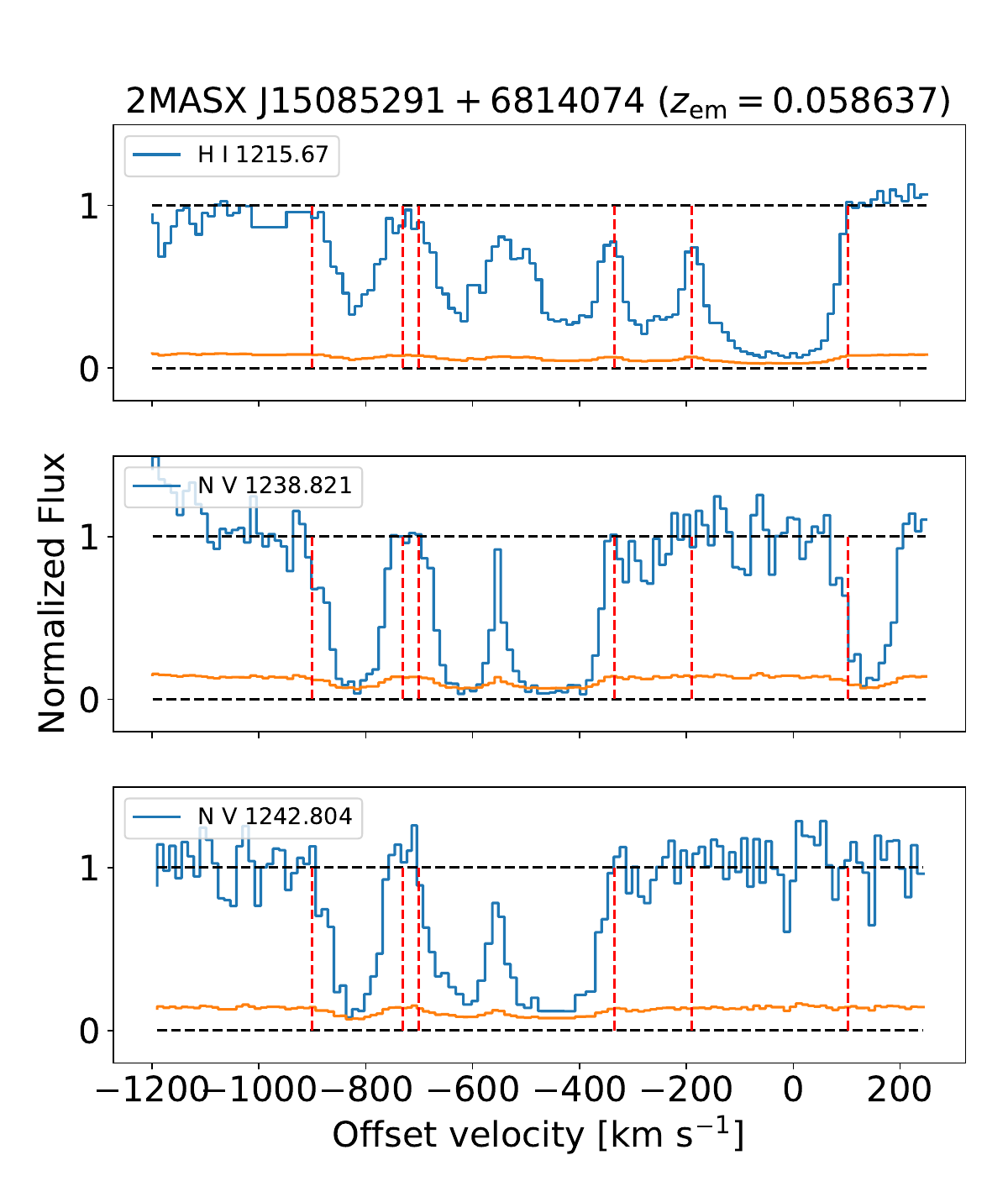}
\caption[2MASX J15085291 System Plot]{Intrinsic Absorption in 2MASX\,J$15085291+6814074$. We show a velocity-stacked plot of the detection transitions for the intrinsic absorber in the archived HST/COS spectra. In each panel, the flux density is shown as a blue histogram, and the uncertainty is shown as an orange histogram. For clarity, spectra are shown with a sampling of 5 pixels per bin. A horizontal dashed black line mark the zero-flux level. Vertical red dashed lines encompass the absorption components at \(-830\)\,\kms, and \(-xxx\)\,\kms.}
\label{fig:npm1g68}
\end{figure}

COS observations for PID 12276 including this target have primarily focussed on intervening absorbers. In Figure~\ref{fig:npm1g68}, we present velocity-stacked spectra of the {\HI} \(\lambda1215\) line and the {\NV} \(\lambda\lambda1238, 1242\) doublet. Three additional components are detected in both {\HI} \(\lambda1215\) line and the {\NV} \(\lambda1238\). However, only one is detected in {\NV} \(\lambda1242\) as the profile is truncated. That component is kinematically isolated while the others are blended. The isolated component lies at \(-830\)\,\kms\ while the blend lies at an equivalent-weighted average velocity of \(-490\)\,\kms. Additional absorption appears in the {\HI} profile at \(v>-190\)\,\kms, however the identity of the absorbing gas is unclear as there is no corroboration by absorption in other ions at the same velocities.

\subsection{2MASS\,J$1521396.7+033729.2$ \(\zem = 0.1265\) PID: 12248}
\begin{figure}
\includegraphics[width=0.5\textwidth]{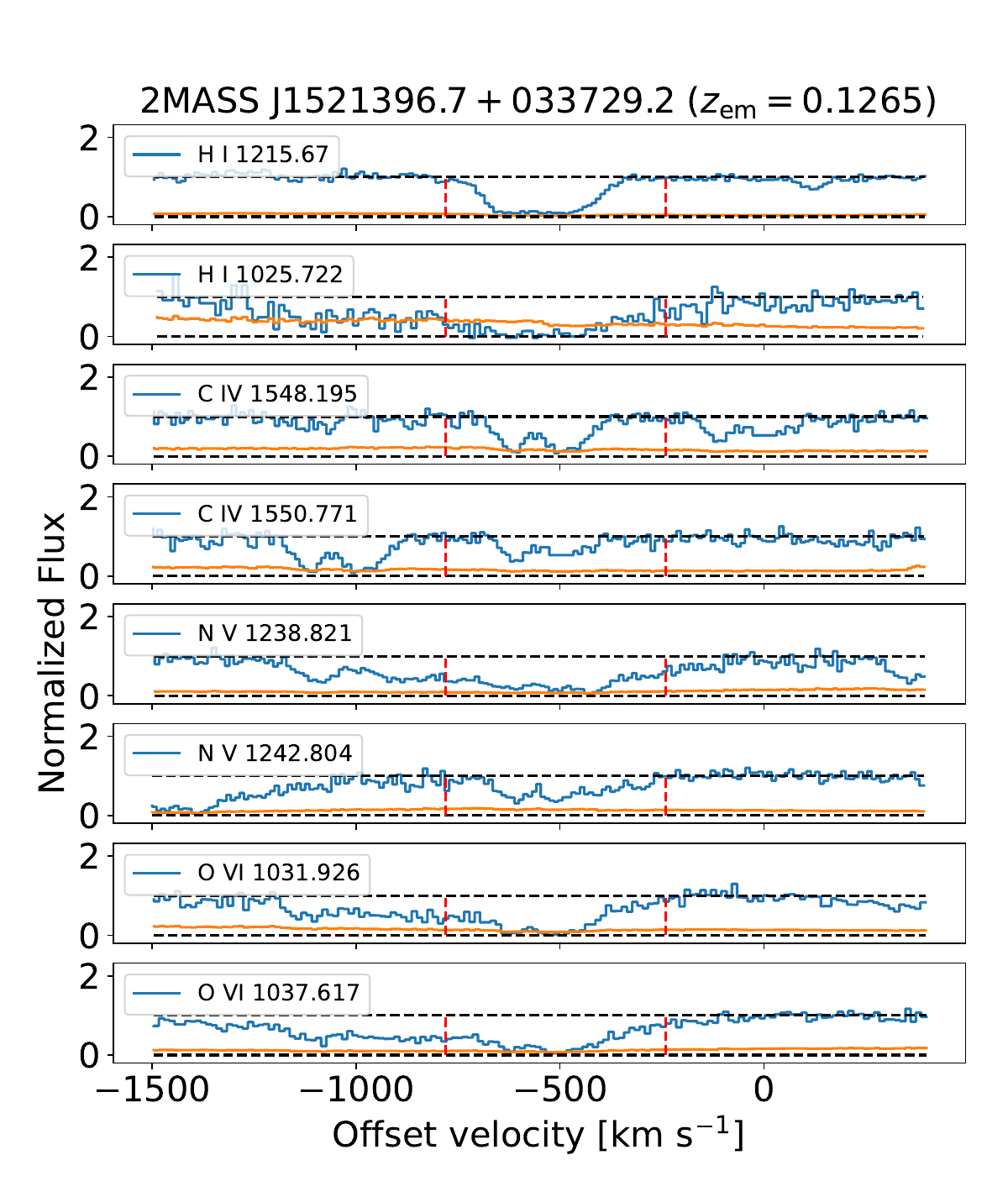}
\caption[2MASS\,J$1521396.7+033729.2$ System Plot]{Intrinsic Absorption in 2MASS\,J$1521396.7+033729.2$. We show a velocity-stacked plot of the detection transitions for the intrinsic absorber in the archived HST/COS spectra. In each panel, the flux density is shown as a blue histogram, and the uncertainty is shown as an orange histogram. For clarity, spectra are shown with a sampling of 5 pixels per bin. A horizontal dashed black line mark the zero-flux level. Vertical red dashed lines encompass the absorption system at \(-516\)\,\kms\ relative to \(\zem = 0.1265\). The {\NV} \(\lambda1238\) intrinsic absorber is blended with Galactic {\SiIV} \(\lambda1393\) absorption.}
\label{fig:q1521}
\end{figure}

COS observations for PID 12248 including this target have primarily focussed on intervening absorbers. In Figure~\ref{fig:q1521}, we present velocity-stack spectra of the {\HI} \(\lambda\lambda1025, 1215\) line, the {\CIV} \(\lambda\lambda1548, 1550\) doublet, the {\NV} \(\lambda\lambda1238, 1242\) doublet, and the {\OVI} \(\lambda\lambda1031, 1037\) doublet. One blended system is detected at \(-516\)\,\kms. The {\NV} \(\lambda1238\) intrinsic absorber is blended with Galactic {\SiIV} \(\lambda1393\) absorption.

\subsection{Mrk\,290 \(\zem = 0.030227\) PID: 11524}
\citet{2015MNRAS.447.2671Z} presented the intrinsic {\NV} absorption in two blended components components (see their Figure~3) at redshifts \(z= 0.028570\)\ and \(0.028789\). We adopt a single equivalent width-weighted average absorber redshift of \(0.0287\). With a fiducial systemic redshift of 0.030227 from SDSS DR13 \citep{Albareti2017,2018A&A...613A..51P}, we adopt an offset velocity of \(-439\)\,\kms.

\subsection{2MASX\,J$18324966+5340219$ \(\zem = 0.045\) PID: 12275}

\begin{figure}
\includegraphics[width=0.5\textwidth]{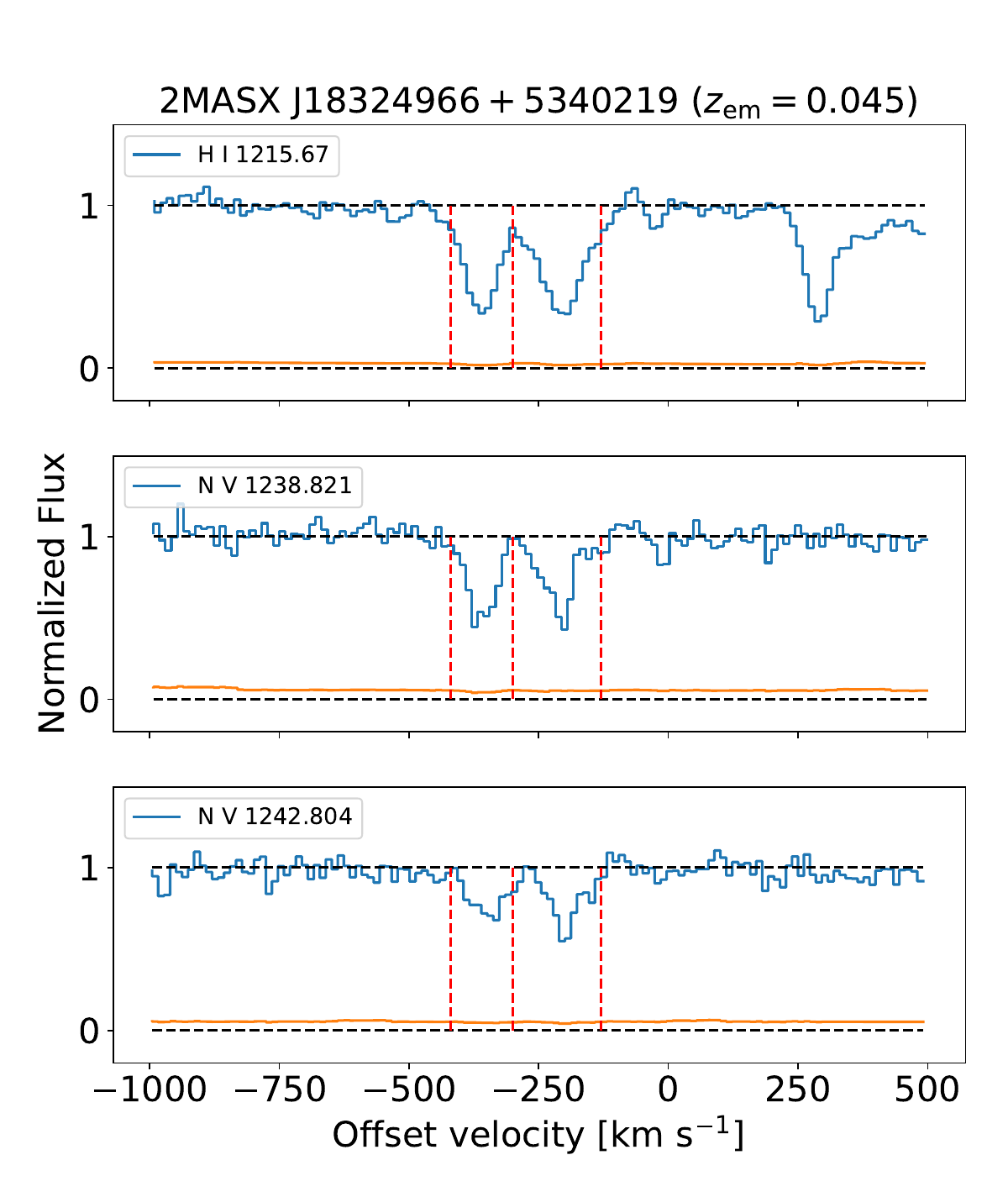}
\caption[HS 1831+5338 System Plot]{Intrinsic Absorption in 2MASX\,J$18324966+5340219$. We show a velocity-stacked plot of the detection transitions for the intrinsic absorber in the archived HST/COS spectra. In each panel, the flux density is shown as a blue histogram, and the uncertainty is shown as an orange histogram. For clarity, spectra are shown with a sampling of 5 pixels per bin. A horizontal dashed black line mark the zero-flux level. Vertical red dashed lines encompass the absorption components at \(-350\)\,\kms, and \(-200\)\,\kms.}
\label{fig:hs1831}
\end{figure}

COS observations for PID 12275 including this target have primarily focussed on intervening absorbers. In Figure~\ref{fig:hs1831}, we present velocity-stack spectra of the {\HI} \(\lambda1215\) line and the {\NV} \(\lambda\lambda1238, 1242\) doublet. Two separated components are detected at \(-350\)\,\kms and \(-200\)\,\kms. These are clearly separated and we treat them as kinematically-isolated systems.

\subsection{Mrk\,509 \(\zem = 0.034397\) PID: 12022}
This target was part of a multiwavelength observing campaign in September through December 2009. The COS observation were presented in \citet{2011A&A...534A..41K} (see their Figure~5) with several components in the velocity range \(-450\)\,\kms\ to \(+250\)\,\kms\ relative to the systemic redshift of \(\zem=0.034397\). There is a clear separation between blends of components with one collection in the range \(-450\)\,\kms\ to \(-200\)\,\kms\ and another collection in the range \(-100\)\,\kms\ to \(+250\)\,\kms. From these, we adopt equivalent-width weighted absorption system velocities of \(-315\)\,\kms\ and \(+39\)\,\kms, respectively.

\subsection{Mrk\,1513 \(\zem = 0.062977\) PID: 11524}
The outflow detected in the COS spectrum of Mrk\,1513 was previously presented/analyzed by \citet{2014MNRAS.439.3649T} who reported absorbers at \(-1521 \pm 20\)\kms, and \(-17 \pm 20\)\,\kms\ (see their Figure~9). We confirm the former and adopt a velocity for the later of \(+11\)\,\kms.

\subsection{Q\,$2135-147$ \(\zem = 0.20047\) PID: 13398}

\begin{figure}
\includegraphics[width=0.5\textwidth]{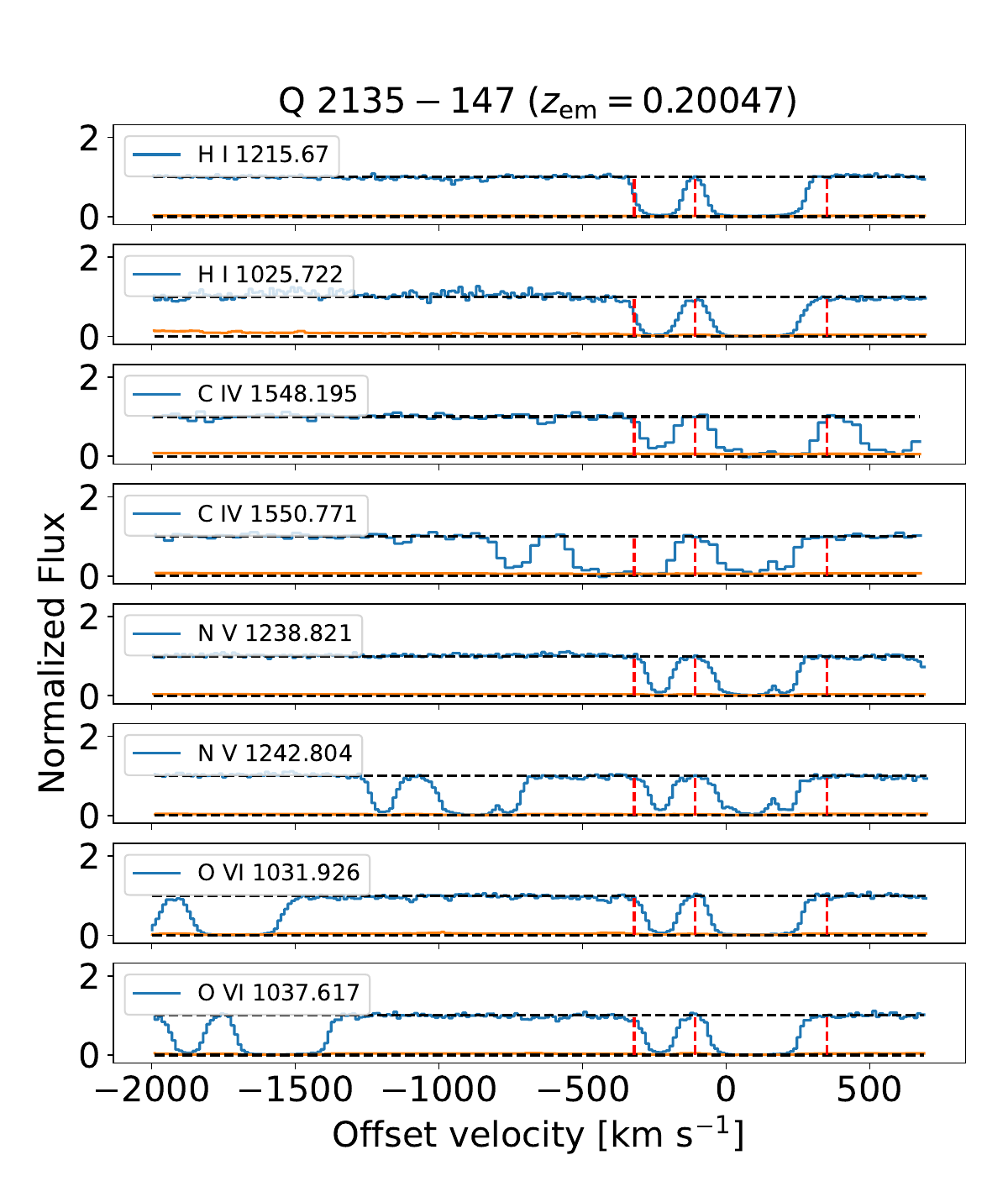}
\caption[Q\,$2135.0-147$ System Plot]{Intrinsic Absorption in Q\,$2135.0-147$. We show a velocity-stacked plot of the detection transitions for the intrinsic absorber in the archived HST/COS spectra. In each panel, the flux density is shown as a blue histogram, and the uncertainty is shown as an orange histogram. For clarity, spectra are shown with a sampling of 5 pixels per bin. A horizontal dashed black line mark the zero-flux level. Vertical red dashed lines encompass the absorption systems at \(-222\)\,\kms\ and \(+108\)\,\kms\ relative to \(\zem = 0.20047\).}
\label{fig:q2135}
\end{figure}

COS observations for PID 13398 including this target have primarily focused on intervening absorbers. In Figure~\ref{fig:q2135}, we present velocity-stack spectra of the {\HI} \(\lambda\lambda1025, 1215\) lines, the {\CIV} \(\lambda\lambda1548, 1550\) doublet, the {\NV} \(\lambda\lambda1238, 1242\) doublet, and the {\OVI} \(\lambda\lambda1031, 1037\) doublet. Several components lie in the velocity range \([-320,350]\)\,\kms, with a clear kinematic boundary at \(-109\)\,\kms. Hence, we adopt two equivalent-width-weighted average velocities of \(-222\)\,\kms\ and \(+108\)\,\kms\ relative to \(\zem = 0.20047\).

\subsection{Mrk\,304 \(\zem = 0.065762\) PID: 12604}

This target was part of the QUEST sample and the COS spectra are presented in \citet[][see their Figure~14s]{2022ApJ...926...60V}. They report five blended components in the velocity range \([-3400,-200]\)\,\kms\ relative to the systemic redshift of \(\zem = 0.066\). This is larger than the separation of the {\NV} doublet and the system blends the two transitions and is often referred to as a mini-BAL. We identify two distinct kinematic regions in the velocity ranges \([-3400,-2813\)\,\kms\ (a NAL) , and \([-2530,-200]\)\,\kms\ (the mini-BAL). These have equivalent-width-weighted velocities of \(-3108\)\,\kms\ and \(-1272\)\,\kms, respectively, relative to \(\zem = 0.065762\).

\subsection{UGC\,12163 \(\zem = 0.0247\) PID: 12212}
\citet{2002ApJ...566..187C} previously analyzed HST/FOS and STIS echelle spectra of this target, revealing absorption in {\HI}, {\NV}, {\CIV}, {\mbox{Si\,{\sc iii}-iv} in the velocity range \(-400\)\,\kms\ to \(+200\)\,\kms, with a centroid of \(-152\)\,\kms\ for {\NV} (see their Figure 3). In Fig.~\ref{fig:ugc12163}, we present a system plot of the COS spectra. We find absorption in the velocity range \(-400\)\,\kms\ to \(+175\)\,\kms, with an equivalent-width weighted average of \(-152\)\,\kms.

\begin{figure}
\includegraphics[width=0.5\textwidth]{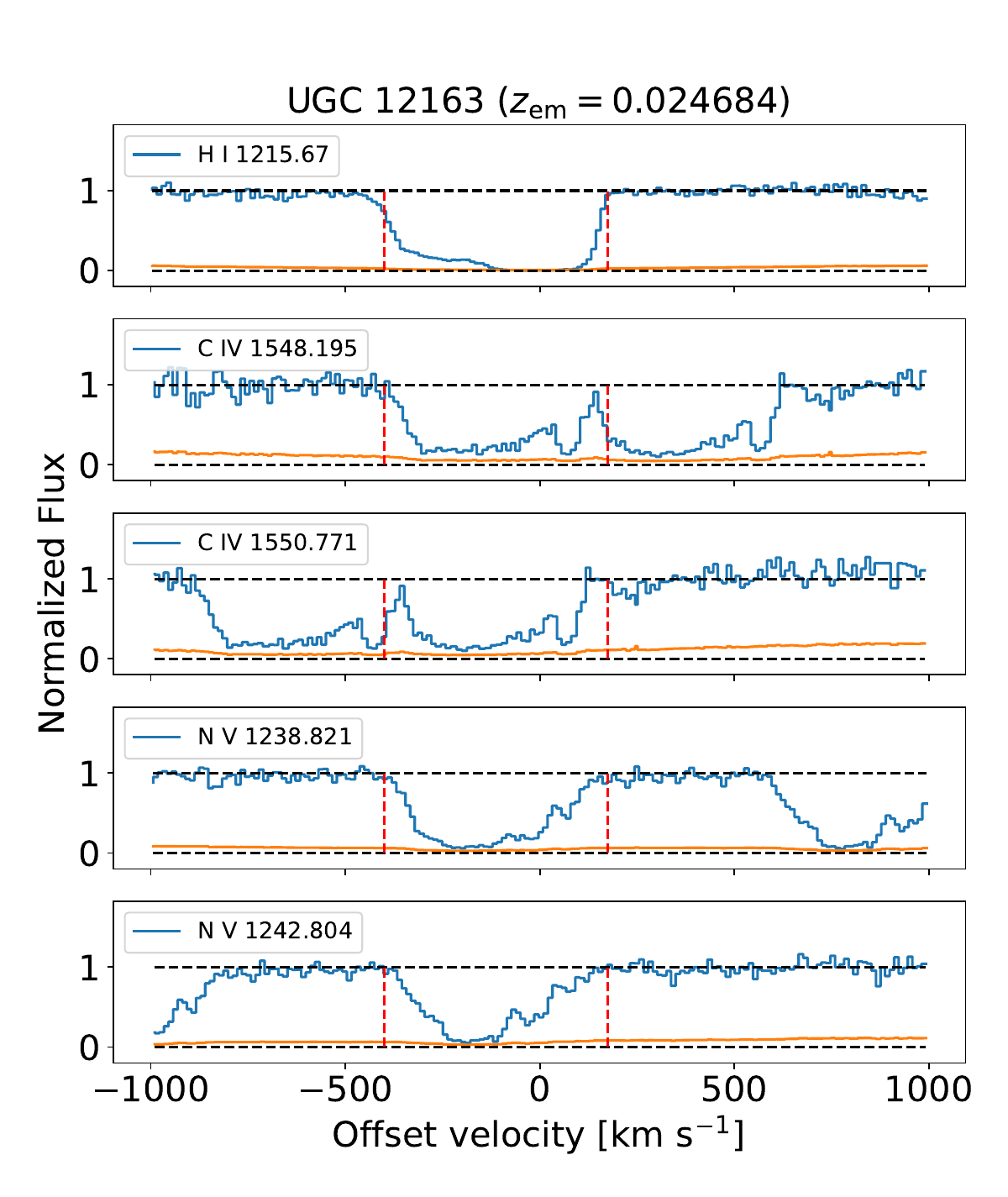}
\caption[UGC 12163 System Plot]{Intrinsic Absorption in UGC\,12163. We show a velocity-stacked plot of the detection transitions for the intrinsic absorber in the archived HST/COS spectra. In each panel, the flux density is shown as a blue histogram, and the uncertainty is shown as an orange histogram. For clarity, spectra are shown with a sampling of 5 pixels per bin. A horizontal dashed black line mark the zero-flux level. Vertical red dashed lines encompass the absorption component at \(-131\)\,\kms.}
\label{fig:ugc12163}
\end{figure}

\subsection{MR\,$2251-178$ \(\zem = 0.06398\) PID: 12029}
MR\,$2251-178$\ was part of a multiwavelength observing campaign in December 2020 and the results/analysis are presented in \citet{2022ApJ...940...41M}. MR\,$2251-178$\ famously has an X-ray warm absorber that is clearly seen in variable UV absorption \citep{2004ApJ...611...68K,2001ApJ...556L...7G}. We present the velocity-stacked profiles of the {\HI} \(\lambda12150\) line, the {\CIV} \(\lambda\lambda1548, 1550\) doublet, and the {\NV} \(\lambda\lambda1238, 1242\) doublet in Figure~\ref{fig:mr2251}. We adopt a velocity of \(-270\)\,\kms\ for the system of warm absorber components.

\begin{figure}
\includegraphics[width=0.5\textwidth]{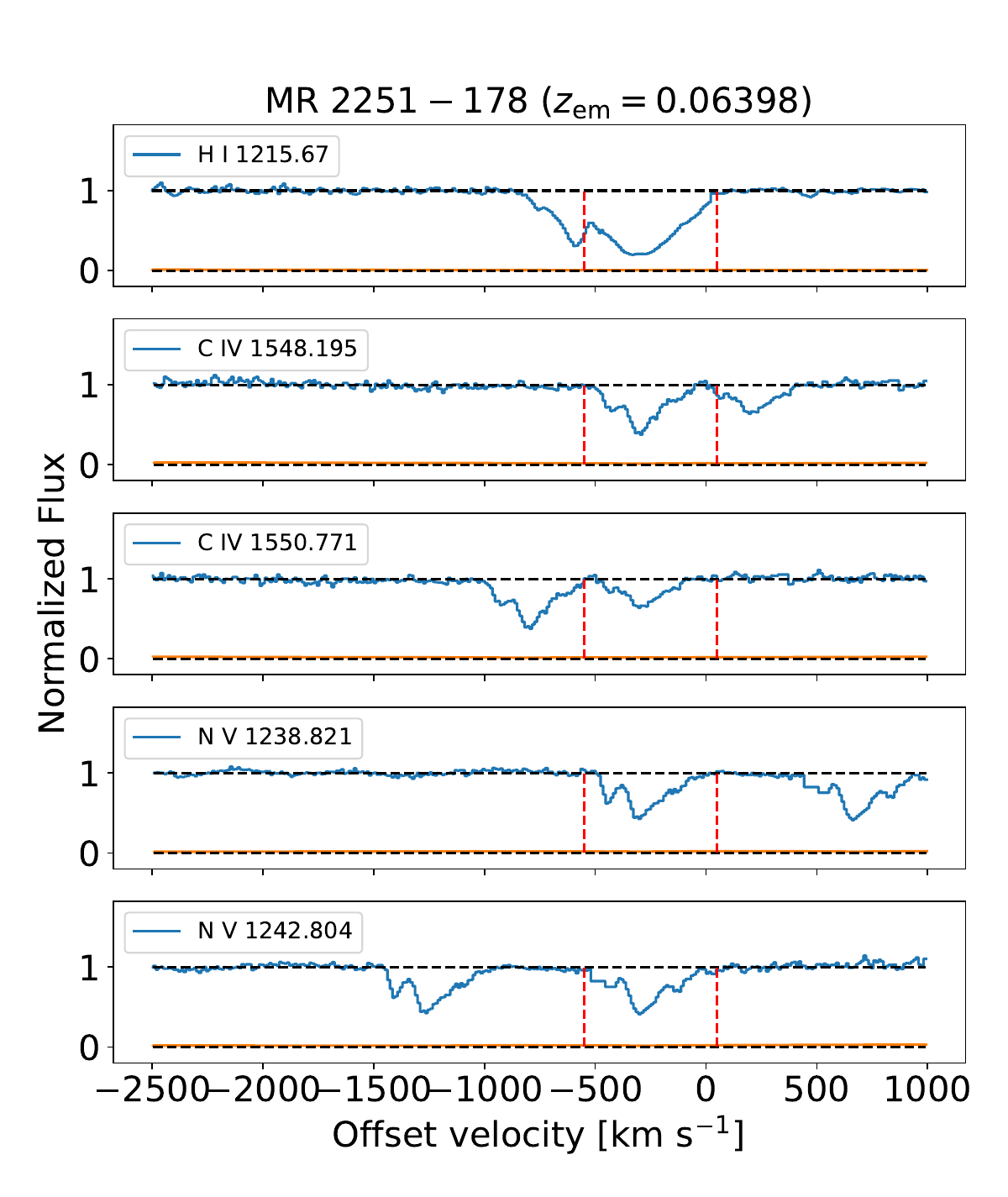}
\caption[MR\,$2251-178$ System Plot]{Intrinsic Absorption in MR\,$2251-178$. We show a velocity-stacked plot of the detection transitions for the intrinsic absorber in the archived HST/COS spectra. In each panel, the flux density is shown as a blue histogram, and the uncertainty is shown as an orange histogram. For clarity, spectra are shown with a sampling of 5 pixels per bin. A horizontal dashed black line mark the zero-flux level. Vertical red dashed lines encompass the absorption system at \(-270\)\,\kms.}
\label{fig:mr2251}
\end{figure}

\subsection{6dF\,J$2301520-55083$ \(\zem = 0.141\) PID: 13444}

The COS observations for this target have previously been used in studies of the Galactic and circum-Galactic media. Two clearly separated components are observed in the intrinsic absorption from {\HI} \(\lambda\lambda1215, 1025, 1215\) lines, the {\NV} \(\lambda\lambda1238, 1242\) doublet, and the {\OVI} \(\lambda\lambda1031, 1037\) doublet (Fig.~\ref{fig:he2258}) in the velocity range \([-800,-380]\)\,\kms. We adopt equivalent-width-weighted velocities of \(-672\)\,\kms\ and \(-487\)\,\kms\ relative to \(\zem = 0.141\).

\begin{figure}
\includegraphics[width=0.5\textwidth]{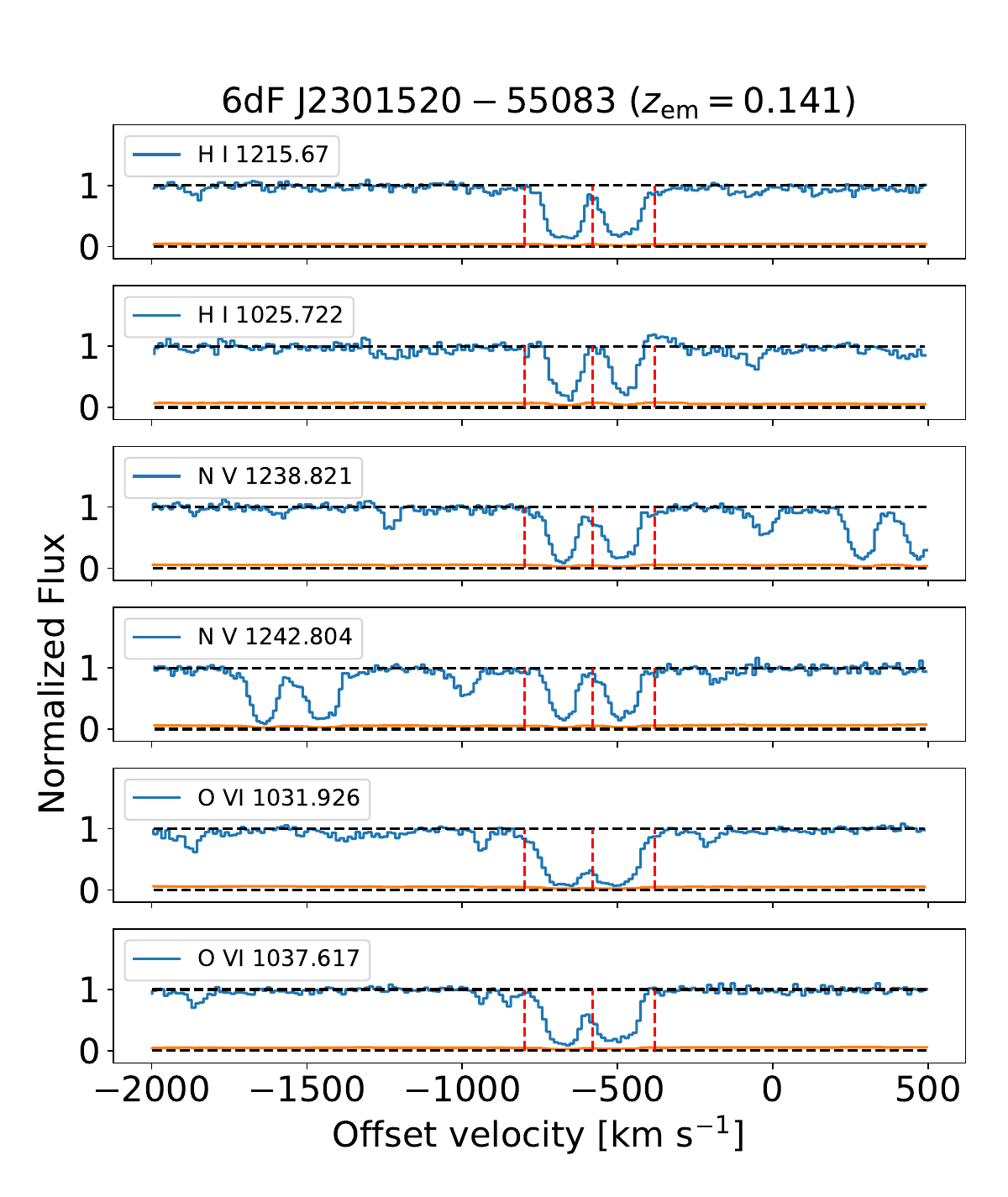}
\caption[6dF\,J$2301520-55083$ System Plot]{Intrinsic Absorption in 6dF\,J$2301520-55083$. We show a velocity-stacked plot of the detection transitions for the intrinsic absorber in the archived HST/COS spectra. In each panel, the flux density is shown as a blue histogram, and the uncertainty is shown as an orange histogram. For clarity, spectra are shown with a sampling of 5 pixels per bin. A horizontal dashed black line mark the zero-flux level. Vertical red dashed lines encompass the absorption systems at \(-672\)\,\kms\ and \(-487\)\,\kms\ relative to \(\zem = 0.141\).}
\label{fig:he2258}
\end{figure}

\subsection{NGC\,7469 \(\zem = 0.0163\) PID: 12212}
\citet{2020A&A...633A..61A} present a detailed analysis of a deep multiwavelength campaign including the COS spectrum. The report three absorption components in the velocity range \([-2070,-410]\)\,\kms\ (see their Figures 1 \& 2, and Table 2). While the components are distinct, we adopt a system velocity at the equivalent-width weighted average of \(-1926\)\,\kms.

\subsection{2MASS\,J$23215113-7026441$  \(\zem = 0.3\) PID: 12936}

COS observation of this object were taken for the purposes of studying Galactic and circum-Galactic gas. Much of the {\HI} \(\lambda1215\) broad emission line and the putative location of associated absorption fall in the detector gap. So, we turn to alternative transitions, specifically {\HI} \(\lambda1025\) and the {\OVI} \(\lambda\lambda1031,1037\) doublet, for corroboration. Furthermore, while the redshift precision if quite poor, blends of detected emission lines prevent more precise determinations. 

In Figure~\ref{fig:q2321}, we present velocity-stacked spectra of the {\HI} \(\lambda1025\)  line and the {\OVI} \(\lambda\lambda1031,1037\) and {\NV} \(\lambda\lambda1238, 1242\) doublets. Two blended components are evident from the {\OVI} \(\lambda\lambda1031,1037\) in the velocity range \([175,380]\)\,\kms. One component of the {\OVI} \(\lambda1037\)\ line is blended with unrelated absorption. In {\NV}, there is excellent alignment between the two doublet transitions as well as the stronger {\OVI} component. The {\HI} \(\lambda1025\)\ in the aforementioned velocity range appears consistent with a blend of the two components. Outside that velocity range, it is blended with Galactic {\CII} \(\lambda1334\). We adopt an equivalent-width-weighted average velocity of \(307\)\,\kms\ relative to \(\zem = 0.3\).

\begin{figure}
\includegraphics[width=0.5\textwidth]{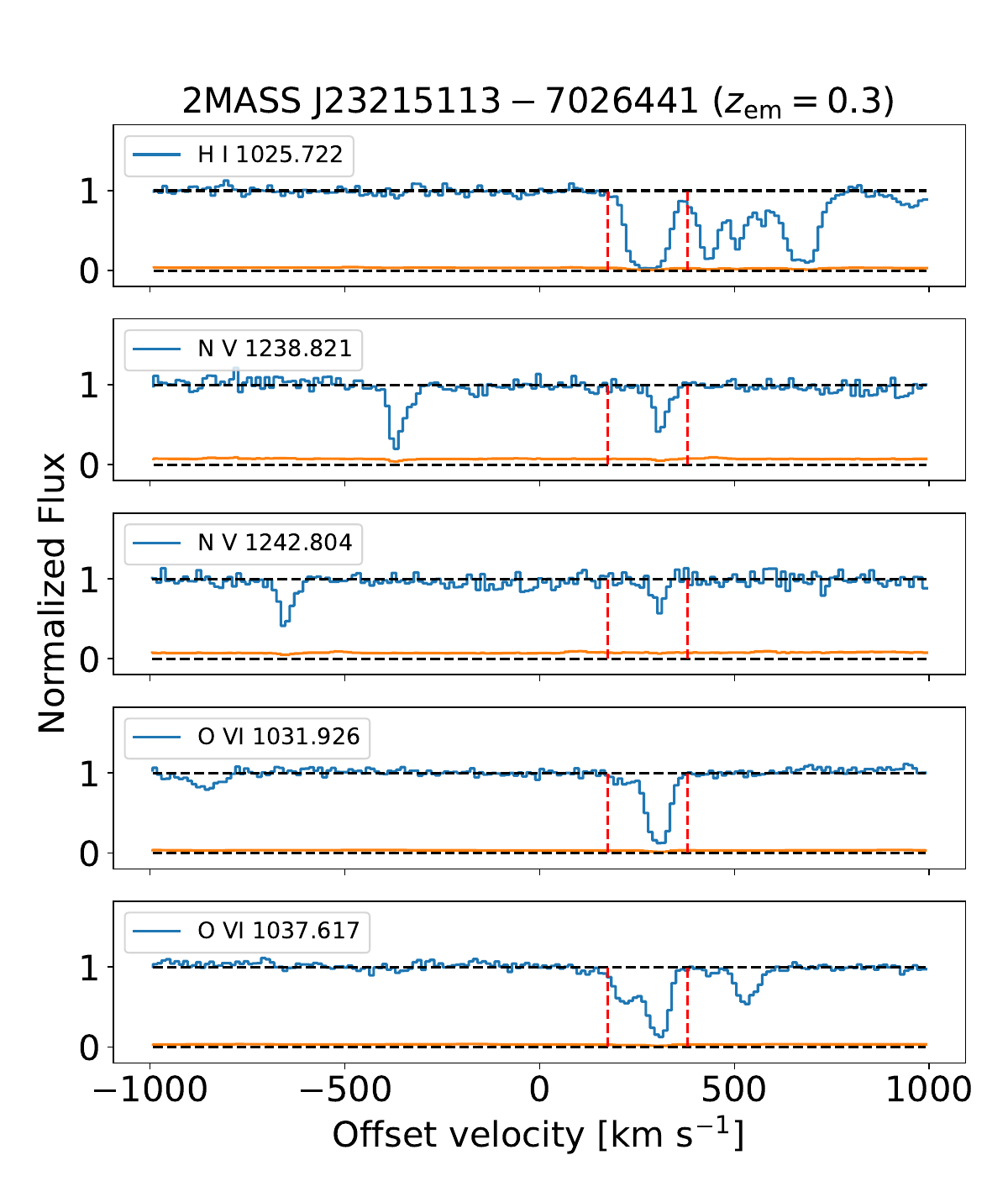}
\caption[2MASS J23215113 System Plot]{Intrinsic Absorption in 2MASS\,J$23215113-7026441$. We show a velocity-stacked plot of the detection transitions for the intrinsic absorber in the archived HST/COS spectra. (The {\HI} \(\lambda1215\) transition falls in a detector gap, so no data were recorded.) In each panel, the flux density is shown as a blue histogram, and the uncertainty is shown as an orange histogram. For clarity, spectra are shown with a sampling of 5 pixels per bin. A horizontal dashed black line mark the zero-flux level. Vertical red dashed lines encompass the absorption component at \(307\)\,\kms\ relative to \(\zem = 0.3\).}
\label{fig:q2321}
\end{figure}


\bsp	
\label{lastpage}
\end{document}